\documentclass[a4paper,11pt]{article}
\pdfoutput=1 

\usepackage{jcappub} 

\usepackage[T1]{fontenc} 
\usepackage[normalem]{ulem}

\title{Detectability of Massive Boson Stars using Gravitational Waves from Fundamental Oscillations}

\author[a,1]{Swarnim Shirke,\note{Corresponding author.}}
\author[a]{Bikram Keshari Pradhan}
\author[a]{Debarati Chatterjee}
\author[b]{Laura Sagunski}
\author[b]{J\"urgen Schaffner-Bielich}

\affiliation[a]{Inter-University Centre for Astronomy and Astrophysics, \\
Post Bag 4, Ganeshkhind, Pune University Campus, Pune - 411007, India}
\affiliation[b]{Institut f\"ur Theoretische Physik,Goethe Universit\"at, \\
Max von Laue Str. 1, 60438 Frankfurt am Main, Germany}

\emailAdd{swarnim@iucaa.in}

\abstract{
Boson Stars are macroscopic self-gravitating configurations made of complex scalar fields. These exotic compact objects would manifest as dark Boson stars and, in the absence of electromagnetic signatures, could mimic properties of compact stars in the gravitational wave spectrum. In a recent study, using the simplest potential for massive Boson stars, we demonstrated that fundamental non-radial oscillations ($f$-modes) obey scaling relations that allow them to be distinguished from neutron stars and black holes. In this work, we provide analytical fits for these scaling relations, valid for the dark matter parameter space compatible with current astrophysical and cosmological data, that can be directly incorporated into future studies of massive Boson stars in the strong coupling regime, avoiding the need for numerical calculations. We also provide analytical fits for empirical and universal relations for gravitational wave asteroseismology, which can be used to infer microscopic dark matter properties following a successful detection. Further, we investigate the possibility of detection of $f$-modes and the dark matter parameter space that can be probed with current and future gravitational wave detectors across multiple frequency bands.
Assuming a burst GW model and demanding a signal-to-noise ratio of 5, we show that the current and future detectors can, in principle, probe Boson star $f$-modes up to cosmological distances: 1 Mpc with aLIGO, 30 Mpc with Cosmic Explorer and Einstein Telescope, and in the best case scenario, about 300 Mpc with LISA.
}

\begin{document}

\maketitle
\flushbottom

\section{Introduction}
\label{sec:intro}

Boson stars (BSs) are stellar configurations of bosonic particles (see~\cite{Visinelli2021, LieblingPalenzuela2023} for an updated review on BSs). The formation channels for BSs include primordial density fluctuations in the scalar field~\cite{LiddleMadsen1992}, gravothermal collapse in an early matter-dominated era of the universe~\cite{Ralegankar2024}, or a process known as \textit{gravitational cooling}~\cite{Seidel1994}. These are some of the possible exotic compact objects (ECOs) that can mimic other compact objects like neutron stars (NSs) and black holes (BHs), acting as BH mimickers, as well as contribute to the total dark matter (DM) budget (see~\cite{CardosoPani2019} for a review on ECOs).
Various types of BSs have been studied in the literature~\cite{Wheeler1955, Feinblum1968, Kaup1968, Ruffini1969, Mielke1981, Colpi1986}. Based on the interaction potential, they are classified into different types of BSs like mini-BS (no interaction), massive BS, axion star, solitonic star, etc. In this work, we will focus on massive BSs, with the simplest form of potential consisting of quartic self-interaction capable of forming massive compact objects as shown in~\cite{Colpi1986}.

Being a candidate for dark matter, these particles are assumed to couple sufficiently weakly to electromagnetic (EM) photons and other Standard Model particles. This means that non-accreting BSs are not observable via conventional EM telescopes. Further, massive BSs are compact with compactness $C \lesssim 0.16$ and can have a wide range of mass depending on the model parameters, overlapping the range of BH mass. This makes them a good candidate for BH mimickers. In the case of accreting systems, the BSs can be distinguished using the EM observations based on the analysis of the shadow region~\cite{Olivares2020, Rosa2022, Rosa2023} and line broadening~\cite{Rosa2024}.
These methods are inaccessible in the case of non-accreting BSs. However, the successful detection of gravitational waves (GWs)~\cite{GWDetection2016} has opened up a new window to probe the dark sectors of the universe. There have been many works and searches for BSs using GWs in the last few decades~\cite{Palenzuela2007, Palenzuela2008, Cardoso2016, Palenzuela2017,Urbano2019,Abbott2020,Sakstein2020,Bustillo2021,Capano2023,Cipriani2024,Evstafyeva2024}. 




Quasinormal oscillation modes of stars are important in the context of binary pre- and post-merger as well as isolated system excitations. \cite{Kojima1991} was the first work to explore the quasinormal modes (QNMs) of BSs. There have been a few studies on BS QNMs following that~\cite{Yoshida1994, Balakrishna1998, Macedo2013a, Flores2019, Kain2021, Celato2025}. 
$f$-modes, the main motivation of this work, are fundamental (zero nodes) non-radial oscillation modes of a star. Typically, oscillation modes of compact stars like NSs are classified based on the restoring force. In the case of $f$-modes, the restoring force is the fluid pressure. Massive BSs in the strong interaction limit have an equation of state (EoS) equivalent to a polytropic fluid, as will be shown in Sec.~\ref{sec:model}. This allows us to study the BS $f$-modes, where the pressure is sourced by the repulsive self-interaction. 

~\cite{Flores2019} is the first work to study the $f$-modes for massive BSs in the strong-interaction limit. 
Another work~\cite{Celato2025} appeared during the completion of this work that explored BS $f$-modes. However, these studies were restricted to select model parameters. In a separate publication~\cite{shirke_BS_fmodes_short}, 
we showed for the first time that scaling relations are obeyed by the $f$-mode frequency and damping time in the strong-interaction limit. This scaling result is significant, as it allows us to comment on $f$-mode characteristics of configurations of arbitrary parameters within the strong-interaction limit ($\Lambda \gg 1$; defined in Eq.~\ref{eqn:Lambda}) and can be directly used for studies of massive BSs. 
In this work, we apply these scaling relations independent of microscopic parameters to obtain analytical fits connecting macroscopic structure properties and $f$-mode characteristics of massive BSs, avoiding the need for numerical calculations and reducing computation costs. We also provide universal relations that can be used to infer stellar properties from GW observations and investigate the prospects of detecting GWs from $f$-modes from massive BSs using the current and future detectors.

The article is organized as follows: In Sec.~\ref{sec:model} we provide the details of the BS model, describing the scaling property of massive BS. We discuss the available scalar DM parameter space in Sec.~\ref{sec:parameter_space}, reviewing all the available constraints. We present all the results in Sec~\ref{sec:results}, discussing the static observables, $f$-modes, and the corresponding universal relations. We also comment on the parameter space observable using GWs from $f$-modes from massive BSs and discuss the detectability by various GW detectors. We summarize our results in Sec.~\ref{sec:discussions}, providing a comparison with earlier works and outlining the future scope. We use the convention $\hbar = c =1$ and hence $G=1/M_{\rm Pl}^{2}$ throughout this work. 

\section{Model}\label{sec:model}
Here, we provide details of the model used to describe Bosonic DM that will be used to study BS $f$-modes in this work. We also show the scaling relations used.

We describe DM by a scalar field self-interacting via quartic self-interactions as elaborated in \cite{Colpi1986,Flores2019}. The interaction strength is given by $\lambda$.
\begin{equation}
    \mathcal{L} = \frac{1}{2}\partial_{\mu}\phi^*\partial^{\mu}\phi + V(\phi)
\end{equation}
\begin{equation}
    V(|\phi|) = \frac{1}{2}m^2|\phi|^2 + \frac{1}{4}\lambda|\phi|^4
\end{equation}

Here $\lambda$ is the self-interaction coupling and $m$ is the mass of the DM particle. A dimensionless parameter is defined as
\begin{equation}\label{eqn:Lambda}
    \Lambda =\frac{\lambda M_{\rm Pl}^2}{4\pi m^2}~.
\end{equation}

For $\Lambda \gg 1$,~\cite{Colpi1986} showed that we could form massive stars whose mass resembles that of a degenerate fermionic star. In this case, $M_{\rm max} = 0.22 \sqrt{\Lambda}M_{\rm Pl}^2/m = 0.06\sqrt{\lambda}M_{\rm Pl}^3/m^2$~\cite{Colpi1986}.
The macroscopic properties are dictated by the effective factor defined as~\cite{Cipriani2024}
\begin{equation}
    \boxed{x \equiv \sqrt{\lambda}/m^2}~.
\end{equation}

Using this, we define the dimensionless mass ($M'$) and dimensionless radius ($R'$) as $M' = M/(xM_{\rm Pl}^3)~,  R' = R/(xM_{\rm Pl})$. The dimensionless energy density and pressure are scaled as $\rho' = \rho x^2~, p' = px^2$. The dimensionless compactness $C=GM/R = M/(M_{\rm Pl}^2R) = M'/R'=C'$ remains unaltered. Thus, $C=C'$ can be used interchangeably.

We consider here this case of self-interacting massive BSs with $\Lambda \gg 1$, for which the scalings hold. We restrict our analysis to $\Lambda > 1000$, referred to as the strong-interaction limit~\cite{Seoane2010}. All the results in this work are applicable only in this strong-interaction regime. This has been marked by the yellow arrow in Fig.~\ref{fig:parameter_space}. In this limit, the Einstein-Klein-Gordon system of equations for the scalar field resembles that of a perfect fluid star with the effective equation of state given by~\cite{Colpi1986, Karkevandi2022, Cipriani2024},
\begin{equation}\label{eqn:eos}
    p' = \frac{1}{9}(\sqrt{1+3\rho'}-1)^2~.
\end{equation}

For $\rho' \ll 1$, $p'=\rho'^2/4$, i.e., we get a polytrope EoS with polytropic index $n=1$. For $\rho' \gg 1$, we get an ultra-relativistic EoS $p'=\rho'/3$.

The TOV equations, when written in terms of scaled mass ($M'$) and radius ($R'$), are known to follow self-similarity, i.e., the equations become independent of model parameter $x$ (see appendix of Maselli et al.~\cite{Maselli2017}). This results in a unique mass-radius relation $M'-R'$, which can be used to obtain the $M-R$ curve for any parameters using the scaling relations. The shape of the solution remains the same. Upon solving these scaled TOV equations, we find that the maximum central density reached for a maximum TOV mass configuration is $\rho'=1.6$. Thus, $\rho' \gg 1$ condition is not reached inside BSs.
We did not find the effect of scaling on the Love equations and on the moment of inertia in the literature. Hence, we show it here in Appendix.~\ref{sec:appendix_love_equations} and~\ref{sec:appendix_I_equations}.

The non-radial fundamental quasi-normal mode ($f$-mode) serves as a primary source of GW emission in compact stars. The relativistic Cowling approximation is often used to find mode frequency by neglecting metric perturbation; however, it neglects metric perturbation and overestimates the frequency as compared to that obtained within the framework of a general relativistic treatment.

In this study, the mode parameters are obtained by solving perturbations within the framework of linearized general relativistic treatment employing a direct integration method, as outlined in previous works~\cite{Detweiler85,Sotani2001,Pradhan2022}, to solve the $f$-mode frequency of BSs. Numerical methods developed in a previous work~\cite{Pradhan2022} are employed for extracting the mode characteristics.

For the $f$-mode frequencies and damping times, we solve the $f$-mode eigenfrequencies using full-GR formalism. The details of the calculation of BS $f$-modes have been outlined in our other work~\cite{shirke_BS_fmodes_short}. There it was shown that if the $f$-mode frequency $f$ and damping time $\tau$ are scaled as:
\begin{align}
 f &= f'/(xM_{\rm Pl}) \\
 \tau &= \tau'(xM_{\rm Pl})~,
\end{align}
Then the $f$-mode equations, too, like the TOV equations, exhibit self-similarity, i.e., become independent of model parameters. The formalism and derivation of scaling have been provided in detail in the Appendix of~\cite{shirke_BS_fmodes_short}.

\section{Parameter Space}\label{sec:parameter_space}
\begin{figure*}[t]
    \centering
    \includegraphics[width=0.98\linewidth]{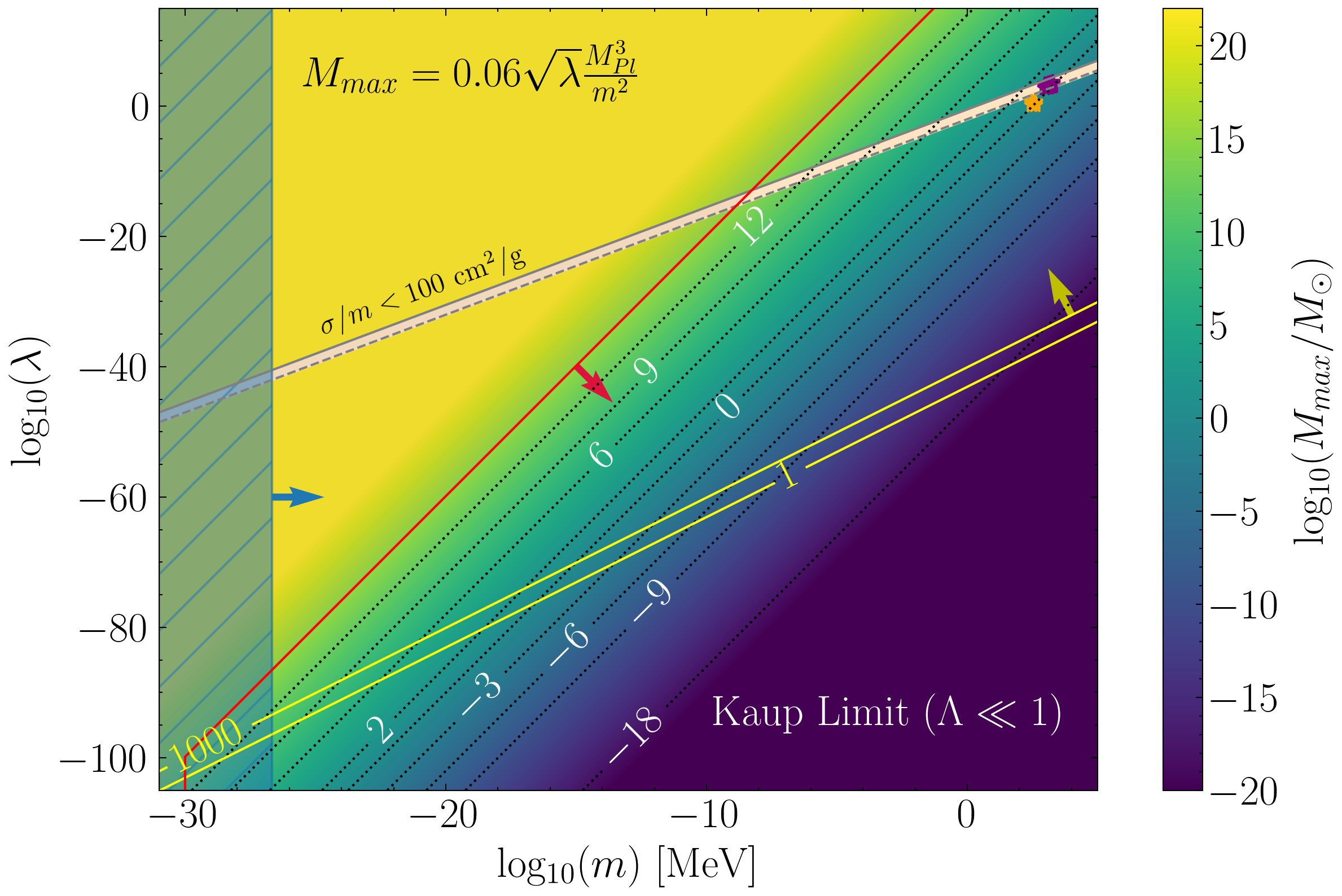}
    \caption{Figure shows the available parameter space in the $\lambda-m$ space. The color indicates the logarithm of the maximum mass of a stable BS for a given pair of ($\lambda$,m). We show select contour lines (black) to get an idea of masses: $\log_{10}M =$ -18 (mountain mass; this is also the lower bound on primordial BHs set by Hawking radiation as lighter PBHs would evaporate within Hubble time); -9 (asteroid belt mass); -6 (Earth-like planet mass); -3 (giant planet mass); 0 (stellar mass); 2 (giant massive star mass); 6 and 9 (SMBH mass); 12 (galaxy mass). 
    The yellow contours represent values of $\Lambda$. We restrict to the strong-interaction limit $\Lambda > 1000$ in this work. All the results in this work, including the colorbar in this figure, are applicable only to this region.  The orange and purple patch denotes the parameters used in earlier works on massive BSs in the strong-interaction regime~\cite{Maselli2017,Flores2019,Celato2025}. See text for details on various constraints.
    }
    \label{fig:parameter_space}
\end{figure*}

Fig.~\ref{fig:parameter_space} shows the available parameter space in the $\lambda-m$ space for scalar DM. For mass, we show the range $10^{-25} \text{eV} < m < 100 \text{GeV}$ covering the mass of fuzzy DM or ultralight ALPs to roughly the mass of WIMPs. We consider $\lambda$ upto $\lambda=15$. This encompasses the upper limit of $\sim 4\pi$ as imposed by~\cite{Eby2016}. All the constraints are discussed in detail in the next section. The values $\lambda \lesssim 10^{-100}$ are irrelevant as lower values consistent with observations do not fall in the strong-interaction limit considered here. The yellow lines denote contours of fixed $\Lambda$, and this analysis is only applicable to $\Lambda > 1000$, i.e., the strong-interaction regime marked by the yellow arrow. The region where $\Lambda \ll 1$ is the Kaup limit, and interactions can be ignored. This is shown just for depiction purposes. The results are not applicable in general below  $\Lambda =1000$. The gray lines show contours where $\sigma/m$ (interaction cross-section per unit mass; see Eq.~\ref{eqn:cross_section}) is constant, and the bisque region between them is the region permitted by other astrophysical observations. We note that this is a very narrow region in the parameter space; however, the lower bound (dashed gray line) is not strict. The red line marks the constraint from CMB and LSS data~\cite{Cembranos2018}. The region above this line is excluded. The blue region is excluded by the recently set lower limit in scalar DM mass~\cite{Zimmerman2024} using dwarf galaxies. 

The maximum mass of the Bosonic dark star is shown in colour. This can be calculated for a given set of parameters $(\lambda, m)$ is given by \cite{Colpi1986}: $M_{\rm max} = 0.22\sqrt{\Lambda}\frac{M_{\rm Pl}^2}{m} = 0.06\frac{\sqrt{\lambda}}{m^2}M_{\rm Pl}^3 = 0.06xM_{\rm Pl}^3$. The colour scale saturates into yellow at $M_{\rm max} =20^{22} M_{\odot}$, which is the rough estimate for the observable universe's mass. We plot the contours of some fixed mass values. We find that the $M_{\rm max}$ ranges a larger region in mass spanning mountain mass to galactic mass (see caption of Fig.~\ref{fig:parameter_space}). The arrows indicate the regions allowed by various constraints. The only region bounded by the gray ($\sigma/m$), red (CMB, LSS), and blue (dwarf galaxies) lines is the allowed region for scalar DM and, hence, BSs. Further, this work is only applicable in the region bounded by the yellow $\Lambda =1000$ (strong-interaction regime) line.

Flores et al.~\cite{Flores2019} restricted their study to stellar mass BSs and used selected values of $m \in \{1,1.25,1.5,1.75,2\}m_n$ and $a \in \{5,10,15,20\}$ fm to study the $f$-modes. The scattering length parameter $a = \lambda/(8\pi m) $ was used to fix the interaction strength instead of $\lambda$ in~\cite{Flores2019}. These are marked by a purple patch in Fig.~\ref{fig:parameter_space}. Maselli et al.~\cite{Maselli2017} studied the $I$-Love-$Q$ relations using a select set of model values. The parameters chosen were $m \in \{300,400\}$ MeV and $\lambda \in \{0,0.5,1.5\}\pi$. Celato et al.~\cite{Celato2025} explored the $f$-mode universal relations for the same parameters.~ These are marked in orange in Fig.~\ref{fig:parameter_space}. It is clear that both these patches lie near the $M_{\rm max} \sim 1 M_{\odot}$ contour and focus on the configurations in the NS mass range. \\

\subsection{Constraints}\label{sec:constraints}
There are various independent constraints on the $\lambda$ and $m$ parameters from various astrophysical observations. We provide a brief review of these constraints below.\\
 \textit{Cross-section:}  
 The self-interaction cross-section ($\sigma$) between the DM particles is related to the $\lambda$ parameter of this model by~\cite{Eby2016} 
 \begin{equation}\label{eqn:cross_section}
     \sigma = \frac{\lambda^2}{64\pi m^2}~.
 \end{equation}
 The small-scale problems in the $\Lambda$CDM model and cold DM-only simulations, like the i) cusp-core problem, ii) missing satellite problem, iii) too big to fail problem, are known to be resolved by allowing for self-interacting DM (SIDM)~\cite{Tulin2018}. These studies demand that the interaction strength be satisfied within a particular range in order to resolve the problems. The bounds on $\sigma/m$ are given by $0.1 < \sigma/m < 100$ cm$^2$/g. The upper limit is usually known to be around 10cm$^2$/g, however, we consider here the least constraining case, allowing for more flexibility and the largest possible parameter space. Thus, we can translate this constraint to $\lambda$ and $m$ as
\begin{align}
    0.1 <  &\frac{\lambda^2}{64\pi m^3} <100 \text{ [cm$^2$/g]}
\end{align}
The upper limits come primarily from observations of merging clusters (for e.g., Bullet Cluster), whereas the lower bounds arise from the assumption that self-interactions alone are responsible for the resolution of the aforementioned problems encountered in CDM-only simulations (see~\cite{Tulin2018} for a comprehensive list of all constraints). There are alternative solutions to these problems, the most obvious one being the inclusion of baryonic feedback at small scales. Thus, the lower bound is not strict, and hence, we denote it by a dashed line in Fig.~\ref{fig:parameter_space}. Only parameters consistent with these constraints were considered in~\cite{Flores2019}. However, we see that a large range of parameter space is allowed, and stellar mass BSs can be produced even when the lower bound is removed.

Seoane et al.~\cite{Seoane2010} added constraints on the parameter space, assuming the object at the centre of our Milky Way is a BS (acting as a BH mimicker). This gave $3.7\times10^4 \le m/\lambda^{1/4} \le 2.9\times10^5$ [eV]. A similar analysis for NGC 4258 gave $6.3 \le m/\lambda^{1/4} \le 9.6\times10^4$ [eV]. However, we do not make such an assumption while exploring the parameter space. Also, it is now established that Sagittarius A* is a supermassive BH (SMBH)~\cite{SagittariusEHT2022}, which is also true for the M87 galactic centre~\cite{M87EHT2019}. Arbey et al.~\cite{Arbey2003} fitted scalar DM to galaxy rotation curves of dwarf spiral galaxies and obtained $50 <m^4/\lambda < 75$ [eV$^4$]. This corresponds to $M_{\rm max}$ of $\mathcal{O}(10^{16}) M_{\odot}$, which is too large to be considered for our study of compact BSs. Eby et al.~\cite{Eby2016} impose the constraint of $\lambda \lesssim 4\pi$ for perturbative reasons, which, when combined with $\sigma/m$ constraint, allow DM of mass of only below 100 MeV approximately. We do not consider this condition here.

\textit{Mass:} Fuzzy DM can be described by ultra-light ALPs roughly in the mass range $m \in [10^{-25}, 10^{-17}]$ eV. Planck Collaboration put a lower bound on the DM mass of fuzzy DM as $m \gtrsim 10^{-24}$ eV using the CMB anisotropies~\cite{Hlozek2015}. This has been updated to $m \gtrsim 10^{-23}$ eV by combining these results with Dark Energy Survey- Year 1 data\cite{Dentler2022} for weak gravitational lensing and to $m \gtrsim 2 \times 10^{-20}$ eV using the Lyman-alpha forest~\cite{Rogers2021}. More bounds were added from the study of the number of satellite galaxies for Milky Way as $m \gtrsim 2.9 \times 10^{-21}$ eV~\cite{Nadler2021} and $m \gtrsim 2.5 \times 10^{-22}$ eV from the study of UV luminosities of high-redshift galaxies using Hubble and James-Webb telescopes~\cite{Winch2024}. These bounds on Bosonic DM assume axion potentials, which mimic a massive BS potential for weak coupling. 

For the quartic potential case,~\cite{Cembranos2018} derived a lower bound of $m \gtrsim 10^{-24}$ eV using CMB and large-scale structure data of Planck Collaboration~\cite{PlanckCollaboration2016} and WiggleZ Dark Energy Survey~\cite{Parkinson2012}. There is a corresponding upper bound on $\lambda$ given as $\lambda/m^4 < 10^{20.14}$ [MeV$^{-4}$]. This corresponds to $M_{\rm max} \sim \mathcal{O}(10^{15}) M_{\odot}$, which is much larger than the masses we are interested in for BSs. We still show these by a red line in Fig.~\ref{fig:parameter_space}. This also corresponds to an upper bound in $x$ given by $x < 10^{10.07}$. Since we consider a quartic potential here, this is the most relevant constraint. These constraints were used to study the admixture of Bosonic DM in NSs in~\cite{Rezaei2023}. 

The bounds discussed so far are model-dependent. Many independent bounds for DM come from studies of fuzzy DM that consider non-interacting scalar particles. Various constraints on DM mass come from observations of dwarf Milky Way satellite galaxies~\cite{Hayashi2021, Zimmerman2024, Teodori2025, Benito2025}, their number~\cite{Nadler2021}, dynamical heating~\cite{Marsh2019, Dalal2022}, stellar streams~\cite{Benito2020}, lensed radio jet~\cite{Powell2023}, lensed quasars~\cite{Laroche2022}, 21-cm observations~\cite{Nebrin2019}, high redshift galaxies~\cite{Kulkarni2022}, SMBH superradiance~\cite{Stott2018, Davoudiasl2019} and the recently discovered stochastic GWs~\cite{Afzal2023Nanograv, Smarra2023} using the International Pulsar Timing Array (IPTA). Each of these methods involves different assumptions and systematics, providing independent and complementary constraints on DM mass. Out of these, Zimmerman et al.~\cite{Zimmerman2024} is a robust bound of $m \gtrsim 2.2 \times 10^{-21}$ eV using dwarf galaxies, which does not depend on the assumed cosmology, microphysics, or dynamics of dark matter. The only assumption is that the DM is composed of a single scalar degree of freedom. This bound was recently improved further to $m \gtrsim 5 \times 10^{-21}$ eV by Teodori et al.~\cite{Teodori2025}, taking into account the dynamics as opposed to using only kinematic information as in the case of the former. However, the former still remains model-independent and robust. The strongest constraints are obtained from the study of dynamical heating of stellar orbits in ultra-faint dwarf galaxies~\cite{Dalal2022}, which imposes $m > 3 \times 10^{-19}$ eV. Note that all these bounds do not consider any self-interactions. We consider the most robust and model-independent one here~\cite{Zimmerman2024} only for the purpose of depiction. This bound is shown in Fig.~\ref{fig:parameter_space} as a blue-hatched exclusion region.

\section{Results}\label{sec:results}

In this section, we first present the solution for the scaled static variables $M'$, $R'$, and $C$. We then focus on the $f$-mode characteristics and their universal relations with various macroscopic observables. Finally, we study the DM parameter space that is sensitive to various gravitational wave detectors and discuss the prospects of detectability.

\subsection{Scaled Relations for static observables}

We first discuss the scaling of the mass-radius relations along with compactness. Although these have been studied numerically before, we provide analytical fits to these curves, which we did not find in any previous literature. Fit relations would enable switching between the quantities $M$, $R$, and $C$ using simple analytic expressions and can be directly used in future studies on massive BSs. These solutions are obtained by considering the scaled EoS (see Eq.~\ref{eqn:eos}) and solving the scaling property of the TOV equation (see appendix of~\cite{Maselli2017, Celato2025} in the strong-interaction limit, which was first shown by Colpi et al.~\cite{Colpi1986}.

\begin{figure*}
    \centering
    \includegraphics[width=0.4\linewidth, height=7cm]{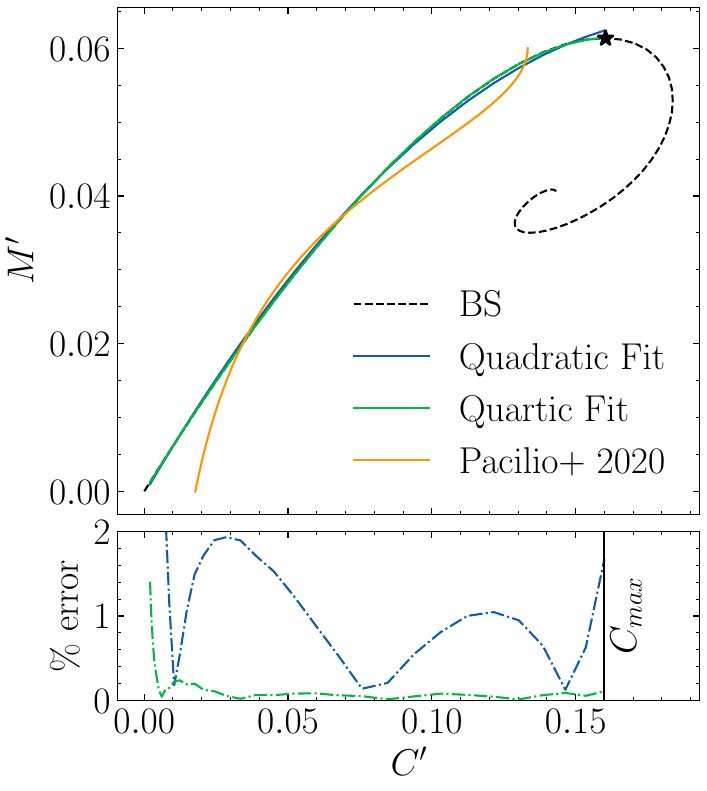}
    \label{fig:m_c_fit}
    \includegraphics[width=0.4\linewidth, height=7cm]{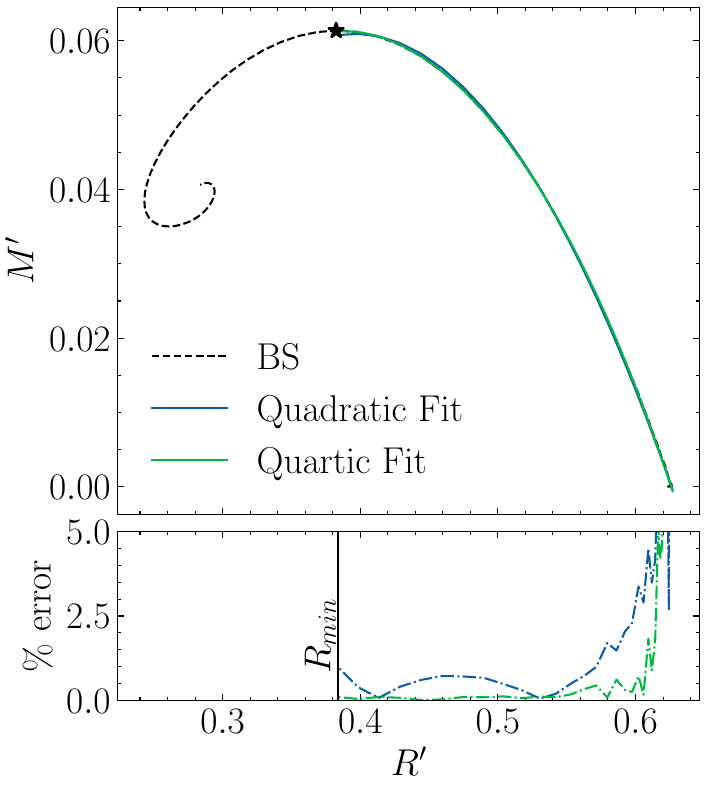}
    \label{fig:m_r_fit}
    \caption{Quadratic (blue) and quartic fits (green) as described in Eqn.~\ref{eqn:quadratic_quartic_relations} for (a) $M'-C'$,
    and (b) $M'-R'$ relations for massive BSs in the strong-interaction limit. The dashed black line is the actual calculated numerical solution. The yellow curve is taken from~\cite{Pacilio2020}, where the fit is derived for non-spinning massive BSs. The percent errors for both fits are shown in the lower panels of each figure. The star marks the point beyond which the BS configurations become unstable. The stable solutions are valid only for compactness (radius) below (above) the star. The fit coefficients are reported in Table.~\ref{tab:quadratic_quartic_fit_coefficients}.}
    \label{fig:m_r_c_fits}
\end{figure*}

In Fig.\ref{fig:m_r_c_fits}(a), we plot the scaled mass $M'$ as a function of compactness $C$. $M'$ increases with $C$ and reaches a value of $M'_{\rm max} \approx 0.06$ at $C'_{\rm max}= C_{\rm max} \approx 0.16$. We perform fitting only up to this critical point, as the region after it, where the mass decreases, is unstable and not of interest. We fit quadratic and quartic functions of the form
\begin{align}\label{eqn:quadratic_quartic_relations}
    Y &= a_0 + a_1X+a_2X^2~, \nonumber \\
    Y &= b_0 + b_1X + b_2X^2 + b_3X^3 + b_4X^4~.
\end{align}
The fit coefficients are given in Table.~\ref{tab:quadratic_quartic_fit_coefficients}. We solve for $f$-modes for a minimum compactness of 0.02 in this work, so we restrict ourselves to $C \gtrsim 0.02$. This is lower than the minimum compactness of $C=0.035$ considered in previous works~\cite{Ryan1997, Vaglio2023}. For $C \gtrsim 0.02$, we observe that the quadratic functions provide a good fit with an accuracy within $2\%$. For the case of the quartic fit, the accuracy is within $0.25\%$. Note that these are fitting errors signifying the deviation of the analytical fit from the numerical solution.

We also plot the relation given in Eq.~A2 of~\cite{Pacilio2020} in yellow. This fit was derived for zero spin while studying the variation of $C^{-1}$ for different mass configurations as a function of dimensionless spin $\chi = J/M^2$, where $J$ is the spin angular momentum of BS. From the plot, we see that this does not fit the $M-C$ curve for a non-rotating BS. This also results in an unexpected behaviour of $k_2$ (Fig. 6 of~\cite{Pacilio2020}).
A possible reason could be that the fit in~\cite{Pacilio2020} was obtained using only five configurations in the low spin limit. Thus, our fit provides a new analytical relation for the same.


We show the well-known $M'-R'$ curve in Fig.\ref{fig:m_r_c_fits}(b).  
At low masses, the radius reaches a constant value of $R'_{\rm max}=0.626$. $R'$ decreases and reaches a minimum value of $R' = 0.38$ as $M'$ increases reaching a maximum value of $M'=0.06$.
Below $R'(C=0.02)=0.6$, the quadratic function fits within an accuracy of $3.5\%$, and the quartic function within $0.75\%$. We also perform the fit for $R'-C$ (not shown here). The fit coefficients for both these relations are provided in Table.~\ref{tab:quadratic_quartic_fit_coefficients}.

 The dimensionless tidal deformability ($\bar{\Lambda}$) does not scale with $x$. This has been derived in Appendix~\ref{sec:appendix_love_equations}. 
 The scaling for the moment of inertia has been derived in Appendix.~\ref{sec:appendix_I_equations}.

\begin{table}[]
    \centering
    \begin{tabular}{c|c|c|c|c|c|c|c|c|c}
         $Y$ & $X$ & $a_0$ & $a_1$ & $a_2$ & $b_0$ & $b_1$ & $b_2$ & $b_3$ & $b_4$ \\  \hline
         $M'$ & $C'$ & -3.545$\times 10^{-4}$ &  0.665 & -1.705 & -3.191$\times 10^{-5}$ & 0.634 & -1.466 & 3.142 & -23.295 \\
         $R'$ & $C'$ & 0.626 & -1.044 & -2.784 & 0.626 & -1.061 & -4.256 & 31.617 & -141.821 \\
         $M'$ & $R'$ & -0.119 &  0.910 & -1.150 & -0.254 & 2.234 &  -5.814 & 7.042 & -3.866 \\
    \end{tabular}
    \caption{Fitting coefficients for the quadratic and quartic function fits to $M'-C'$, $M'-R'$, and $R'-C'$ as given in Eqs.~\ref{eqn:quadratic_quartic_relations}. These coefficients are used to plot the fitting functions in Figs.~\ref{fig:m_c_fit} and~\ref{fig:m_r_fit}.}
    \label{tab:quadratic_quartic_fit_coefficients}
\end{table}

\subsection{Scaled Relations for fundamental modes}

We now move to $f$-mode characteristics, which are the focus of our work. We showed in~\cite{shirke_BS_fmodes_short} that upon applying the scaling $f' = fxM_{\rm Pl}$ and $\tau' =\tau/xM_{\rm Pl}$, we get unique solutions for $f$-mode characteristics as a function of $M'$ and $C$. This is because the $f$-mode equations and perturbation equations are completely independent of microscopic model parameters when written in scaled coordinates. We have shown this explicitly in the appendix of~\cite{shirke_BS_fmodes_short}. Just like the TOV solutions, for every central density $\rho'$, we get a value for $f'$ and $\tau'$ and obtain a curve as we vary the central density. Here, we again provide analytical fits to the solutions numerically obtained in~\cite{shirke_BS_fmodes_short} for easier future computations. We can then use scaling to obtain the $f-M$ curve for any set of $(\lambda, m)$ parameters. 
 
\begin{figure*}
    \includegraphics[width=0.49\linewidth]{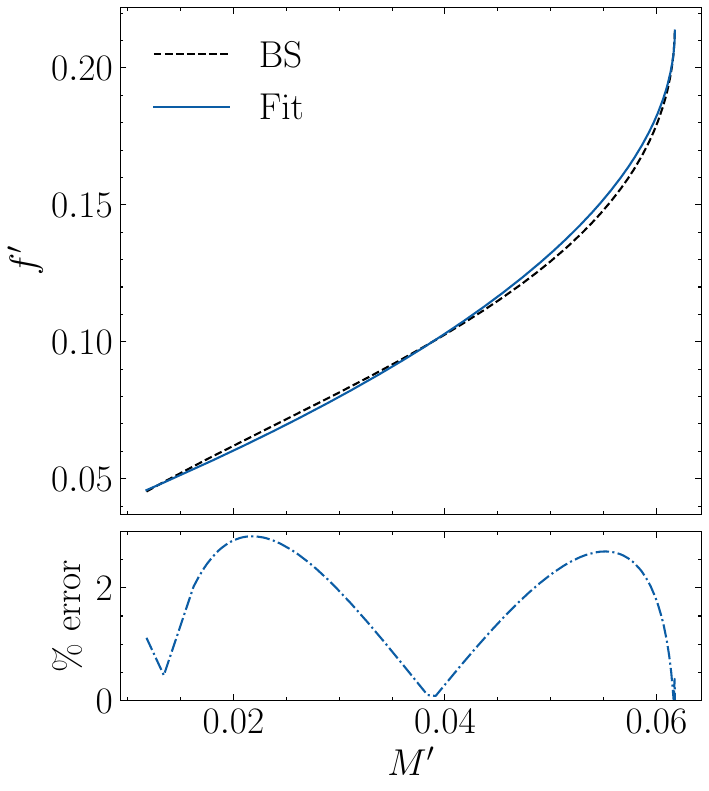}
    \label{fig:f_m_fit}
    \includegraphics[width=0.49\linewidth]{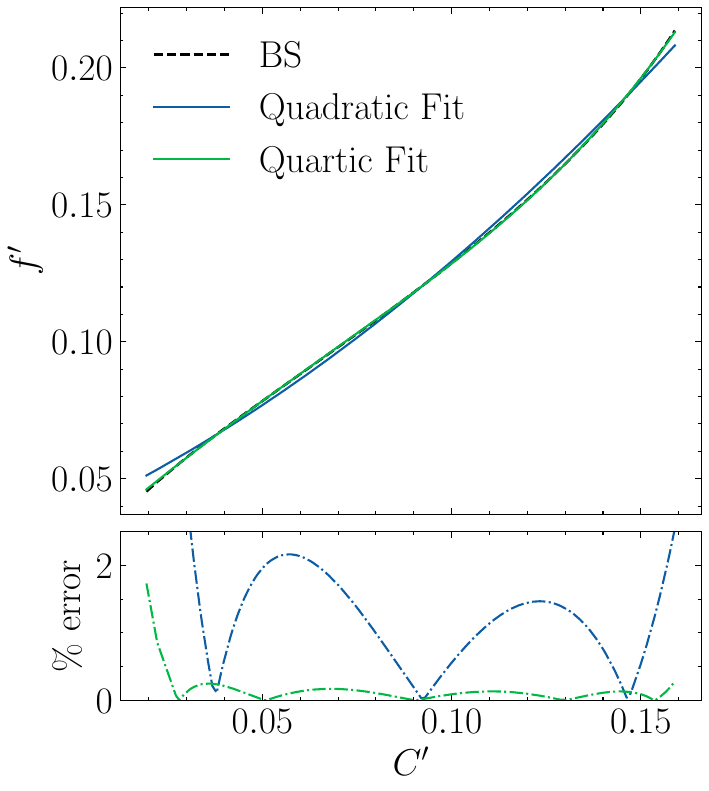}
    \label{fig:f_c_fit}
    \caption{The scaled dimensionless $f$-mode frequency ($f'$) as a function of a) scaled mass and b) compactness of BSs. The blue curve in the left panel is the fitted hyperbola as given in Eq.~\ref{eqn:f_m_fit}. The blue and the green curves in the right panel are the fits for the quadratic and quartic functions, respectively, as given in Eqn.~\ref{eqn:quadratic_quartic_relations}. The fit coefficients are given in Table.~\ref{tab:quadratic_quartic_fit_coefficients_fmodes}. The lower panel shows the percent error for each fit.}
    \label{fig:fmode_freq_fit}
\end{figure*}

 We show this curve as obtained numerically in~\cite{shirke_BS_fmodes_short} in Fig.~\ref{fig:fmode_freq_fit}(a) with a black-dashed line. This is obtained by solving for $f$-mode characteristics as discussed in Sec.~\ref{sec:model}. $f'$ increases with $M'$ and reaches a maximum value of $f'_{\rm max}=0.21$ for the maximum mass configuration of the BS. We observe that the curve begins with a constant slope at low mass, and the slope diverges as it reaches the maximum mass. Hence, we fit a hyperbola to this relation given by 
 \begin{equation}\label{eqn:f_m_fit}
     f' = -7.297\sqrt{\left(\frac{M'-M'_{\rm max} -189.0116}{189.0116}\right)^2 - 1} + f'_{\rm max}~.
 \end{equation}
Here we have imposed the condition that $f'(M'=M_{\rm max}) = f'_{\rm max}$. 
This analytical fit agrees with the numerical solution within $3\%$. We also show the solution for $f'-C'$ in Fig.~\ref{fig:fmode_freq_fit}(b) as compactness is a more natural variable. 
The quartic fit is accurate to $1\%$ for $C \gtrsim 0.02$. 
Similar fits can be performed for the damping time (figures not shown). The fit coefficients are provided in Table.~\ref{tab:quadratic_quartic_fit_coefficients_fmodes}.
 


\begin{table}[]
    \centering
    \scalebox{0.8}{
    \begin{tabular}{c|c|c|c|c|c|c|c|c|c}
         $Y$ & $X$ & $a_0$ & $a_1$ & $a_2$ & $b_0$ & $b_1$ & $b_2$ & $b_3$ & $b_4$ \\  \hline
         $f'$ & $C'$ & 3.747 $\times 10^{-2}$ & 0.656 &  2.627 & 2.363 $\times 10^{-2}$ & 1.203 & -2.232 & -4.417 & 113.160 \\
         $M'$ & $1/\log{\tau'}$ & $-$ & $-$ &  $-$ & -9.718 $\times 10^{-2}$ & 4.457 & -84.357 & 7.771 $\times 10^{2}$ & -2.460 $\times 10^{3}$ \\
         $\tau'$ & $1/C'$ & $-$ & $-$ &  $-$ & 4.124 $\times 10^{2}$ & -1.147 $\times 10^{2}$ & 24.602 & 5.030 & 3.856 $\times 10^{-3}$ \\
    \end{tabular}}
    \caption{Fitting coefficients for the fits between the quantities $f'$, $\tau'$, $M'$, and $C'$ as given in Eqs.~\ref{eqn:quadratic_quartic_relations}. The fit for $f'-M'$ is given in Eq.~\ref{eqn:f_m_fit}.}
    \label{tab:quadratic_quartic_fit_coefficients_fmodes}
\end{table}

\subsection{Universal Relations}

Compact stars are known to follow some empirically observed universal relations, which can be used to infer stellar properties from observations of stellar oscillations. Such inference of stellar properties using observation of pulsations comes under the study of asteroseismology. 
In this section, we discuss some known universal relations
focusing on the $f$-mode characteristics. 
Other known relations for static stars, in particular the $I$-Love-$C$, are discussed in Appendix.~\ref{sec:appendix_I_Love_C} where we use the scaling derived in Appendix.~\ref{sec:appendix_love_equations} and Appendix.~\ref{sec:appendix_I_equations} to show the $I-$Love$-C$ universal relations are exact for massive BSs.

\subsubsection{Empirical Fits}
The $f$-mode frequency is known to scale with the square root of average density ($M/R^3$) of NSs and was first proposed by~\cite{AnderssonKokkotas1996, AnderssonKokkotas1998}. The relation is of the form~
\begin{equation}\label{eqn:f_density}
    f(\text{kHz}) = a + b \sqrt{\frac{GM}{R^3}}~.
\end{equation}
Here, $f$ is the $f$-mode frequency, $M$ is the mass of the star, and $R$ is its radius. $a$ and $b$ are the fitting coefficients. 
This relation was checked for BSs for select values of $\lambda$ and $m$ in~\cite{Flores2019} that form BSs with masses in the range of NSs 
and the coefficients reported were $a=-0.0195 \pm 0.0008$ kHz and $b= 62.997 \pm 0.058$ kHz km. The relation was also explored in~\cite{Celato2025} for select BS cases and reported the coefficients $a=-0.01461$ kHz and $b=42.11$ kHz km.

\begin{figure*}
    \label{fig:f'_dens'_fit}
    \includegraphics[width=0.49\linewidth]{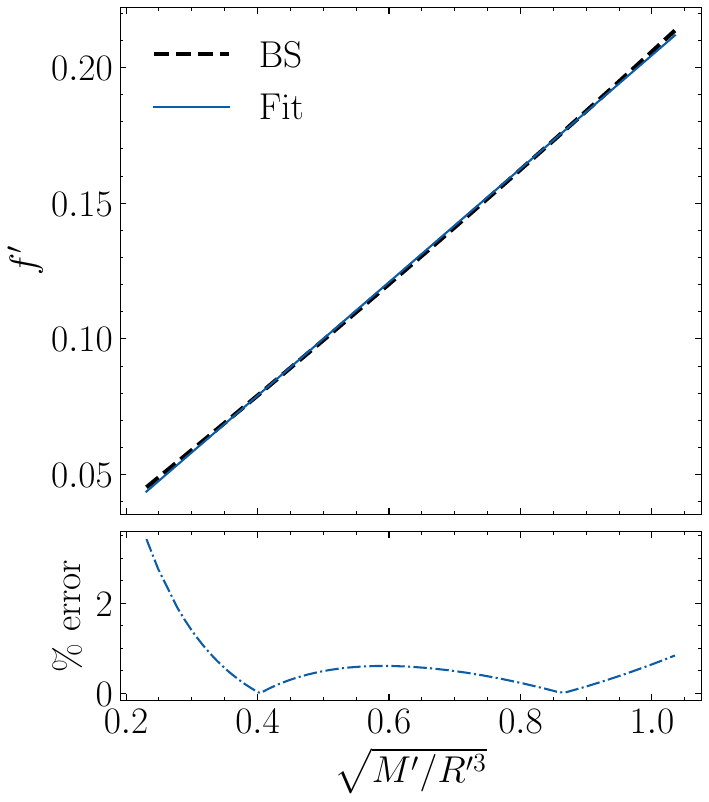}
    \label{fig:tau'_compactness'_fit}
    \includegraphics[width=0.49\linewidth]{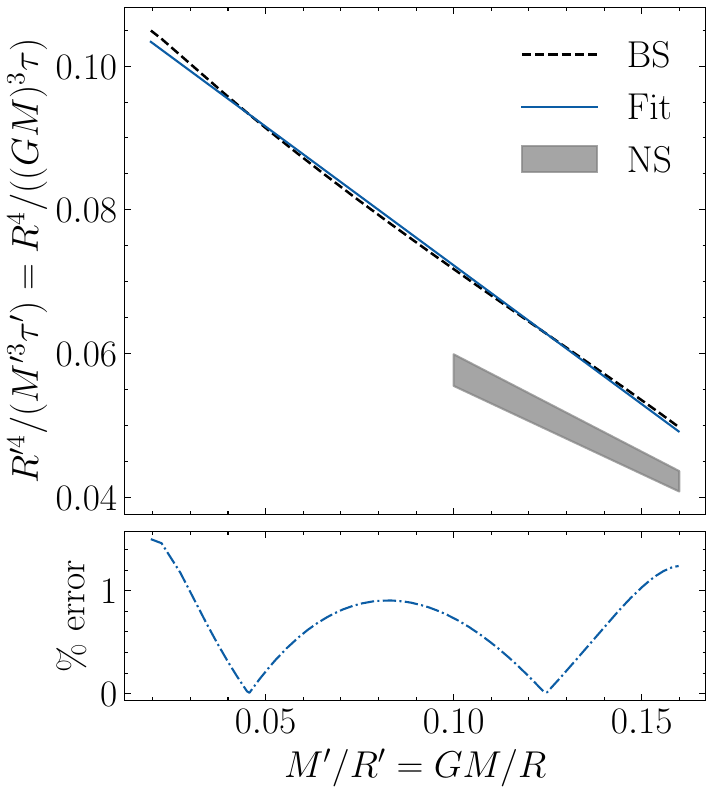}
    \caption{Universal relations connecting (a) $f$-mode frequency in primed coordinates to the average density of BS and (b) scaled damping time $(R'^4/M'^3\tau')$ to compactness as suggested in~\cite{AnderssonKokkotas1998}. The grey band is the uncertainty in NSs with and without hyperonic matter. The band spans the fit relations reported in~\cite{AnderssonKokkotas1998, Benhar2004, Pradhan2022}. The blue curves are fits given in Eqs.~\ref{eqn:f_density_prime} and~\ref {eqn:tau_c_fit}. The errors in BS fits are shown in the lower panels.}
    \label{fig:UR_f_dens_tau_c}
\end{figure*}

We reproduced the results from Fig.3 of~\cite{Flores2019}, connecting $f$ with average density (not shown here), which reported a spread as the curves did not identically overlap. We checked that upon scaling the parameters and plotting the scaled quantities, all curves coincide to result in a unique curve, independent of DM model parameters.
This is shown in Fig.~\ref{fig:UR_f_dens_tau_c}(a). We perform a linear fit to this curve given by
\begin{equation}\label{eqn:f_density_prime}
    f' = -0.0045 + 0.21 \sqrt{\frac{M'}{R'^3}}~.
\end{equation}
The fit agrees within $2.6\%$. We refer to the fit coefficients in Eq.~\ref{eqn:f_density_prime} as $a' = -0.0045$ and $b'=0.21$. Writing back in unprimed quantities as given in Eq.~\ref{eqn:f_density}, we get $b=b'=0.21 = 63.0$ kHz km. 
This value is consistent with the one obtained in~\cite{Flores2019} but is higher compared to the one obtained in~\cite{Celato2025}. 
This could be due to the difference in the density range for which the fit was performed and the small number of parameters and configurations that were considered in these works.
In this work, we have derived this equation in an EoS-independent way rather than choosing some select model parameters and fitting them. 

$a = a'/(xM_{\rm Pl}) = 5.67 \times 10^{-7}/(x/\text{MeV}^{-2})$ kHz depends on $x$. 
This dependence on $x$ can explain the variation obtained in~\cite{Flores2019} in the slope parameter.
They also report an error in $b$; however, from this analysis, we expect $b$ to remain the same for all $x$, since $b'=b$. The residual variations could be attributed to numerical error.

Another empirical relation for the $f$-mode damping time exists as a function of compactness when it is scaled by mass and radius~\cite{AnderssonKokkotas1998}. The fit for a BS is given by
\begin{equation}\label{eqn:tau_c_fit}
    \frac{R^4}{(GM)^3\tau} = 0.111 - 0.386C~.
\end{equation}
The coefficients are in close agreement with the recent fit provided by~\cite{Celato2025}$: \frac{R^4}{(GM)^3\tau} = 0.1105 - 0.3764C~.$ The relation is invariant under the change to primed quantities. Thus, we directly plot $R^4/((GM)^3\tau)$ as a function of $C$ in Fig.~\ref{fig:UR_f_dens_tau_c}(b). This is truly a universal relation for massive BSs in the strong-interaction limit and is followed regardless of the underlying model parameters. This unique curve (black curve in Fig.~\ref{fig:UR_f_dens_tau_c}(b)) is a result of the invariance of the plotted quantities with $x$. We show the linear fit to this numerically obtained curve given by Eqn.~\ref{eqn:tau_c_fit} with a blue line. The deviation of the fit from the numerical solution is within $1.5\%$. Since the quantities plotted are invariant, we compare them to the corresponding band obtained for NSs~\cite{AnderssonKokkotas1998, Benhar2004, Pradhan2022}. 
Thus, we conclude that BSs have a different relation and this can be used to distinguish them from NSs.
We cannot make a comparison with Fig. 3 of~\cite{Flores2019}, as their compactness goes beyond 0.18, as opposed to our case, which only goes up to 0.16. and as they use a different fitting function. 

\subsubsection{$f-$Love$-C$}
There are also other relations involving mass-scaled $f$-mode characteristics that are universal in nature. The complex eigenfrequency, when scaled with mass, shows a tight correlation with compactness~\cite{AnderssonKokkotas1998, Pradhan2022} and the dimensionless tidal deformability~\cite{Chan2014, SotaniKumar2021, Pradhan2023}. The fits to these are of the quadratic and quartic forms, respectively. Here, $\text{Re}(\omega_f) = 2 \pi f$ is the $f$-mode angular frequency, and $\text{Im}(\omega_f)=1/\tau$ is the inverse of damping time.

\begin{figure*}
    \includegraphics[width=0.32\linewidth]{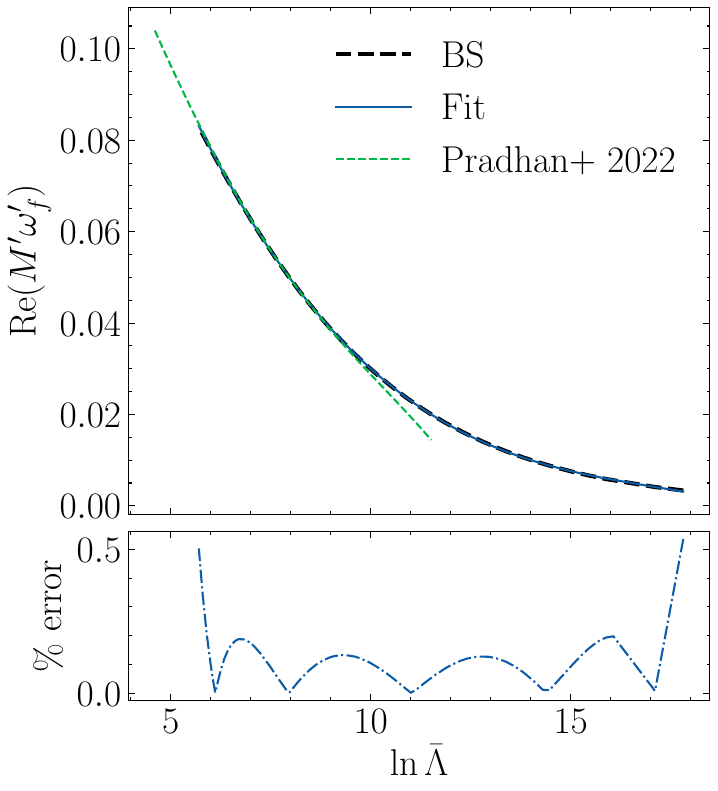}
    \includegraphics[width=0.32\linewidth]{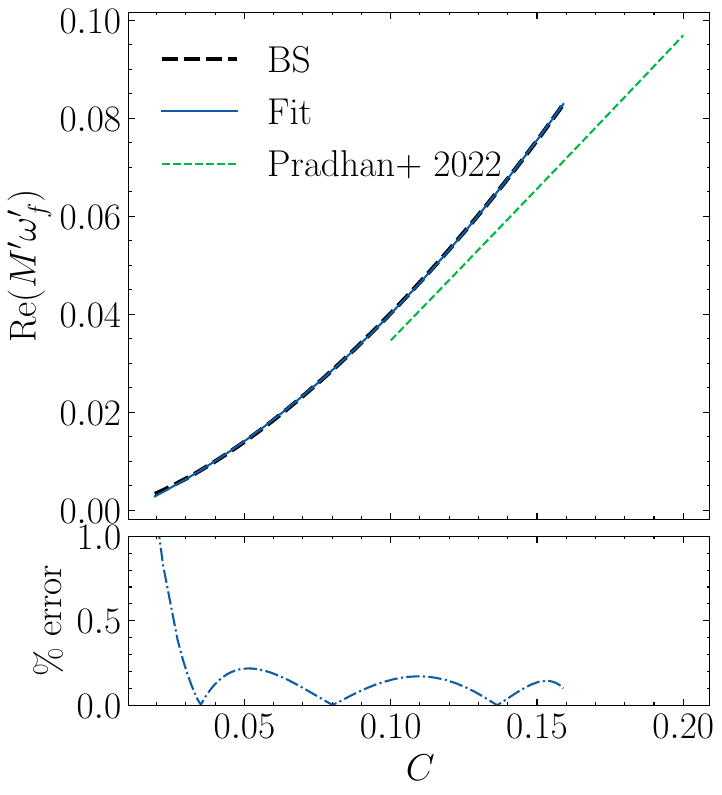}
    \includegraphics[width=0.32\linewidth]{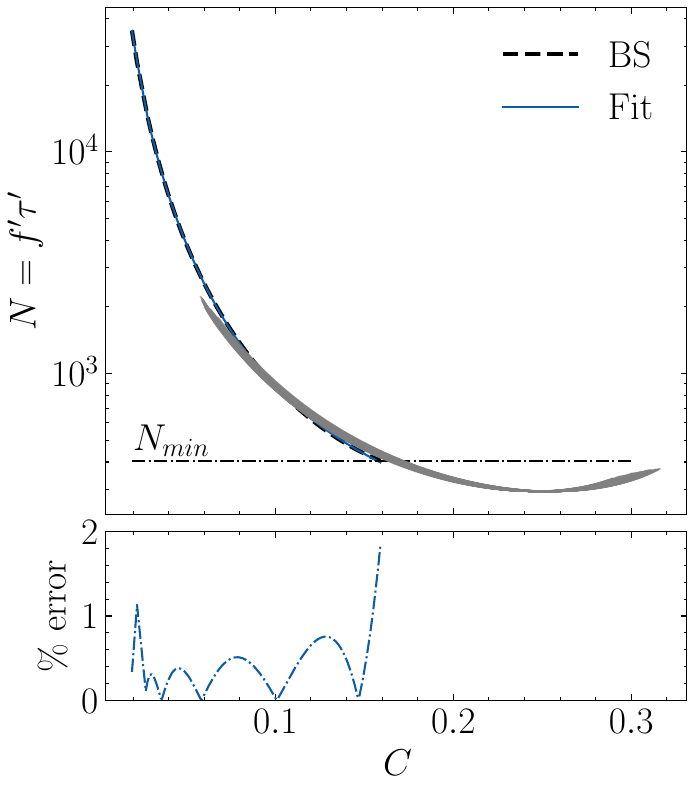}
    \caption{Universal relations for mass-scaled $f$-mode frequency as a function of the (a) logarithm of the dimensionless tidal deformability and (b) compactness. The relations followed by NSs and hyperonic NSs are shown in green~\cite{Pradhan2022}. (c) Number of cycles ($N=f\tau=N'\tau'$) completed by BSs within the damping time. BSs complete a minimum of $N_{\rm min}=406$ oscillations for the case of the highest compactness configuration. The $N-C$ curves for pure neutron stars~\cite{Pradhan2022} are shown in grey for comparison.
    The analytical fits to the numerical solutions are shown in blue, whose fitting coefficients are provided in Table.~\ref{tab:quadratic_quartic_fit_coefficients_UR}. The corresponding percentage errors are shown in the bottom panels.}
    \label{fig:UR_f}
\end{figure*}

We plot these the relations for $\text{Re}(M\omega_f)$ in Fig.~\ref{fig:UR_f} (a) and (b) and for $\text{Im}(M\omega_f)$ (not shown). It can be checked that $M\omega_f/M_{\rm Pl}^2= M'\omega'_f$. Thus, $M'\omega'_f$, just like $\bar{\Lambda}$ and $C$ do not scale with $x$. Since all these plotted quantities are independent of the model parameters, these relations are again universal, making the curves unique, i.e., these are followed by all massive BSs in the strong interaction limit. Ref.~\cite{Pradhan2022} also obtained such fits using a large set of nuclear and hyperonic EoS. We show these fits for comparison. The relations with the tidal deformability are consistent with the case of NSs. The ones containing compactness, however, show deviation and follow different relations than NSs.
We see the $f$-Love and the $f-C$ fits have an error of less than $0.5\%$ and $1\%$ respectively. For the case of $\tau$ relations, the errors are higher, within $4\%$ for $\tau-C$ and as high as $50\%$ for $\tau$-Love.

\begin{table}[]
    \centering
    \scalebox{0.7}{
    \begin{tabular}{c|c|c|c|c|c|c|c|c|c}
         $Y$ & $X$ & $a_0$ & $a_1$ & $a_2$ & $b_0$ & $b_1$ & $b_2$ & $b_3$ & $b_4$ \\  \hline
         $\text{Re}(M\omega_f)$ & $\ln{\bar{\Lambda}}$ & - & - &  - & 2.365 $\times 10^{-1}$ & -3.753 $\times 10^{-2}$ & 2.103 $\times 10^{-3}$ & -4.204 $\times 10^{-5}$ & 5.420$\times 10^{-8}$ \\
         $\text{Re}(M\omega_f)$ & $C$ & -0.0024 & 0.237 &  1.886 & - & - & - & - & - \\
         $\text{Im}(M\omega_f)$ & $\ln{\bar{\Lambda}}$ & - & - &  - & 2.024 $\times 10^{-4}$ & -5.189 $\times 10^{-5}$ & 5.000 $\times 10^{-6}$ & -2.141 $\times 10^{-7}$ & 3.428$\times 10^{-9}$ \\
         $\text{Im}(M\omega_f)/C^4$ & $C$ & 0.110 & -0.362 &  -0.056 & - & - & - & - & - \\
         $N$ & $1/C$ & - & - &  - & 1.833 & 1.442 $\times 10^{-1}$ & -3.901 $\times 10^{-3}$ & 6.121 $\times 10^{-5}$ & -3.879 $\times 10^{-7}$ \\
    \end{tabular}}
    \caption{Fitting coefficients for the $f$-Love-$C$ universal relations and $N-C$. The fit equations are given in Eqs.~\ref{eqn:quadratic_quartic_relations}.}
    \label{tab:quadratic_quartic_fit_coefficients_UR}
\end{table}

The quantity $N =f\tau = f'\tau'$ gives the number of oscillations that the BS undergoes before its amplitude damps by a factor of $e$. Again, this quantity does not scale with $x$. We plot it against $C$ in Fig.~\ref{fig:UR_f}(c) (taken from~\cite{shirke_BS_fmodes_short}). We see that $N$ decreases with compactness, i.e., more compact configurations undergo fewer oscillations before damping out. For $C\approx 0.02$, $N$ can be as high as 10,000. For the case of the most compact configuration, we get a minimum number of such oscillations given by $N_{\rm min} =406$. We show a quartic polynomial (Eqn.~\ref{eqn:quadratic_quartic_relations}) fit, with a fitting error below $2\%$.  
Since $f'\tau'=f\tau$, we compare them to the corresponding band obtained for NSs~\cite{Pradhan2022}.


\subsection{Observable DM parameter space}

In this section, we focus on the scalar DM parameter space. We discussed various constraints in Sec.~\ref{sec:parameter_space}, and the available parameter space is visually shown in Fig.~\ref{fig:parameter_space}. For this entire region, it would be interesting to obtain the information of $f$-mode characteristics and check which regions can be probed by various GW detectors.

\begin{figure*}
    \centering
    \includegraphics[width=0.49\linewidth]{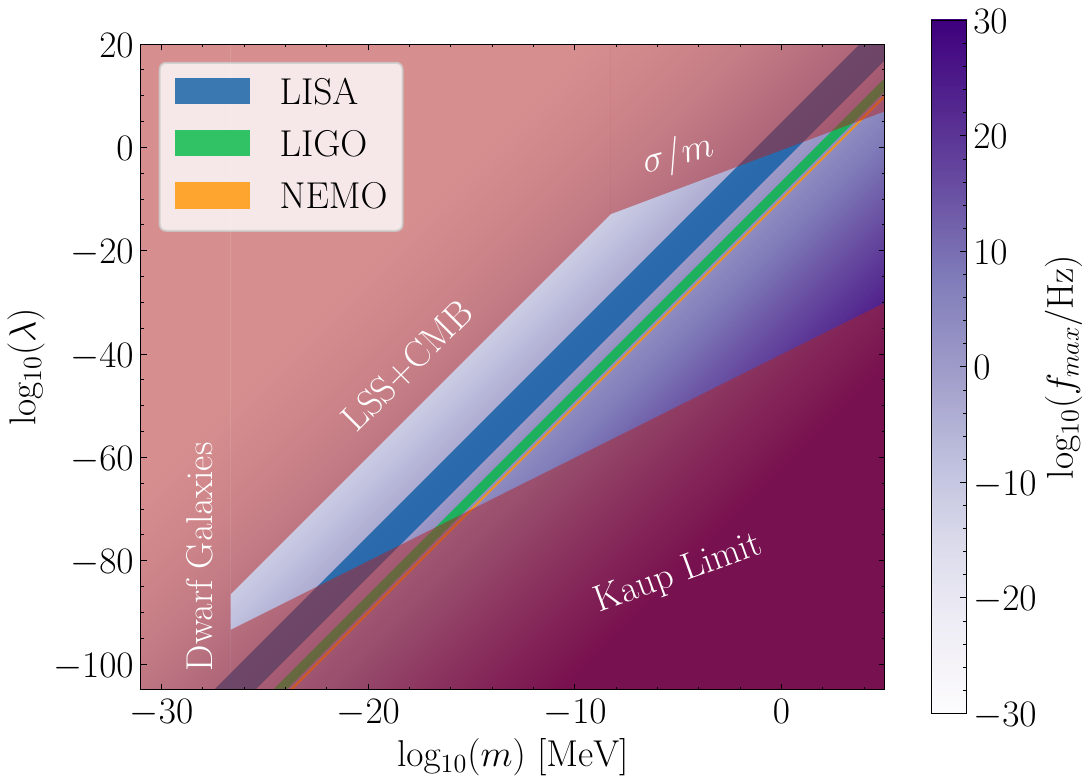}
    \includegraphics[width=0.49\linewidth]{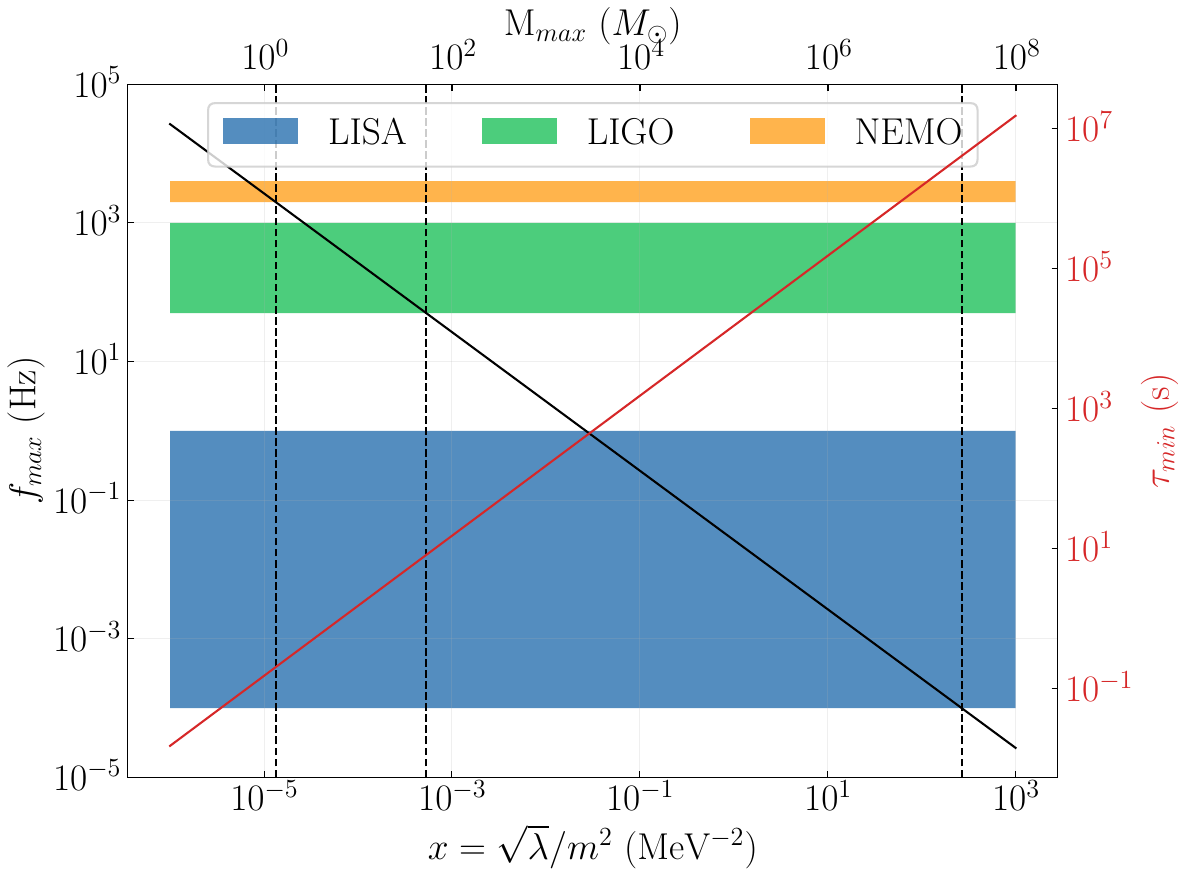}
    \caption{ a) The maximum $f$-mode frequency for the entire parameter space for BSs. The colourbar indicates the $\log_{10}(f_{\rm max})$ in Hz. The constraints from Fig.~\ref{fig:parameter_space} are used to mask the excluded region in red. Only the unmasked region is relevant to the analysis. b) Maximum $f$-mode frequency (black) and minimum damping time (red) as a function of scaling parameter. The corresponding maximum mass possible is shown on the upper axis. The vertical dashed lines indicate the upper limit on $x$ that can be probed by NEMO, LIGO, and LISA, respectively, as we go from left to right. 
    The frequency bands for LIGO and future planned GW detectors, LISA and NEMO, are overlaid. See text for more details.}
    \label{fig:fmode_paramter_space}
\end{figure*}

First, we show the same parameter space with the constraints in Fig.~\ref{fig:fmode_paramter_space}(a). For every set of parameters ($m$, $\lambda$), we can compute the maximum possible $f$-mode frequency ($f_{\rm max}$), corresponding to the most compact configuration. We can see that $f_{\rm max}$ is higher for larger DM mass ($m$) and for smaller values of self-interaction ($\lambda$). 
This trend is clear from the scaling relation $f_{\rm max} = 0.21/(xM_{\rm Pl}) = (0.21/M_{\rm Pl})(m^2/\sqrt{\lambda})$. 

From an observation point of view, it would be of great interest to know what region of this vast parameter space could actually be accessible in the near future. $f$-mode oscillations of BSs would result in the emission of GWs of $f$-mode frequencies. Hence, for reference, we plot the frequency band sensitive to various current and future GW detectors. We show the bands for three detectors, LISA, LIGO, and NEMO, in Fig.~\ref{fig:fmode_paramter_space}(a). The frequency range over which the detectors have high sensitivity for various detectors is taken to be [0.1 mHz, 1 Hz] for LISA, [50, 1000] Hz for LIGO, and [2, 4] kHz for NEMO. These bands are used as iso-contours for $f_{\rm max}$ in the plot. The bands for Cosmic Explorer (CE) and Einstein Telescope (ET) are not explicitly shown as their sensitivity bands roughly encompass those of LIGO and NEMO. Thus, all the GW detectors together roughly cover the region bounded by the blue and orange bands. 

It is important to note that since we are plotting the maximum $f$-mode frequency here, it does not mean that these are the only bands that can be probed. The region towards the right of these bands has $f_{\rm max}$ larger than the sensitivity range, i.e., the frequency of a configuration with lower compactness for these parameters can still fall in the detection band. Hence, in principle, the entire parameter space falling to the right of the band for each detector can be probed by that detector. 

Since we know that $f_{\rm max}$ scales with a particular combination of ($m$, $\lambda$), given by $f_{\rm max} \propto m^2/\sqrt{\lambda} = 1/x$, we focus on the parameter $x$, as we would like to infer microscopic DM parameters from future GWs from $f$-modes, and obtain the range of $x$ that can be probed. In Fig.~\ref{fig:fmode_paramter_space}(b), we show this relation of $f_{\rm max}$ on $x$ in a black line. Note that the black line marks the upper limit on frequency. We see that at $x=10^3$ MeV$^{-2}$, the black line is already below the frequency range of all detectors and is not observable. The point where the black line goes below this detection threshold is $x_{\rm thresh} = 2.66 \times 10^2$ MeV$^{2}$. This is shown by the rightmost black-dashed line in the figure. This is an absolute upper threshold on $x$, such that $x > x_{\rm thresh}$ is inaccessible to the currently planned GW detectors. This threshold is set by the one observing in the lowest frequency range, i.e., LISA. The Pulsar Timing Array (PTA) observes GWs in nano-Hz frequencies but is restricted to stochastic GWs. Hence, we have not considered it here. 


Similar upper thresholds can be obtained for other detectors as well. For LIGO, the threshold is $x \le 5.32 \times 10^{-4}$ MeV$^{-2}$ while for NEMO it is $x \le 1.33 \times 10^{-5}$ MeV$^{-2}$. If we consider the proposed deci-Hz detectors~\cite{decigo2011, tiango2020} that are expected to be sensitive in $[0.1,10]$ Hz range, we obtain $x \le 0.266$ MeV$^{-2}$. In principle, the region $x \le 1.33 \times 10^{-5}$ MeV$^{-2}$ is accessible to all detectors.

 On the same plot, we also show the minimum value of the damping time for a given $x$ as $\tau_{\rm min} = 1900xM_{\rm Pl}$ in red. Since the damping time is not easily measurable, we focus on frequency in this study. On the upper axis, we show the values of maximum mass corresponding to the values of $x$ given by $M_{\rm max} = 0.06xM_{\rm Pl}^3$. The range of $x$ sensitive to a particular detector would change if we considered BSs of lower compactness. In general, we would like to see the actual ranges of mass and $x$ that can be probed.

\begin{figure*}
    \includegraphics[width=0.49\linewidth]{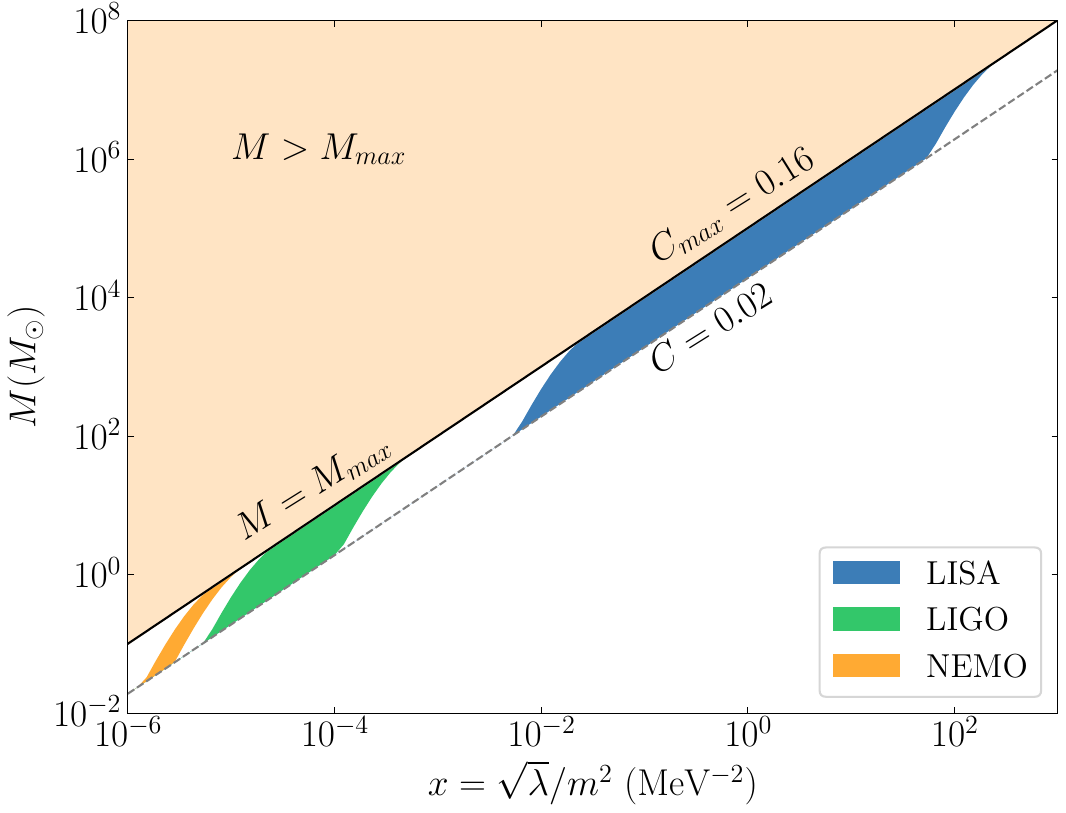} \includegraphics[width=0.49\linewidth]{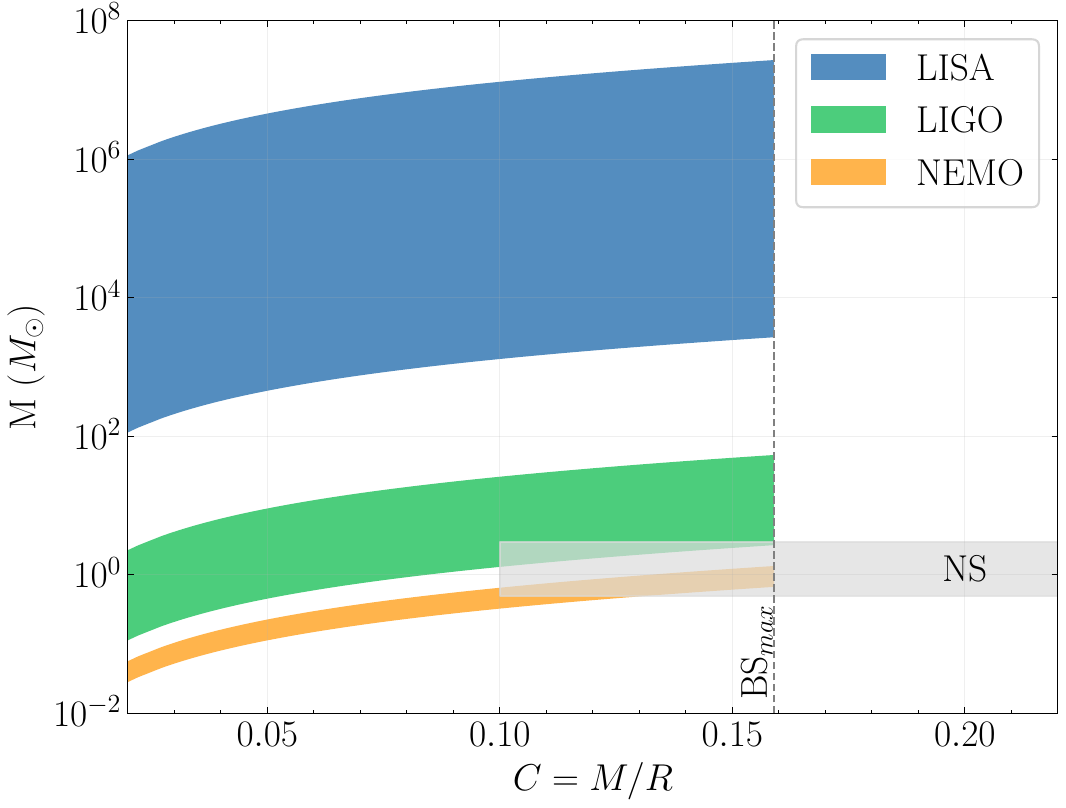}
    \caption{(a) Masses of BSs falling in various GW bands for a given set of model parameters $x$. (b) Mass and compactness of BSs that could be probed by LIGO and planned future detectors, LISA and NEMO. The mass-compactness range for NSs is marked by the silver patch.}
    \label{fig:fmode_x_m_c}
\end{figure*}

In Fig.~\ref{fig:fmode_x_m_c}(a), we show the mass range corresponding to various values of $x$ that can be probed by the same GW detectors. The black line marks the maximum possible mass for a given value of $x$ ($M=0.06xM_{\rm Pl}^3$), which also corresponds to the maximum compactness case. The region above this is inaccessible as we enter the unstable branch. These masses are expected to collapse into a BH. The dashed line marks the mass of BSs having a fixed $C=0.02$. Since we solve for $f$-modes (as mentioned in Sec.~\ref{sec:model}) only for $C \gtrsim 0.02$, we only focus on the regions between these two lines. Using the $f'-M'$ relation (Eq.~\ref{eqn:f_m_fit}), we can use the scaling to write $f=f(x,M)$. Doing this, for each $(x, M)$ in this region, we can check if the $f$-mode frequency falls in any of the GW detector bands. Such regions in $(x, M)$ falling in LISA, LIGO, and NEMO bands are shown in blue, green, and orange, respectively. 
We see that BSs roughly in the mass range of $10^2-10^7 M_{\odot}$ are observable by LISA, $0.1-10 M_{\odot}$ by LIGO and $0.01-1 M_{\odot}$ by NEMO. Note that for each case, we can probe even lower masses if we consider BSs with compactness lower than 0.02.

To get a more physical picture, in Fig.~\ref{fig:fmode_x_m_c}(b), we show the same information in the $M-C$ plane. We make use of Eq.14 of~\cite{shirke_BS_fmodes_short} to obtain $f=f(C,M)$ and repeat the process. 
We conclude that a mass range lower than that corresponding to $C_{\rm max}$ can be probed for low compactness configurations. We can make a direct comparison with NSs as they typically lie within 1-3 $M_{\odot}$ and have compactness roughly in the range 0.1-0.3. BHs fall on the extreme right of this diagram at $C=0.5$.

\subsection{Detectability}\label{sec:detectability}
We now see the prospects of detectability of these $f$-modes from massive BSs. Since these are quasinormal oscillations which damp out due to GW emission, we assume a model for burst GW as considered in~\cite{Ho2020}, regardless of the damping time, where oscillations are excited in a BS at $t=0$. The possible mechanisms to excite these could be i) during the formation of a BS (like gravitational cooling) or ii) nearby astrophysical objects that hit the BS or interact closely (external perturbations). We analyze the detectability assuming $f$-modes are excited. The oscillation frequency is its fundamental frequency $f$, which damps with a timescale of $\tau$ as it radiates GWs. The GW strain amplitude $h(t)$ for such a signal is given by~
\begin{equation}
    h(t) =
    \begin{cases}
        0 & t<0 \\
        h_0e^{-t/\tau}\sin(2\pi f t)\qquad & t\ge 0
    \end{cases}
\end{equation}
Here, $h_0$ is the peak amplitude measured by the GW detectors. It can be related to the total energy in the oscillations that is radiated away $E$, the distance of the detectors from the source of burst $d$, and the $f$-mode characteristics $f$ and $\tau$. This is given by~\cite{Ho2020, PradhanPathak2023}~
\begin{align}
    h_0 &= \frac{1}{\pi d f}\left(\frac{5G}{c^3}\frac{E}{\tau}\right)^{1/2} = 1.53\times 10^{-17}\left(\frac{1\mbox{ kpc}}{d}\right)\left(\frac{E}{M_{\odot}}\right)^{1/2}\left(\frac{1\mbox{ kHz}}{f}\right)\left(\frac{1\mbox{ s}}{\tau}\right)^{1/2}.
\end{align}

The signal-to-noise ratio or SNR ($\rho$) for such a burst signal is derived as~\cite{Echeverria1989, Kokkotas2001}
\begin{equation}
\rho = \sqrt{\frac{4Q^2}{1+4 Q^2}}h_0 \sqrt{\frac{\tau}{2S_n}}~.
\label{SNR}
\end{equation}
Here, $Q$ is the quality factor of the signal given by $Q=\pi f \tau$. This roughly gives the number of cycles ($N$) of the burst oscillations before the amplitude falls by a factor of $e$. We saw in Fig.~\ref{fig:UR_f}(c) that $N_{\rm min}$ for BSs is 406. This means that $Q \ge 406\pi$, which means $4Q^2 \gg 1$. Thus, the SNR for BSs simply reduces to
\begin{equation}\label{eqn:snr_BS}
    \rho = h_0 \sqrt{\frac{\tau}{2S_n}}~.
\end{equation}
$S_n$ is the noise spectral density of the detector at the frequency of evaluation $f$.
We use Eq.~\ref{eqn:snr_BS} to calculate the SNR. 
\begin{align}
    \rho &= 4.23 \times 10^{-23} E_d \left(\frac{M_{\rm BS}}{M_{\odot}}\right)^{3/2}\left(\frac{0.21}{f'(C)}\right)\left(\frac{0.06}{M'(C)}\right)\left(\frac{\mbox{Hz}^{-1/2}}{\sqrt{S_n}}\right)~.
\end{align}
We have used the relation reported in~\cite{shirke_BS_fmodes_short} to write $f$ in terms of $M$ and $C$. This relation of BS SNR is completely in terms of $E$, $d$, $M$, and $C$ and does not include any explicit dependence on the $f$-mode characteristics. We have defined the parameter $E_d$ as
\begin{equation}
    E_d \equiv \frac{\sqrt{E/(10^{-10}M_{\rm BS}})}{d/(1\text{kpc})}~.
\end{equation}
We have used a factor of $10^{-10}M_{\rm BS}$ as a normalization of the energy emitted by the $f$-modes. Thus, $E_d=1$ means an energy corresponding to $10^{-10}$ times that of the mass-energy of the BS is radiated at a distance of $1$ kpc. Or alternatively, an energy or of $n\times 10^{-10}M_{\rm BS}$ radiated at a distance of $\sqrt{n}$ kpc. We also note that the SNR for a fixed $E_d$ grows with $M_{\rm BS}$. The primary reason for this is that the $f$-mode frequency for fixed compactness goes as the inverse of mass, and lower frequency results in larger signal strength.

In a realistic scenario, BSs in the universe might have different compactness depending on the initial formation conditions, but the DM parameters are expected to be fixed. Thus, DM would have a fixed value of $x$, and compactness could be anything depending on the mass of the BS that formed. Hence, it is more natural to think in terms of $C$-$x$ parameters. So we also provide the expression for $\rho(x, C)$ here. We can write $M_{\rm BS} = M'(C)xM_{\rm Pl}^3$. Doing this we get
\begin{equation}
    \rho = 1.30 \times 10^{-15} E_d \left(\frac{x}{\text{MeV}^{-2}}\right)^{3/2}\left(\frac{0.21}{f'(C)}\right)\left(\frac{M'(C)}{0.06}\right)^{1/2}\left(\frac{\mbox{Hz}^{-1/2}}{\sqrt{S_n}}\right)~.
\end{equation}

Now, we set $\rho \ge 5$ as a threshold for detection. This gives us
\begin{equation}
    E_d \ge 3.85\times 10^{15}\left(\frac{\text{MeV}^{-2}}{x}\right)^{3/2}\left(\frac{f'(C)}{0.21}\right)\left(\frac{0.06}{M'(C)}\right)^{1/2}\left(\frac{\sqrt{S_n(f(x,C))}}{\mbox{Hz}^{-1/2}}\right)~.
\end{equation}
The equality gives the minimum values of $E_d$ ($(E_d)_{\rm min}$) required for the BS burst source of compactness $C$ made of DM with parameter $x$ that is detectable with an SNR of 5 by a GW detector with the noise spectral density $S_n$. The lower the value of $(E_d)_{\rm min}$, the better the detectability.

\begin{figure*}
    \label{fig:detectability_aligo}
    \includegraphics[width=0.49\linewidth]{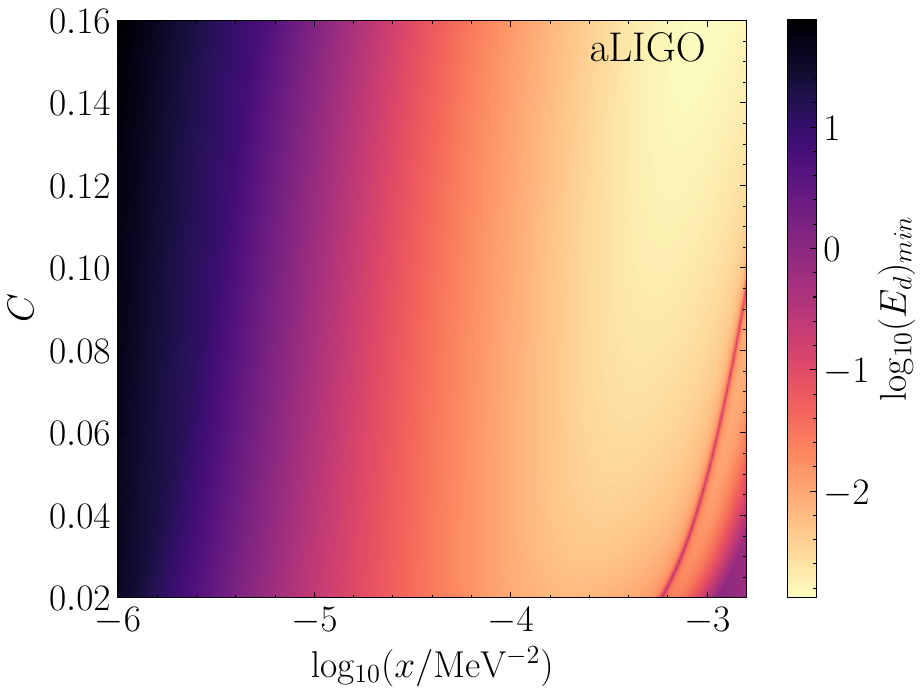}
    \label{fig:detectability_lisa}
    \includegraphics[width=0.49\linewidth]{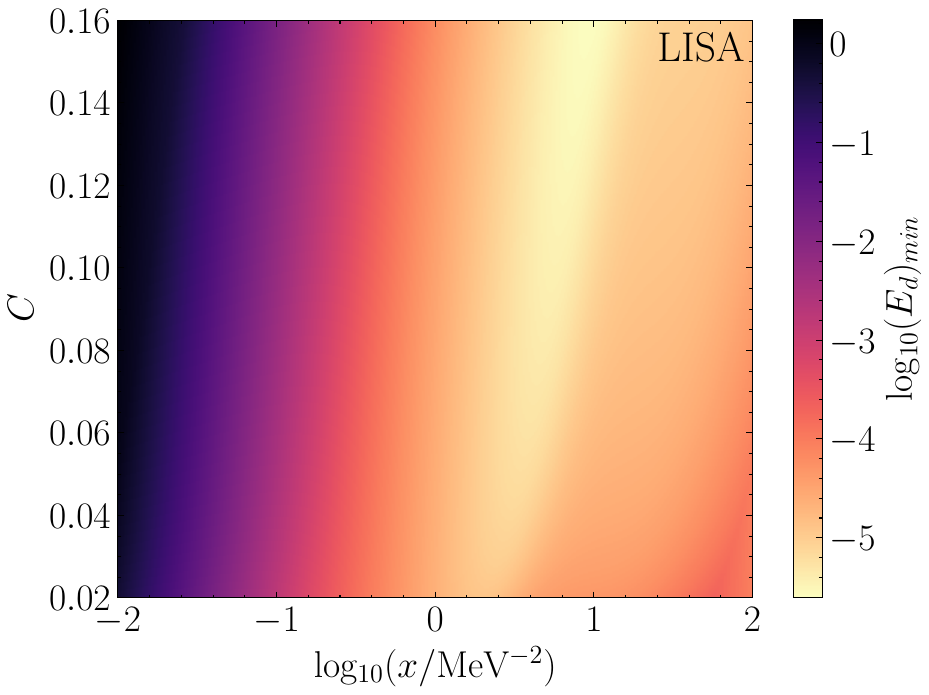}
    \caption{The detectability of BSs using (a) aLIGO and (b) LISA GW detectors. The colours indicate the minimum values of $E_d$ required for the detection with $\rho \ge 5$. The sensitivity curve for aLIGO is obtained from~\url{https://dcc.ligo.org/LIGO-T1500293/public}. For LISA, the sensitivity curve was constructed using the analytical fit provided in~\cite{Robson2019}. The range for $x$ probed is different, as the noise spectral density range used is 5-5000 Hz for aLIGO and 0.1 mHz - 1 Hz for LISA.}
    \label{fig:detectability_lisa_aLIGO}
\end{figure*}

In Fig.~\ref{fig:detectability_lisa_aLIGO}(a) we plot this $(E_d)_{\rm min}$ as a function of $x$ and $C$ for the case of advanced LIGO (aLIGO). 
We observe that detectability is better for lower frequencies (high $x$ and low $C$). The dark strip seen at the bottom right is due to the higher noise for aLIGO around 9.7 Hz. This worsens the detectability. 
We observe that $\log_{10}(E_d)_{\rm min} > -3$ in the best case scenario. Thus, for $E = 10^{-10} M_{\rm BS}$, BSs up to about 1 Mpc would be observable. Thus, aLIGO would be sensitive to BSs within and around our galaxy. For sources with $(x,C)$ such that the $f$-mode frequency is at the extreme limits of the observing range, BSs only within 100 pc ($\log_{10}(E_d)_{\rm min} \sim 1$) would be observable.

As the SNR is better for lower frequencies and higher BS mass, the detectability is expected to be the best in the case of LISA, as it is planned to be observed in the millihertz range. We make the same plot for the LISA detector (see Fig.~\ref{fig:detectability_lisa_aLIGO}(b)). 
We see that LISA will be able to probe about a hundred times deeper with $\log_{10}(E_d)_{\rm min} \gtrsim -5.5$. This means that for $E = 10^{-10} M_{\rm BS}$, the distance of the BS has to be within $\sim300$ Mpc. Since these are cosmological distances associated with burst energies, this corresponds to the luminosity distance. The BS masses probed in this case are over $100M_{\odot}$. For the extreme case of $M_{\rm BS} \approx 100M_{\odot}$, corresponding to $x=10^{-2}$MeV$^{-2}$, the $f$-mode frequency lies at the higher frequency end of 1 Hz of LISA, and the observable distance is only about a kpc ($\log_{10}(E_d)_{\rm min} \sim 0$).

\begin{figure*}[]
    \includegraphics[width=0.49\linewidth]{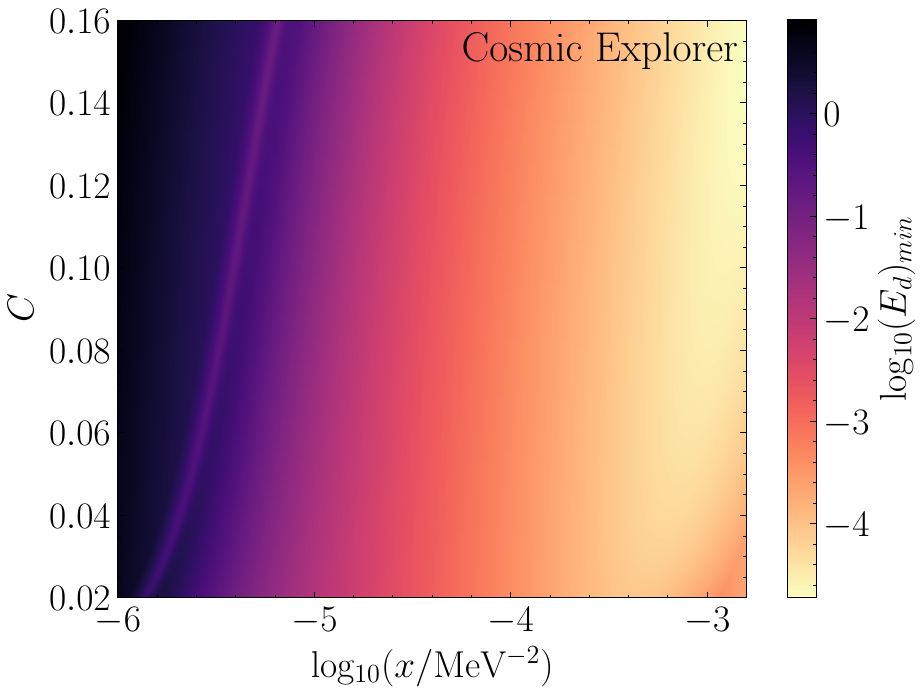}
    \includegraphics[width=0.49\linewidth]{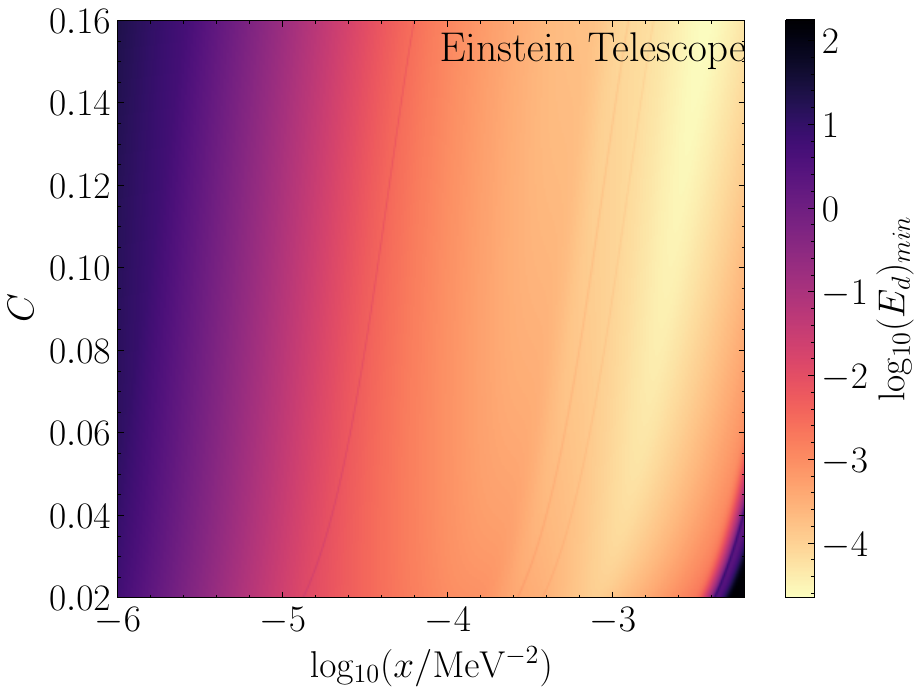}
    \caption{The detectability of BSs using (a) Cosmic Explorer and (b) Einstein Telescope GW detectors. The colours indicate the minimum values of $E_d$ required for the detection with $\rho \ge 5$. The sensitivity curve for CE is obtained from~\url{https://dcc.cosmicexplorer.org/CE-T2000017/public} and from~\url{https://dcc.ligo.org/LIGO-T1500293/public} for ET. The range for $x$ probed is different as the CE noise spectral density used is in the range 5-5000 Hz and 1 Hz - 10 kHz for ET.}
    \label{fig:detectability_ce_et}
\end{figure*}

The same frequency range as that of aLIGO is probed by the CE and ET detectors. We expect better sensitivity with next-generation detectors as they have an order of magnitude improved sensitivity. We show the same plots of CE and ET in Fig.~\ref{fig:detectability_ce_et}. A fainter band in the dark region on the CE plot is due to a dip in the noise at around 4kHz, improving the detectability. We also see a few very fine dark lines in the case of ET, all being due to narrow spikes in the noise spectral density. 
We see that detectabilities of CE and ET are comparable with $\log_{10}(E_d)_{\rm min} \gtrsim -4.5$, which corresponds to a distance of $\sim 30$ Mpc for $E = 10^{-10} M_{\rm BS}$. This improvement in more than an order of magnitude compared to aLIGO is expected as i) $(E_d)_{\rm min} \propto \sqrt{S_n}$, and $\sqrt{S_n}$ is lower by an order of magnitude for ET and CE for $f\gtrsim100$ Hz and ii) the SNR is higher for lower frequencies, and the sensitivity at low frequencies $f\lesssim50$ Hz is better by more than an order of magnitude for CE and ET. For these detectors, we have seen that relevant BS masses are 0.1-10 $M_{\odot}$. For BSs with masses (and frequencies) on extreme ends of the frequency bands, the detectability is again about $\log_{10}(E_d)_{\rm min} \sim 1$ corresponding to 100 pc. All the distances quoted are only for the assumed energy. For bursts with higher energies, farther distances can be probed and vice versa.

\section{Discussions}\label{sec:discussions}

\subsection{Summary}

In a recent work~\cite{shirke_BS_fmodes_short}, we investigated $f$-modes of massive BSs considering the $\phi^4$ self-interaction potential and demonstrated that they obey scaling relations independent of DM model parameters. In this work, we apply these relations to obtain useful analytical fits for global properties and universal relations important for GW asteroseismology, and comment on how to infer microscopic DM parameters with future GW observations. These results are summarized below.

\textit{Scalar DM parameter space:} We display the entire parameter space available for massive BSs and review all the available and relevant constraints. The astrophysical observations provide upper limits on $\lambda$ as a function of mass. A recent model-independent study~\cite{Zimmerman2024} gave a lower limit of the scalar DM mass. As we focus on massive BSs in the strong-interaction limit in this study, we stay in the $\Lambda > 1000$  regime, which defines a lower bound on $\lambda$ for a given scalar DM mass.

\textit{Static properties:} Mass and radius of massive BSs scale with the parameter $x=\sqrt{\lambda}/m^2$. 
We provide analytical fit functions (quadratic and quartic polynomials) along with fit coefficients and errors for the known $M'-C'$, $R'-C'$, and $M'-R'$ curves and report the corresponding errors. The quartic fits have an accuracy of $\lesssim 0.75\%$. The fits are very useful for future works on massive BSs, as direct analytical expressions can be used to calculate static observables instead of numerical solutions, saving computation time. We also find that the maximum density that can be reached at the centre of a stable BS is $\rho_c = \rho'_c /x^2=1.6/x^2$.

\textit{$f$-modes:} We then evaluate the $f$-mode characteristics for massive BSs using a full general-relativistic setup. We use the $f$-mode frequency and damping time scaling relations as reported in~\cite{shirke_BS_fmodes_short}. We report analytical fits for the $f'-M'$, $f'-C'$, and $\tau'-M'$, $\tau'-C'$ numerical relations, which can be used to evaluate the $f$-mode characteristics for any parameter values in the available parameter space. 
All the fits are accurate to within $5\%$ and can be directly used in future works concerning $f$-modes of massive BSs without the need for any numerical calculations.

\textit{Empirical relations:} We study the known empirical and quasi-universal relations for $f$-modes that are essential to probe the properties of interior matter. We showed that since scaling takes away the EoS dependency, we get a unique relation for $f'=f'(\sqrt{M'/R'^3})$ and $R'^4\tau'/M'^3 = \tau'(M'/R')$. We provide linear fits to both that agree within $3\%$. The $\tau'$ relation is found to be universal for BSs; however, it deviates from that for NSs.

\textit{Universal relations:} We also check the well-known universal relation of the mass-scaled $f$-mode characteristics with the dimensionless tidal deformability and compactness.
We report unique relations $M'\omega_f-\bar{\Lambda}$ and $M'\omega_f-C$ for both the real and imaginary parts of the complex eigenfrequency, making it an exact relation for BSs, along with corresponding analytical fits and errors. We also compare it with NSs and find that the relations involving compactness deviate from those for BSs.
We also report an analytical fit for $N=f\tau=f'\tau'$ numerical solution as a function of compactness, representing the number of cycles before the QNM damps $e$-folds. 

\textit{Observable DM parameter space:} Focussing on the observational aspects of these $f$-modes for BSs, we show the maximum possible $f$-mode frequency for each set of parameters $(\lambda, m)$ of DM, spanning the entire available parameter space. Using the frequency range of current and future GW detectors, we show what DM parameter space can be probed by the LISA, LIGO, and NEMO detectors. 
We derive an upper limit of $x$ given by $x_{\rm thresh} = 2.66 \times 10^2$ MeV$^{2}$ such that model parameters with $x > x_{\rm thresh}$ cannot be probed by any GW detector.

\textit{Detectability:} To check the detectability, we explore the $C-x$ parameters space, which represents a BS of compactness $C$ given the DM model parameters. For each point in this space, we calculate the lower bound on a quantity $E_d \equiv \sqrt{E}/d$ assuming a GW burst model for all damping timescales, demanding that the SNR for the $f$-mode burst is $\gtrsim 5$. 
We find that for the $f$-mode energy of $E \approx 10^{-10} M_{\rm BS}$, GW detectors can probe BSs out to cosmological distances. In particular, aLIGO can detect $f$-modes of BSs up to a distance of 1 Mpc and up to 30 Mpc with the next-generation GW detectors like CE and ET. Higher distances, however, can still be probed for stronger bursts with larger $E$. The detectability of BS $f$-modes is much better in the case of LISA as compared to ET and CE, with an observable distance of about 300 Mpc. 

\subsection{Comparison with other works}
\textit{Static observables:} Colpi et al.~\cite{Colpi1986} first showed that in the strong-interaction limit, the EoS and, hence, the mass-radius curves are independent of the DM model parameters. Following this, many works have used this fact, but simple analytical expressions connecting mass, radius, and compactness were not provided. Tang et al.~\cite{Yang2024} provided multiple fits for different mass regions for mini BSs, i.e., in the Kaup limit. 
Pacilio et al.~\cite{Pacilio2020} reported an analytical relation between mass-compactness. None of these works derived analytical fits for macroscopic properties of massive BSs, as provided in this work. 

Maselli et al.~\cite{Maselli2017} showed that massive BSs follow the $I$-Love-$Q$ relations for select values of model parameters in the NS mass range. Wu et al.~\cite{Wu2023} also showed this for a few BS models. In this work, we explicitly prove that such universal relations are expected as the quantities do not scale with model parameters. We show this for $I$-Love-$C$ relations and do not consider the spin-induced quadrupole moment here as we restrict the non-spinning case, also providing a comparison with NSs.

Sennett et al.~\cite{Sennett2017} reported that $\bar{\Lambda}_{\rm min}$ for BSs is 280. A recent work by Cipriani et al.~\cite{Cipriani2024} reported $\bar{\Lambda}_{\rm min}\approx290$ and $k_{2, \rm min}\approx0.045$. Our results are consistent as we obtain $\bar{\Lambda}_{\rm min}=280$ and $k_{2, \rm min}=0.045$. Cardoso et al.~\cite{Cardoso2017} studied tidal deformability in detail. The minimum value for the BSs considered here was not explicitly reported.

\textit{BS QNMs:} There are several studies of BS QNMs in the literature with which a comparison cannot be made. We mention these studies here for completeness, along with the reason why the comparison is infeasible.~\cite{Kojima1991} studied BS QNMs for odd parity, which have been shown not to couple to gravitational radiation. \cite{Yoshida1994} explored the even parity QNMs in detail; however, the work is restricted to the non-interacting case ($\Lambda=0$). \cite{Balakrishna1998} explores QNMs of self-interacting BSs, restricting themselves to the weak-interaction limit ($\Lambda < 200$), and does not focus on the damping times. \cite{Macedo2013a} carries out a similar analysis but for three types of cases (Mini BS, Massive BS, and Solitonic BS). For the case of massive BSs, as discussed in this work, they fix $\Lambda =200$, again in the weak interaction limit. The dimensionless quantity used was $\omega'=\omega/m$ as opposed to $\omega'=\omega\sqrt{\lambda}/m$ in our case. Thus, we cannot compare our results with those of these works.~\cite{Kain2021} studied radial QNMs, which do not give out GWs.

\textit{Massive BS QNMs in the strong-interaction limit:} Flores et al.~\cite{Flores2019} first explored non-radial fundamental modes for massive BSs in the strong-coupling limit in detail. They considered select parameter values consistent with the $\sigma/m$ constraints on self-interacting DM and only restricted the work to BSs falling in the mass range of 1-6$M_{\odot}$ and found that the $f$-mode characteristics of BSs in this mass range follow the empirical universal relations that are different from those followed by NSs. These are consistent with our findings. We showed in~\cite{shirke_BS_fmodes_short} that scaling relations can be used to compute $f$-mode characteristics for any model parameters using a single solution. There is no need to study $f$-modes separately for different parameter values. Using this, we also show why the empirical fits are expected to follow in this case and compare it with the case of NSs. We also include the $f$-Love-C universal relations in our work and extend the work to comment on the range of parameter space that current and future GW detectors can probe and explore detectability. Another work~\cite{Celato2025} appeared recently as this work was being carried out. It explored the $f$-mode universal relations for select model parameters. They also explored the $f$-mode universal relations involving the moment of inertia.

\textit{Observations:} On the observational front,~\cite{Guidice2016, Bezares2018} provided an in-depth review of various observational signatures of ECOs using GWs, focussing on the inspiral phase of binary systems, and the detectability using LIGO was discussed. These were extended to include additional interaction terms~\cite{Croon2019} and for extreme mass ratio inspiral~\cite{Guo2019}.~\cite{shirke_BS_fmodes_short} extended this list to include quasinormal oscillations. Here, we discuss the detectability as well as parameter space and mass-compactness regions that can be probed using BS $f$-modes using LIGO as well as future next-generation GW detectors.

\subsection{Future Scope}
Having derived the scaling and fit relations for non-radial $f$-mode oscillations, these can be used for the searches of massive BSs in the strong-interaction limit where the effect of $f$-modes is important. These would be relevant in the future when $f$-mode oscillations become detectable. $f$-modes for BSs with different interaction potentials, like $\phi^n$~\cite{Pitz2023} or axion potentials, vector bosons, as well as other types of ECOs~\cite{CardosoPani2019} need to be explored further in a similar way. Our analysis on detectability also implies that the non-detection of such bursts can, in turn, be used to rule out BS bursts with parameters indicated in Figs.~\ref{fig:detectability_lisa_aLIGO} and~\ref{fig:detectability_ce_et}.


Stochastic GW background from binary systems of ECOs~\cite{Barausse2018}, BSs~\cite{Croon2018}, and other exotic sources~\cite{Banks2023} are also interesting prospects. The recent detection of stochastic GWs by the IPTA has restored interest in stochastic GW sources. We found that we could have long-lived low-frequency $f$-modes for massive BSs depending on the DM model parameters. These BS oscillations could act as an additional source of the stochastic background.

\appendix
\section{Scaling of Love equations}\label{sec:appendix_love_equations}
It is known that the Tolman-Oppenheimer-Volkoff (TOV) equations, when written in terms of scaled parameters, are independent of $m$ and $\lambda$ (see Appendix of~\cite{Maselli2017}). The same holds for tidal deformability. This was shown in~\cite{Sennett2017} that in the strong-interaction limit, the $\bar{\Lambda}-M'$ curves remain unchanged. We explicitly show and prove here that the solutions of electric and magnetic tidal Love numbers are independent of model parameters when scaled dimensionless quantities are used. We show this by considering the usual definition of Love numbers and the formalism used to compute them for any compact object given the TOV profile $\rho(r)$, $p(r)$, $m(r)$ as shown below. Note that the scaling is valid only in the strong interaction limit.

The tidal field consists of two parts~\cite{Perot2021} i) electric (even parity/polar) ($\epsilon_L$) and ii) magnetic (odd parity/axial) ($M_L$). The corresponding mass multipole ($Q_L$) and current multipole ($S_L$), a pure GR effect, are related by the equations
\begin{align}
    Q_L &= \bar{\lambda_l} \epsilon_L~,~ S_L = \bar{\sigma_l} M_L~,
\end{align}
where, $L$ are the space indices and $l$ is the order of multipole moment. Here, $\bar{\lambda_l}$ and $\bar{\sigma_l}$ are called the gravitoelectric and gravitomagnetic tidal deformability. Dimensionless tidal deformabilities ($\bar{\Lambda_l}, \bar{\Sigma_l}$) are defined by dividing them by a factor $GM^{2l+1}$. These are given by
\begin{align}
    \bar{\Lambda_l} &= \frac{2}{(2l-1)!!}k_l\left(\frac{R}{GM}\right)^{2l+1} = \frac{2}{(2l-1)!!}\frac{k_l}{C^{2l+1}}~, \\
    \bar{\Sigma_l} &= \frac{1}{4(2l-1)!!}j_l\left(\frac{R}{GM}\right)^{2l+1} = \frac{1}{4(2l-1)!!}\frac{j_l}{C^{2l+1}} \\
\end{align}
where $k_l$ and $j_l$ are the gravitoelectric and gravitomagnetic tidal Love numbers. These dimensionless tidal deformabilities depend on the EoS and on the mass configuration via the Love numbers and $C$. We want to see the effect of EoS on this, so we write the equations in terms of scaled quantities. We have already seen that when written in terms of scaled quantities $C=C'$. The dimensionless tidal Love numbers are purely a function of dimensionless compactness and a dimensionless quantity $y$ given by~\cite{Perot2021} $k_l = k_l(C, y_l);~j_l = j_l(C, \tilde{y}_l)~.$ $y_l$ and $\tilde{y}_l$ are in turned evaluated as $y_l = \frac{R}{H_l(R)}\frac{dH_l(R)}{dR};~
    \tilde{y}_l = \frac{R}{\tilde{H}_l(R)}\frac{d\tilde{H}_l(R)}{dR}$, where the profiles for $H_l$ and $\tilde{H}_l$ are computed by solving the differential Eqs. 10 and 11 from~\cite{Perot2021}. Here, $R$ is the radius of the star. By applying the scaling relations $r = r'xM_{\rm Pl}$, $m = m'xM_{\rm Pl}^3$, $\rho = \rho'/x^2$, and $p = p'/x^2$ as discussed in Sec.~\ref{sec:model} for massive BSs in the strong-interaction limit, these equations become

\begin{align}
    \frac{d^2H_l(r')}{dr'^2x^2M_{\rm Pl}^2} &+ \frac{dH_l(r')}{dr'x^2M_{\rm Pl}^2}\left(1-2\frac{m'(r')}{r'}\right)^{-1}\left[\frac{2}{r'} - \frac{2m'(r')}{r'^2} - 4\pi r'(\rho'(r') - p'(r'))\right]  \nonumber \\
    &+\frac{H_l(r')}{x^2M_{\rm Pl}^2}\left(1-2\frac{m'(r')}{r'}\right)^{-1}\Bigg(\Bigg.4\pi \left[5\rho'(r') + 9p'(r') + \frac{d\rho'}{dp'}(\rho'(r') + p'(r'))\right]   \nonumber \\
    &-\frac{l(l+1)}{r'^2} - 4\left(1-\frac{2m'(r')}{r'}\right)^{-1} \left( \frac{m'(r')}{r'^2} + 4\pi r'p'(r')\right)^2\Bigg.\Bigg) =0 ,
\end{align}
\begin{align}
    \frac{d^2\Tilde{H}_l(r')}{dr'^2x^2M_{\rm Pl}^2} &- \frac{d\Tilde{H}_l(r')}{dr'x^2M_{\rm Pl}^2}\left(1-2\frac{m'(r')}{r'}\right)^{-1} 4\pi r'(\rho'(r') + p'(r'))  \nonumber \\
    &- \frac{\Tilde{H}_l(r')}{x^2M_{\rm Pl}^2}\left(1-2\frac{m'(r')}{r'}\right)^{-1}\Bigg(\Bigg.
    \frac{l(l+1)}{r'^2} - \frac{4m'(r')}{r'^3} + 8\pi \theta(p'(r') + \rho'(r'))\Bigg.\Bigg) &= 0~.
\end{align}

Thus, if we define $H'_l = H_l / x^2M_{\rm Pl}^2$ and $\Tilde{H'}_l = \tilde{H}_l/x^2M_{\rm Pl}^2$, the equations become completely free of $x$, as we know from self-similar TOV that $m'(r)$, $p'(r')$, $\epsilon'(r')$ are completely independent of $x$ and only depend on the central density, i.e, we obtain the same Eqs. 10 and 11 of~\cite{Perot2021} in term of $H'_l$ and $\Tilde{H'}_l$. 

Thus, profiles $H'_l(r')$ and $\tilde{H'}(r')$ are independent of $x$. This also means that $y'_l(R')$ and $\tilde{y}'_l(R')$ are independent of $x$. Thus, like $C$, we get a single value for both $y'_l(R')$ and $\tilde{y}'_l(R')$, independent of $x$ for a given central density $\rho'$. This translates to single value of $k_l$, $j_l$ and $\bar{\Lambda}_l$, $\bar{\Sigma}_l$. We can then vary the central density to get a unique relation between the dimensionless tidal deformabilities and $C$ or $M'$.
In other words, for a fixed $M'/R'/C$, the $y_l$, $\tilde{y}_l$ and consequently $k_l$, $j_l$, $\bar{\Lambda_l}$ and  $\bar{\Sigma_l}$ are fixed and do no depend on $x$. $\bar{\Lambda_l}(M')$ and $\bar{\Sigma}(M')$ are thus EoS independent and can be used to obtain the corresponding $\bar{\Lambda_l}(M)$ and $\bar{\Sigma}(M)$ for arbitrary EoS using mass scaling. For example, the quadrupolar dimensionless tidal deformability of arbitrary EoS (parametrized by $x$) and mass configuration can be obtained as $\bar{\Lambda_2}(x, M) = \bar{\Lambda_2}(M' = M/xM_{\rm Pl}^3)$. We have denoted $\bar{\Lambda_2}$ by $\bar{\Lambda}$ throughout the paper. We have mathematically shown that the tidal Love numbers and dimensionless tidal deformability of all orders and both types do not scale with $x$.

\section{Scaling for Moment of Inertia}\label{sec:appendix_I_equations}

The moment of inertia is defined as~\cite{YagiYunes2013}
\begin{equation}
    I \equiv \frac{S}{\Omega}~,
\end{equation}
where $S$ and $\Omega$ are the angular momentum and the angular spin frequency of the star, BS in our case. This is related to the metric function $\omega_1$ outside the star as
\begin{equation}
    \omega_1^{ext} = \Omega - \frac{2S}{R^3} = \Omega \left( 1- \frac{2I}{R^3} \right)~.
\end{equation}
$\omega_1$ is the metric function $(t,\phi)$component of the metric in the Boyer-Lindquist-type coordinates~\cite{YagiYunes2013}. 
We define the scaling $I = I'x^3M_{\rm Pl}^3$. Thus, $\Omega$ and $\omega_1$ remain invariant under this transformation. The complete expression for the moment of inertia used for computation is given by
\begin{align}
    I &= \frac{8\pi}{3} \frac{1}{\Omega} \int_0^{R} \frac{e^{-(\hat{\nu} + \hat{\lambda} )/2} R^5 G(\rho + p) \omega_1}{R-2GM(R)} dR~, \\
    \implies I &= \frac{8\pi}{3} \frac{1}{\Omega'} \int_0^{R'} \frac{e^{-(\hat{\nu}' + \hat{\lambda}' )/2} R'^5 (\rho' + p') \omega'_1 (x^3M_{\rm Pl}^3)}{R'-2M'(R')} dR'
\end{align}
Here, $\hat{\nu}$ and $\hat{\lambda}$ are metric function. Using $I = I'x^3M_{\rm Pl}^3$, the equations become independent of $x$.
The dimensionless moment of inertia ($\Bar{I}$) is defined as
 \begin{equation}
     \bar{I} \equiv \frac{I}{(GM)^3} = \frac{I'}{M'^3} = \bar{I'}~.
 \end{equation}
 Thus, the dimensionless moment of inertia does not scale with model parameters.

\section{The $I-$Love$-C$ Relations}\label{sec:appendix_I_Love_C}
The $I$-Love-$Q$ universal relations connecting the moment of inertia, the tidal deformability, and the spin-induced quadrupole moments for NSs have been known for a long time~\cite {YagiYunes2013}. This means that they are followed regardless of the underlying EoS. These relations were shown to be followed by dark BSs by Maselli et al.~\cite{Maselli2017} for select values of $\lambda \in {0.5,1.0,1.5}\pi$ and $m \in {300,400}$ MeV. These are shown in the orange patch in Fig.~\ref{fig:parameter_space}. They found that dark BSs with these model parameters admit the $I$-Love-$Q$ universal relations, i.e., all BS EoSs followed it regardless of the model parameters. This was confirmed by another recent work for a few BS models~\cite{Wu2023}. This was an empirical observation. Here, we prove in theory that BSs follow these universal relations. We do not consider the spin-induced quadrupole moments as we restrict them to non-spinning BSs. Note that spinning scalar BSs without interactions are known to lose sphericity from torus/donut-shaped configurations~\cite{Sanchis-Gual2019}.

\begin{figure*}
    \includegraphics[width=0.32\linewidth, height =6cm]{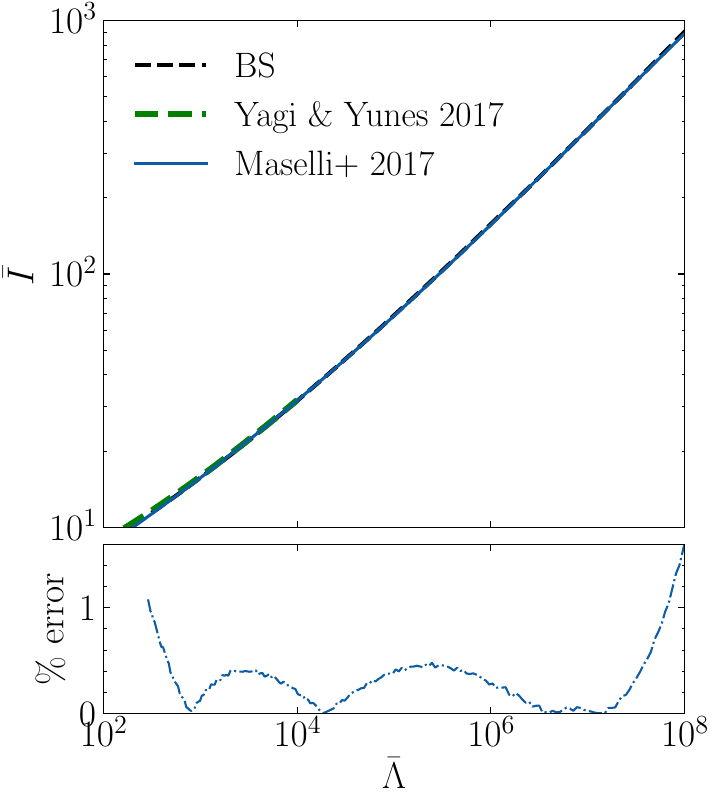}
    \label{fig:UR_I_love_fit}
    \includegraphics[width=0.32\linewidth, height =6cm]{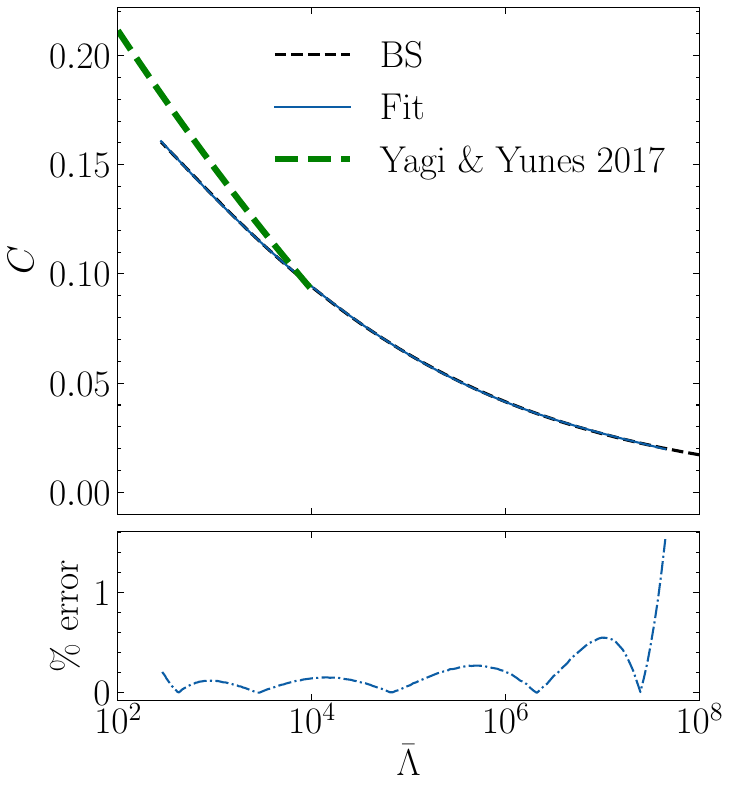}
    \label{fig:UR_C_love_fit}
    \includegraphics[width=0.32\linewidth, height =6cm]{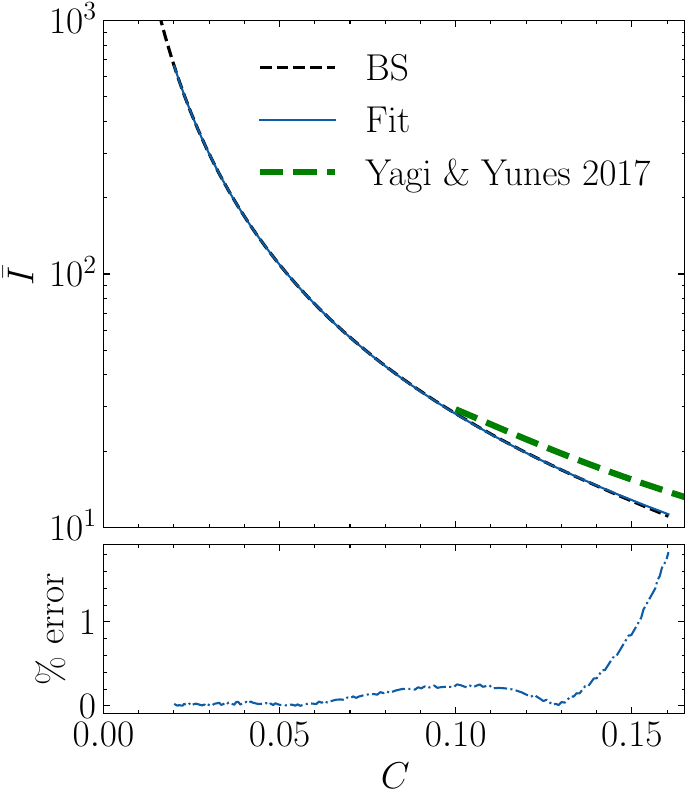}
    \label{fig:UR_I_C_fit}
    \caption{The (a) $\bar{I}$-Love (b) $C$-Love and (c) $\bar{I}$-$C$ relations for massive BSs in the strong-interaction limit. The blue curve in (a) is the fit provided in~\cite{Maselli2017}. For the other two relations, we provide the fits we performed as discussed in the text. The $I$-Love-$C$ fits as provided in~\cite{Yagi2017} for NSs are shown in green. We have shown this up to $C=0.1$ ($\bar{\Lambda}=10^4$) as appropriate for NSs. The errors corresponding to the fits are shown in the lower panels in each case. }
    \label{fig:i_love_c_fits}
\end{figure*}

In Appendix~\ref{sec:appendix_love_equations} and Appendix.~\ref{sec:appendix_I_equations} we explicitly show that $\bar{\Lambda}$ and $\Bar{I}$ are independent of model parameters ($\lambda$, $m$). Thus, we can obtain the $\bar{I}-\bar{\Lambda}$ curve for any BS EoS, and this will be followed by all the EoSs. This is also the reason why all the curves $I$-Love-$Q$ obtained for different EoS parameters in~\cite{Maselli2017} exactly overlap without having any spread. 
We show this $\bar{I}-\bar{\Lambda}$ in Fig.\ref{fig:i_love_c_fits}(a). The black dashed curve is the numerical solution obtained for the scaled EoS. We further plot the fit relation obtained for NSs by~\cite{YagiYunes2013} for the sake of comparison in green. We conclude there is a degeneracy with NSs, and this relation does not break it. The blue curve shows the $I$-Love fit as obtained in~\cite{Maselli2017} for BSs (coefficients are given in Table.~\ref{tab:quadratic_quartic_fit_coefficients_iLoveC}) with $X=\log{\bar{\Lambda}}$, and $Y = \log{\bar{I}}$. This fit (blue) is in excellent agreement with our numerical results as it fits within an accuracy of $1\%$, thus providing a veracity check for our results. However, the same fit had an error of about $10\%$ in~\cite{Maselli2017} (see Fig. 7 of~\cite{Maselli2017}). Thus, we find the fit to be better than what was claimed before and do not provide any new relation here. Note that~\cite{Wu2023} also reported a similar fit relation using select BS models.~\cite{Maselli2017} also found a universal relation for $k_2-C$.  In Appendix.~\ref{sec:appendix_love_equations}, we have proved the same, i.e., we have shown that $k_2$ is also EoS independent, i.e., they do not scale with $x$. 

We also show the $C$-Love and $\bar{I}$-C relations in Figs.~\ref{fig:i_love_c_fits}(b) and~\ref{fig:i_love_c_fits}(c), respectively. We did not find these relations in the literature for BSs. 
We use the same functional forms for BSs as used in~\cite{Yagi2017} for NSs and strange quark stars. For the $C$-Love relation we use
\begin{equation}
    C = \sum_{k=0}^4 b_k(\ln{\bar{\Lambda}})^k~,
\end{equation}
while for the $\bar{I}$-$C$ relation we use
\begin{equation}
    \bar{I} = \sum_{k=1}^{4} b_k C^{-k}~.
\end{equation}
The fit coefficients for all the cases are given in Table.~\ref{tab:quadratic_quartic_fit_coefficients_iLoveC}. These analytical fits to the numerically obtained relations have an error below $2\%$. The corresponding fit relations for NSs reported in~\cite{Yagi2017} are shown in green dashed curves. We plot it in only those regimes for which they were derived, i.e., for $C\ge 0.1$ and $\bar{\Lambda} \le 10^4$. We find that both relations differ from the BS relation at higher compactness (or higher mass, or lower $\bar{\Lambda}$). 

In the case of NSs, the relations involving compactness have an uncertainty of up to $10\%$  owing to uncertainty in the EoS~\cite{Yagi2017}, while the $I$-Love relation is much more accurate for NSs with an error under $1\%$~\cite{Yagi2017}. In the case of BSs, however, we have shown mathematically that each of these relations is a unique curve in the strong interaction limit and does not depend on the BS model parameters. Hence, these relations are exactly the same for all massive BSs in the strong interaction limit.

\begin{table}[]
    \centering
    \begin{tabular}{c|c|c|c|c|c|c|c|c|c}
         $Y$ & $X$ &  $b_0$ & $b_1$ & $b_2$ & $b_3$ & $b_4$ \\  \hline
         $\log{\bar{I}}$ & $\log{\bar{\Lambda}}$  & 1.38 & 9.46 $\times 10^{-2}$ & 1.84$\times 10^{-2}$ & -5.84$\times 10^{-4}$ & 5.51$\times 10^{-6}$ \\
         $C$ & $\log{\bar{\Lambda}}$  & 2.859 $\times 10^{-1}$ & -2.068 $\times 10^{-2}$ & -8.276$\times 10^{-4}$ & 1.139$\times 10^{-4}$ & -2.781$\times 10^{-6}$ \\
         $\bar{I}$ & $1/C$ & $-$ & 1.532 $\times 10^{-1}$ & 2.664 $\times 10^{-1}$ & -1.317$\times 10^{-4}$ & 1.535 $\times 10^{-6}$ \\
    \end{tabular}
    \caption{Fitting coefficients for the fits between the quantities $\bar{I}$, $\bar{\Lambda}$, and $C$ for massive BSs in strong-interaction limit. The fitting function is given in Eqs.~\ref{eqn:quadratic_quartic_relations}. The coefficients for the $\bar{I}$-Love relation are the same as they were reported in~\cite{Maselli2017}.}
    \label{tab:quadratic_quartic_fit_coefficients_iLoveC}
\end{table}

\acknowledgments
L.S. and J.S.B. acknowledge support by the
Deutsche Forschungsgemeinschaft (DFG, German Research Foundation) through the CRC-TR 211 ``Strong-interaction matter under extreme conditions'' – project
no. 315477589 – TRR 211. S.S. acknowledges discussions with Surhud More, with Paolo Pani at the International Centre for Theoretical Sciences (ICTS) program - Beyond the Horizon: Testing the black hole paradigm (code: ICTS/BTH2025/03), and the online workshop ``Dynamical Tracers of the Nature of Dark Matter" which contributed to the improvement of the manuscript. The authors thank the anonymous referee for their comments, which helped improve the manuscript significantly and enhance the interpretation of the results.

\bibliographystyle{JHEP}
\bibliography{refs}

@ARTICLE{Flores2019,
       author = {{V{\'a}squez Flores}, C. and {Parisi}, Alessandro and {Chen}, Chian-Shu and {Lugones}, Germ{\'a}n},
        title = "{Fundamental oscillation modes of self-interacting bosonic dark stars}",
      journal = {Journal of Cosmology and Astroparticle Physics},
         year = 2019,
        month = jun,
       volume = {2019},
       number = {6},
          eid = {051},
        pages = {051},
          doi = {10.1088/1475-7516/2019/06/051},
archivePrefix = {arXiv},
       eprint = {1901.07157},
 primaryClass = {hep-ph},
       adsurl = {https://ui.adsabs.harvard.edu/abs/2019JCAP...06..051V}
}

@ARTICLE{Pradhan2022,
       author = {{Pradhan}, Bikram Keshari and {Chatterjee}, Debarati and {Lanoye}, Michael and {Jaikumar}, Prashanth},
        title = "{General relativistic treatment of f -mode oscillations of hyperonic stars}",
      journal = {Physical Review C},
         year = 2022,
        month = jul,
       volume = {106},
       number = {1},
          eid = {015805},
        pages = {015805},
          doi = {10.1103/PhysRevC.106.015805},
       adsurl = {https://ui.adsabs.harvard.edu/abs/2022PhRvC.106a5805P}
}

@ARTICLE{Pradhan2023,
       author = {{Pradhan}, Bikram Keshari and {Vijaykumar}, Aditya and {Chatterjee}, Debarati},
        title = "{Impact of updated multipole Love numbers and f -Love universal relations in the context of binary neutron stars}",
      journal = {\prd},
         year = 2023,
        month = jan,
       volume = {107},
       number = {2},
          eid = {023010},
        pages = {023010},
          doi = {10.1103/PhysRevD.107.023010},
archivePrefix = {arXiv},
       eprint = {2210.09425},
 primaryClass = {astro-ph.HE},
       adsurl = {https://ui.adsabs.harvard.edu/abs/2023PhRvD.107b3010P}
}

@ARTICLE{YagiYunes2013,
       author = {{Yagi}, Kent and {Yunes}, Nicol{\'a}s},
        title = "{I-Love-Q relations in neutron stars and their applications to astrophysics, gravitational waves, and fundamental physics}",
      journal = {\prd},
         year = 2013,
        month = jul,
       volume = {88},
       number = {2},
          eid = {023009},
        pages = {023009},
          doi = {10.1103/PhysRevD.88.023009},
archivePrefix = {arXiv},
       eprint = {1303.1528},
 primaryClass = {gr-qc},
       adsurl = {https://ui.adsabs.harvard.edu/abs/2013PhRvD..88b3009Y}
}

@ARTICLE{AnderssonKokkotas1996,
       author = {{Andersson}, Nils and {Kokkotas}, Kostas D.},
        title = "{Gravitational Waves and Pulsating Stars: What Can We Learn from Future Observations?}",
      journal = {\prl},
         year = 1996,
        month = nov,
       volume = {77},
       number = {20},
        pages = {4134-4137},
          doi = {10.1103/PhysRevLett.77.4134},
archivePrefix = {arXiv},
       eprint = {gr-qc/9610035},
 primaryClass = {gr-qc},
       adsurl = {https://ui.adsabs.harvard.edu/abs/1996PhRvL..77.4134A}
}

@ARTICLE{AnderssonKokkotas1998,
       author = {{Andersson}, Nils and {Kokkotas}, Kostas D.},
        title = "{Towards gravitational wave asteroseismology}",
      journal = {\mnras},
         year = 1998,
        month = oct,
       volume = {299},
       number = {4},
        pages = {1059-1068},
          doi = {10.1046/j.1365-8711.1998.01840.x},
archivePrefix = {arXiv},
       eprint = {gr-qc/9711088},
 primaryClass = {gr-qc},
       adsurl = {https://ui.adsabs.harvard.edu/abs/1998MNRAS.299.1059A}
}

@ARTICLE{Chan2014,
       author = {{Chan}, T.~K. and {Sham}, Y. -H. and {Leung}, P.~T. and {Lin}, L. -M.},
        title = "{Multipolar universal relations between f -mode frequency and tidal deformability of compact stars}",
      journal = {\prd},
         year = 2014,
        month = dec,
       volume = {90},
       number = {12},
          eid = {124023},
        pages = {124023},
          doi = {10.1103/PhysRevD.90.124023},
archivePrefix = {arXiv},
       eprint = {1408.3789},
 primaryClass = {gr-qc},
       adsurl = {https://ui.adsabs.harvard.edu/abs/2014PhRvD..90l4023C}
}

@ARTICLE{SotaniKumar2021,
       author = {{Sotani}, Hajime and {Kumar}, Bharat},
        title = "{Universal relations between the quasinormal modes of neutron star and tidal deformability}",
      journal = {\prd},
         year = 2021,
        month = dec,
       volume = {104},
       number = {12},
          eid = {123002},
        pages = {123002},
          doi = {10.1103/PhysRevD.104.123002},
       adsurl = {https://ui.adsabs.harvard.edu/abs/2021PhRvD.104l3002S}
}

@ARTICLE{Colpi1986,
       author = {{Colpi}, Monica and {Shapiro}, Stuart L. and {Wasserman}, Ira},
        title = "{Boson stars: Gravitational equilibria of self-interacting scalar fields}",
      journal = {Physical Review Letters},
         year = 1986,
        month = nov,
       volume = {57},
       number = {20},
        pages = {2485-2488},
          doi = {10.1103/PhysRevLett.57.2485},
       adsurl = {https://ui.adsabs.harvard.edu/abs/1986PhRvL..57.2485C}
}

@ARTICLE{Seoane2010,
       author = {{Amaro-Seoane}, Pau and {Barranco}, Juan and {Bernal}, Argelia and {Rezzolla}, Luciano},
        title = "{Constraining scalar fields with stellar kinematics and collisional dark matter}",
      journal = {Journal of Cosmology and Astroparticle Physics},
         year = 2010,
        month = nov,
       volume = {2010},
       number = {11},
          eid = {002},
        pages = {002},
          doi = {10.1088/1475-7516/2010/11/002},
archivePrefix = {arXiv},
       eprint = {1009.0019},
 primaryClass = {astro-ph.CO},
       adsurl = {https://ui.adsabs.harvard.edu/abs/2010JCAP...11..002A}
}

@ARTICLE{Eby2016,
       author = {{Eby}, Joshua and {Kouvaris}, Chris and {Nielsen}, Niklas Gr{\o}nlund and {Wijewardhana}, L.~C.~R.},
        title = "{Boson stars from self-interacting dark matter}",
      journal = {Journal of High Energy Physics},
         year = 2016,
        month = feb,
       volume = {2016},
          eid = {28},
        pages = {28},
          doi = {10.1007/JHEP02(2016)028},
archivePrefix = {arXiv},
       eprint = {1511.04474},
 primaryClass = {hep-ph},
       adsurl = {https://ui.adsabs.harvard.edu/abs/2016JHEP...02..028E}
}

@ARTICLE{Kaup1968,
       author = {{Kaup}, David J.},
        title = "{Klein-Gordon Geon}",
      journal = {Physical Review},
         year = 1968,
        month = aug,
       volume = {172},
       number = {5},
        pages = {1331-1342},
          doi = {10.1103/PhysRev.172.1331},
       adsurl = {https://ui.adsabs.harvard.edu/abs/1968PhRv..172.1331K}
}

@ARTICLE{Ruffini1969,
       author = {{Ruffini}, Remo and {Bonazzola}, Silvano},
        title = "{Systems of Self-Gravitating Particles in General Relativity and the Concept of an Equation of State}",
      journal = {Physical Review},
         year = 1969,
        month = nov,
       volume = {187},
       number = {5},
        pages = {1767-1783},
          doi = {10.1103/PhysRev.187.1767},
       adsurl = {https://ui.adsabs.harvard.edu/abs/1969PhRv..187.1767R}
}

@ARTICLE{Feinblum1968,
       author = {{Feinblum}, David A. and {McKinley}, William A.},
        title = "{Stable States of a Scalar Particle in Its Own Gravational Field}",
      journal = {Physical Review},
         year = 1968,
        month = apr,
       volume = {168},
       number = {5},
        pages = {1445-1450},
          doi = {10.1103/PhysRev.168.1445},
       adsurl = {https://ui.adsabs.harvard.edu/abs/1968PhRv..168.1445F}
}

@ARTICLE{Mielke1981,
       author = {{Mielke}, Eckehard W. and {Scherzer}, Reinhard},
        title = "{Geon-type solutions of the nonlinear Heisenberg-Klein-Gordon equation}",
      journal = {\prd},
         year = 1981,
        month = oct,
       volume = {24},
       number = {8},
        pages = {2111-2126},
          doi = {10.1103/PhysRevD.24.2111},
       adsurl = {https://ui.adsabs.harvard.edu/abs/1981PhRvD..24.2111M}
}

@ARTICLE{Detweiler85,
       author = {{Detweiler}, S. and {Lindblom}, L.},
        title = "{On the nonradial pulsations of general relativistic stellar models}",
      journal = {\apj},
         year = 1985,
        month = may,
       volume = {292},
        pages = {12-15},
          doi = {10.1086/163127},
       adsurl = {https://ui.adsabs.harvard.edu/abs/1985ApJ...292...12D}
}

@article{Sotani2001,
  title = {Density discontinuity of a neutron star and gravitational waves},
  author = {Sotani, Hajime and Tominaga, Kazuhiro and Maeda, Kei-ichi},
  journal = {Phys. Rev. D},
  volume = {65},
  issue = {2},
  pages = {024010},
  numpages = {12},
  year = {2001},
  month = {Dec},
  publisher = {American Physical Society},
  doi = {10.1103/PhysRevD.65.024010},
  url = {https://link.aps.org/doi/10.1103/PhysRevD.65.024010}

}

@ARTICLE{Macedo2013a,
       author = {{Macedo}, Caio F.~B. and {Pani}, Paolo and {Cardoso}, Vitor and {Crispino}, Lu{\'\i}s C.~B.},
        title = "{Astrophysical signatures of boson stars: Quasinormal modes and inspiral resonances}",
      journal = {\prd},
         year = 2013,
        month = sep,
       volume = {88},
       number = {6},
          eid = {064046},
        pages = {064046},
          doi = {10.1103/PhysRevD.88.064046},
archivePrefix = {arXiv},
       eprint = {1307.4812},
 primaryClass = {gr-qc},
       adsurl = {https://ui.adsabs.harvard.edu/abs/2013PhRvD..88f4046M}
}

@ARTICLE{Croon2019,
       author = {{Croon}, Djuna and {Fan}, JiJi and {Sun}, Chen},
        title = "{Boson star from repulsive light scalars and gravitational waves}",
      journal = {\jcap},
         year = 2019,
        month = apr,
       volume = {2019},
       number = {4},
          eid = {008},
        pages = {008},
          doi = {10.1088/1475-7516/2019/04/008},
archivePrefix = {arXiv},
       eprint = {1810.01420},
 primaryClass = {hep-ph},
       adsurl = {https://ui.adsabs.harvard.edu/abs/2019JCAP...04..008C}
}

@ARTICLE{Sennett2017,
       author = {{Sennett}, Noah and {Hinderer}, Tanja and {Steinhoff}, Jan and {Buonanno}, Alessandra and {Ossokine}, Serguei},
        title = "{Distinguishing boson stars from black holes and neutron stars from tidal interactions in inspiraling binary systems}",
      journal = {\prd},
         year = 2017,
        month = jul,
       volume = {96},
       number = {2},
          eid = {024002},
        pages = {024002},
          doi = {10.1103/PhysRevD.96.024002},
archivePrefix = {arXiv},
       eprint = {1704.08651},
 primaryClass = {gr-qc},
       adsurl = {https://ui.adsabs.harvard.edu/abs/2017PhRvD..96b4002S}
}

@ARTICLE{Barausse2018,
       author = {{Barausse}, Enrico and {Brito}, Richard and {Cardoso}, Vitor and {Dvorkin}, Irina and {Pani}, Paolo},
        title = "{The stochastic gravitational-wave background in the absence of horizons}",
      journal = {Classical and Quantum Gravity},
         year = 2018,
        month = oct,
       volume = {35},
       number = {20},
          eid = {20LT01},
        pages = {20LT01},
          doi = {10.1088/1361-6382/aae1de},
archivePrefix = {arXiv},
       eprint = {1805.08229},
 primaryClass = {gr-qc},
       adsurl = {https://ui.adsabs.harvard.edu/abs/2018CQGra..35tLT01B}
}

@ARTICLE{Guo2019,
       author = {{Guo}, Huai-Ke and {Sinha}, Kuver and {Sun}, Chen},
        title = "{Probing boson stars with extreme mass ratio inspirals}",
      journal = {\jcap},
         year = 2019,
        month = sep,
       volume = {2019},
       number = {9},
          eid = {032},
        pages = {032},
          doi = {10.1088/1475-7516/2019/09/032},
archivePrefix = {arXiv},
       eprint = {1904.07871},
 primaryClass = {hep-ph},
       adsurl = {https://ui.adsabs.harvard.edu/abs/2019JCAP...09..032G}
}

@ARTICLE{Croon2018,
       author = {{Croon}, Djuna and {Gleiser}, Marcelo and {Mohapatra}, Sonali and {Sun}, Chen},
        title = "{Gravitational radiation background from boson star binaries}",
      journal = {Physics Letters B},
         year = 2018,
        month = aug,
       volume = {783},
        pages = {158-162},
          doi = {10.1016/j.physletb.2018.03.055},
archivePrefix = {arXiv},
       eprint = {1802.08259},
 primaryClass = {hep-ph},
       adsurl = {https://ui.adsabs.harvard.edu/abs/2018PhLB..783..158C}
}

@ARTICLE{Arbey2003,
       author = {{Arbey}, Alexandre and {Lesgourgues}, Julien and {Salati}, Pierre},
        title = "{Galactic halos of fluid dark matter}",
      journal = {\prd},
         year = 2003,
        month = jul,
       volume = {68},
       number = {2},
          eid = {023511},
        pages = {023511},
          doi = {10.1103/PhysRevD.68.023511},
archivePrefix = {arXiv},
       eprint = {astro-ph/0301533},
 primaryClass = {astro-ph},
       adsurl = {https://ui.adsabs.harvard.edu/abs/2003PhRvD..68b3511A}
}

@ARTICLE{Cardoso2016,
       author = {{Cardoso}, Vitor and {Hopper}, Seth and {Macedo}, Caio F.~B. and {Palenzuela}, Carlos and {Pani}, Paolo},
        title = "{Gravitational-wave signatures of exotic compact objects and of quantum corrections at the horizon scale}",
      journal = {\prd},
         year = 2016,
        month = oct,
       volume = {94},
       number = {8},
          eid = {084031},
        pages = {084031},
          doi = {10.1103/PhysRevD.94.084031},
archivePrefix = {arXiv},
       eprint = {1608.08637},
 primaryClass = {gr-qc},
       adsurl = {https://ui.adsabs.harvard.edu/abs/2016PhRvD..94h4031C}
}

@ARTICLE{Palenzuela2017,
       author = {{Palenzuela}, Carlos and {Pani}, Paolo and {Bezares}, Miguel and {Cardoso}, Vitor and {Lehner}, Luis and {Liebling}, Steven},
        title = "{Gravitational wave signatures of highly compact boson star binaries}",
      journal = {\prd},
         year = 2017,
        month = nov,
       volume = {96},
       number = {10},
          eid = {104058},
        pages = {104058},
          doi = {10.1103/PhysRevD.96.104058},
archivePrefix = {arXiv},
       eprint = {1710.09432},
 primaryClass = {gr-qc},
       adsurl = {https://ui.adsabs.harvard.edu/abs/2017PhRvD..96j4058P}
}

@ARTICLE{Palenzuela2007,
       author = {{Palenzuela}, C. and {Olabarrieta}, I. and {Lehner}, L. and {Liebling}, S.~L.},
        title = "{Head-on collisions of boson stars}",
      journal = {\prd},
         year = 2007,
        month = mar,
       volume = {75},
       number = {6},
          eid = {064005},
        pages = {064005},
          doi = {10.1103/PhysRevD.75.064005},
archivePrefix = {arXiv},
       eprint = {gr-qc/0612067},
 primaryClass = {gr-qc},
       adsurl = {https://ui.adsabs.harvard.edu/abs/2007PhRvD..75f4005P}
}

@ARTICLE{Palenzuela2008,
       author = {{Palenzuela}, C. and {Lehner}, L. and {Liebling}, S.~L.},
        title = "{Orbital dynamics of binary boson star systems}",
      journal = {\prd},
         year = 2008,
        month = feb,
       volume = {77},
       number = {4},
          eid = {044036},
        pages = {044036},
          doi = {10.1103/PhysRevD.77.044036},
archivePrefix = {arXiv},
       eprint = {0706.2435},
 primaryClass = {gr-qc},
       adsurl = {https://ui.adsabs.harvard.edu/abs/2008PhRvD..77d4036P}
}

@ARTICLE{LieblingPalenzuela2023,
       author = {{Liebling}, Steven L. and {Palenzuela}, Carlos},
        title = "{Dynamical boson stars}",
      journal = {Living Reviews in Relativity},
         year = 2023,
        month = dec,
       volume = {26},
       number = {1},
          eid = {1},
        pages = {1},
          doi = {10.1007/s41114-023-00043-4},
       adsurl = {https://ui.adsabs.harvard.edu/abs/2023LRR....26....1L}
}

@ARTICLE{CardosoPani2019,
       author = {{Cardoso}, Vitor and {Pani}, Paolo},
        title = "{Testing the nature of dark compact objects: a status report}",
      journal = {Living Reviews in Relativity},
         year = 2019,
        month = jul,
       volume = {22},
       number = {1},
          eid = {4},
        pages = {4},
          doi = {10.1007/s41114-019-0020-4},
archivePrefix = {arXiv},
       eprint = {1904.05363},
 primaryClass = {gr-qc},
       adsurl = {https://ui.adsabs.harvard.edu/abs/2019LRR....22....4C}
}

@ARTICLE{Cipriani2024,
       author = {{Cipriani}, Lorenzo and {Mannarelli}, Massimo and {Nesti}, Fabrizio and {Trabucco}, Silvia},
        title = "{Superfluid dark stars}",
      journal = {arXiv e-prints},
         year = 2024,
        month = mar,
          eid = {arXiv:2403.03833},
        pages = {arXiv:2403.03833},
          doi = {10.48550/arXiv.2403.03833},
archivePrefix = {arXiv},
       eprint = {2403.03833},
 primaryClass = {astro-ph.CO},
       adsurl = {https://ui.adsabs.harvard.edu/abs/2024arXiv240303833C}
}

@ARTICLE{Tulin2018,
       author = {{Tulin}, Sean and {Yu}, Hai-Bo},
        title = "{Dark matter self-interactions and small scale structure}",
      journal = {\physrep},
         year = 2018,
        month = feb,
       volume = {730},
        pages = {1-57},
          doi = {10.1016/j.physrep.2017.11.004},
archivePrefix = {arXiv},
       eprint = {1705.02358},
 primaryClass = {hep-ph},
       adsurl = {https://ui.adsabs.harvard.edu/abs/2018PhR...730....1T}
}

@ARTICLE{Kojima1991,
       author = {{Kojima}, Y. and {Yoshida}, S. and {Futamase}, T.},
        title = "{Non-Radial Pulsation of a Boson Star. I ---Formulation---}",
      journal = {Progress of Theoretical Physics},
         year = 1991,
        month = aug,
       volume = {86},
       number = {2},
        pages = {401-410},
          doi = {10.1143/ptp/86.2.401},
       adsurl = {https://ui.adsabs.harvard.edu/abs/1991PThPh..86..401K}
}

@ARTICLE{Yoshida1994,
       author = {{Yoshida}, Shijun and {Eriguchi}, Yoshiharu and {Futamase}, Toshifumi},
        title = "{Quasinormal modes of boson stars}",
      journal = {\prd},
         year = 1994,
        month = nov,
       volume = {50},
       number = {10},
        pages = {6235-6246},
          doi = {10.1103/PhysRevD.50.6235},
       adsurl = {https://ui.adsabs.harvard.edu/abs/1994PhRvD..50.6235Y}
}

@ARTICLE{Balakrishna1998,
       author = {{Balakrishna}, Jayashree and {Seidel}, Edward and {Suen}, Wai-Mo},
        title = "{Dynamical evolution of boson stars. II. Excited states and self-interacting fields}",
      journal = {\prd},
         year = 1998,
        month = nov,
       volume = {58},
       number = {10},
          eid = {104004},
        pages = {104004},
          doi = {10.1103/PhysRevD.58.104004},
archivePrefix = {arXiv},
       eprint = {gr-qc/9712064},
 primaryClass = {gr-qc},
       adsurl = {https://ui.adsabs.harvard.edu/abs/1998PhRvD..58j4004B}
}

@ARTICLE{Guidice2016,
       author = {{Giudice}, Gian F. and {McCullough}, Matthew and {Urbano}, Alfredo},
        title = "{Hunting for dark particles with gravitational waves}",
      journal = {\jcap},
         year = 2016,
        month = oct,
       volume = {2016},
       number = {10},
          eid = {001},
        pages = {001},
          doi = {10.1088/1475-7516/2016/10/001},
archivePrefix = {arXiv},
       eprint = {1605.01209},
 primaryClass = {hep-ph},
       adsurl = {https://ui.adsabs.harvard.edu/abs/2016JCAP...10..001G}
}

@ARTICLE{Visinelli2021,
       author = {{Visinelli}, Luca},
        title = "{Boson stars and oscillatons: A review}",
      journal = {International Journal of Modern Physics D},
         year = 2021,
        month = jan,
       volume = {30},
       number = {15},
          eid = {2130006-293},
        pages = {2130006-293},
          doi = {10.1142/S0218271821300068},
archivePrefix = {arXiv},
       eprint = {2109.05481},
 primaryClass = {gr-qc},
       adsurl = {https://ui.adsabs.harvard.edu/abs/2021IJMPD..3030006V}
}

@ARTICLE{Pitz2023,
       author = {{Pitz}, Sarah Louisa and {Schaffner-Bielich}, J{\"u}rgen},
        title = "{Generating ultracompact boson stars with modified scalar potentials}",
      journal = {\prd},
         year = 2023,
        month = nov,
       volume = {108},
       number = {10},
          eid = {103043},
        pages = {103043},
          doi = {10.1103/PhysRevD.108.103043},
archivePrefix = {arXiv},
       eprint = {2308.01254},
 primaryClass = {astro-ph.HE},
       adsurl = {https://ui.adsabs.harvard.edu/abs/2023PhRvD.108j3043P}
}

@ARTICLE{Dentler2022,
       author = {{Dentler}, Mona and {Marsh}, David J.~E. and {Hlo{\v{z}}ek}, Ren{\'e}e and {Lagu{\"e}}, Alex and {Rogers}, Keir K. and {Grin}, Daniel},
        title = "{Fuzzy dark matter and the Dark Energy Survey Year 1 data}",
      journal = {\mnras},
         year = 2022,
        month = oct,
       volume = {515},
       number = {4},
        pages = {5646-5664},
          doi = {10.1093/mnras/stac1946},
archivePrefix = {arXiv},
       eprint = {2111.01199},
 primaryClass = {astro-ph.CO},
       adsurl = {https://ui.adsabs.harvard.edu/abs/2022MNRAS.515.5646D}
}

@ARTICLE{Cembranos2018,
       author = {{Cembranos}, J.~A.~R. and {Maroto}, A.~L. and {N{\'u}{\~n}ez Jare{\~n}o}, S.~J. and {Villarrubia-Rojo}, H.},
        title = "{Constraints on anharmonic corrections of fuzzy dark matter}",
      journal = {Journal of High Energy Physics},
         year = 2018,
        month = aug,
       volume = {2018},
       number = {8},
          eid = {73},
        pages = {73},
          doi = {10.1007/JHEP08(2018)073},
archivePrefix = {arXiv},
       eprint = {1805.08112},
 primaryClass = {astro-ph.CO},
       adsurl = {https://ui.adsabs.harvard.edu/abs/2018JHEP...08..073C}
}

@ARTICLE{PlanckCollaboration2016,
       author = {{Planck Collaboration} and {Ade}, P.~A.~R. and {Aghanim}, N. and {Arnaud}, M. and {Ashdown}, M. and {Aumont}, J. and {Baccigalupi}, C. and {Banday}, A.~J. and {Barreiro}, R.~B. and {Bartlett}, J.~G. and others},
        title = "{Planck 2015 results. XIII. Cosmological parameters}",
      journal = {\aap},
         year = 2016,
        month = sep,
       volume = {594},
          eid = {A13},
        pages = {A13},
          doi = {10.1051/0004-6361/201525830},
archivePrefix = {arXiv},
       eprint = {1502.01589},
 primaryClass = {astro-ph.CO},
       adsurl = {https://ui.adsabs.harvard.edu/abs/2016A&A...594A..13P}
}

@ARTICLE{Parkinson2012,
       author = {{Parkinson}, David and {Riemer-S{\o}rensen}, Signe and {Blake}, Chris and {Poole}, Gregory B. and {Davis}, Tamara M. and {Brough}, Sarah and {Colless}, Matthew and {Contreras}, Carlos and {Couch}, Warrick and {Croom}, Scott and {Croton}, Darren and {Drinkwater}, Michael J. and {Forster}, Karl and {Gilbank}, David and {Gladders}, Mike and {Glazebrook}, Karl and {Jelliffe}, Ben and {Jurek}, Russell J. and {Li}, I. -hui and {Madore}, Barry and {Martin}, D. Christopher and {Pimbblet}, Kevin and {Pracy}, Michael and {Sharp}, Rob and {Wisnioski}, Emily and {Woods}, David and {Wyder}, Ted K. and {Yee}, H.~K.~C.},
        title = "{The WiggleZ Dark Energy Survey: Final data release and cosmological results}",
      journal = {\prd},
         year = 2012,
        month = nov,
       volume = {86},
       number = {10},
          eid = {103518},
        pages = {103518},
          doi = {10.1103/PhysRevD.86.103518},
archivePrefix = {arXiv},
       eprint = {1210.2130},
 primaryClass = {astro-ph.CO},
       adsurl = {https://ui.adsabs.harvard.edu/abs/2012PhRvD..86j3518P}
}

@ARTICLE{Rezaei2023,
       author = {{Rezaei}, Zeinab},
        title = "{Fuzzy dark matter in relativistic stars}",
      journal = {\mnras},
         year = 2023,
        month = sep,
       volume = {524},
       number = {2},
        pages = {2015-2024},
          doi = {10.1093/mnras/stad1975},
archivePrefix = {arXiv},
       eprint = {2306.17665},
 primaryClass = {astro-ph.HE},
       adsurl = {https://ui.adsabs.harvard.edu/abs/2023MNRAS.524.2015R}
}

@ARTICLE{Banks2023,
       author = {{Banks}, Hannah and {Grabowska}, Dorota M. and {McCullough}, Matthew},
        title = "{Gravitational wave backgrounds from colliding exotic compact objects}",
      journal = {\prd},
         year = 2023,
        month = aug,
       volume = {108},
       number = {3},
          eid = {035017},
        pages = {035017},
          doi = {10.1103/PhysRevD.108.035017},
archivePrefix = {arXiv},
       eprint = {2302.07887},
 primaryClass = {gr-qc},
       adsurl = {https://ui.adsabs.harvard.edu/abs/2023PhRvD.108c5017B}
}

@ARTICLE{Urbano2019,
       author = {{Urbano}, Alfredo and {Veerm{\"a}e}, Hardi},
        title = "{On gravitational echoes from ultracompact exotic stars}",
      journal = {\jcap},
         year = 2019,
        month = apr,
       volume = {2019},
       number = {4},
          eid = {011},
        pages = {011},
          doi = {10.1088/1475-7516/2019/04/011},
archivePrefix = {arXiv},
       eprint = {1810.07137},
 primaryClass = {gr-qc},
       adsurl = {https://ui.adsabs.harvard.edu/abs/2019JCAP...04..011U}
}

@ARTICLE{Karkevandi2022,
       author = {{Rafiei Karkevandi}, Davood and {Shakeri}, Soroush and {Sagun}, Violetta and {Ivanytskyi}, Oleksii},
        title = "{Bosonic dark matter in neutron stars and its effect on gravitational wave signal}",
      journal = {\prd},
         year = 2022,
        month = jan,
       volume = {105},
       number = {2},
          eid = {023001},
        pages = {023001},
          doi = {10.1103/PhysRevD.105.023001},
archivePrefix = {arXiv},
       eprint = {2109.03801},
 primaryClass = {astro-ph.HE},
       adsurl = {https://ui.adsabs.harvard.edu/abs/2022PhRvD.105b3001R}
}

@ARTICLE{Capano2023,
       author = {{Capano}, Collin D. and {Cabero}, Miriam and {Westerweck}, Julian and {Abedi}, Jahed and {Kastha}, Shilpa and {Nitz}, Alexander H. and {Wang}, Yi-Fan and {Nielsen}, Alex B. and {Krishnan}, Badri},
        title = "{Multimode Quasinormal Spectrum from a Perturbed Black Hole}",
      journal = {\prl},
         year = 2023,
        month = dec,
       volume = {131},
       number = {22},
          eid = {221402},
        pages = {221402},
          doi = {10.1103/PhysRevLett.131.221402},
archivePrefix = {arXiv},
       eprint = {2105.05238},
 primaryClass = {gr-qc},
       adsurl = {https://ui.adsabs.harvard.edu/abs/2023PhRvL.131v1402C}
}

@ARTICLE{Bustillo2021,
       author = {{Bustillo}, Juan Calder{\'o}n and {Sanchis-Gual}, Nicolas and {Torres-Forn{\'e}}, Alejandro and {Font}, Jos{\'e} A. and {Vajpeyi}, Avi and {Smith}, Rory and {Herdeiro}, Carlos and {Radu}, Eugen and {Leong}, Samson H.~W.},
        title = "{GW190521 as a Merger of Proca Stars: A Potential New Vector Boson of 8.7 {\texttimes}10$^{-13}$ eV}",
      journal = {\prl},
         year = 2021,
        month = feb,
       volume = {126},
       number = {8},
          eid = {081101},
        pages = {081101},
          doi = {10.1103/PhysRevLett.126.081101},
archivePrefix = {arXiv},
       eprint = {2009.05376},
 primaryClass = {gr-qc},
       adsurl = {https://ui.adsabs.harvard.edu/abs/2021PhRvL.126h1101B}
}

@ARTICLE{Sakstein2020,
       author = {{Sakstein}, Jeremy and {Croon}, Djuna and {McDermott}, Samuel D. and {Straight}, Maria C. and {Baxter}, Eric J.},
        title = "{Beyond the Standard Model Explanations of GW190521}",
      journal = {\prl},
         year = 2020,
        month = dec,
       volume = {125},
       number = {26},
          eid = {261105},
        pages = {261105},
          doi = {10.1103/PhysRevLett.125.261105},
archivePrefix = {arXiv},
       eprint = {2009.01213},
 primaryClass = {gr-qc},
       adsurl = {https://ui.adsabs.harvard.edu/abs/2020PhRvL.125z1105S}
}

@ARTICLE{Abbott2020,
       author = {{Abbott}, R. and {Abbott}, T.~D. and {Abraham}, S. and {Acernese}, F. and {Ackley}, K. and {Adams}, C. and {Adhikari}, R.~X. and {Adya}, V.~B. and {Affeldt}, C. and {Agathos}, M. and {Agatsuma}, K. and {Aggarwal}, N. and {Aguiar}, O.~D. and {Aich}, A. and {Aiello}, L. and {Ain}, A. and {Ajith}, P. and {Akcay}, S. and {Allen}, G. and {Allocca}, A. and {Altin}, P.~A. and {Amato}, A. and {Anand}, S. and {Ananyeva}, A. and {Anderson}, S.~B. and {Anderson}, W.~G. and {Angelova}, S.~V. and {Ansoldi}, S. and {Antier}, S. and {Appert}, S. and {Arai}, K. and {Araya}, M.~C. and {Areeda}, J.~S. and {Ar{\`e}ne}, M. and {Arnaud}, N. and {Aronson}, S.~M. and {Arun}, K.~G. and {Asali}, Y. and {Ascenzi}, S. and {Ashton}, G. and {Aston}, S.~M. and {Astone}, P. and {Aubin}, F. and {Aufmuth}, P. and {AultONeal}, K. and {Austin}, C. and {Avendano}, V. and {LIGO Scientific Collaboration} and {Virgo Collaboration}},
        title = "{GW190521: A Binary Black Hole Merger with a Total Mass of 150 M$_{{\ensuremath{\odot}}}$}",
      journal = {\prl},
         year = 2020,
        month = sep,
       volume = {125},
       number = {10},
          eid = {101102},
        pages = {101102},
          doi = {10.1103/PhysRevLett.125.101102},
archivePrefix = {arXiv},
       eprint = {2009.01075},
 primaryClass = {gr-qc},
       adsurl = {https://ui.adsabs.harvard.edu/abs/2020PhRvL.125j1102A}
}

@ARTICLE{Pacilio2020,
       author = {{Pacilio}, Costantino and {Vaglio}, Massimo and {Maselli}, Andrea and {Pani}, Paolo},
        title = "{Gravitational-wave detectors as particle-physics laboratories: Constraining scalar interactions with a coherent inspiral model of boson-star binaries}",
      journal = {\prd},
         year = 2020,
        month = oct,
       volume = {102},
       number = {8},
          eid = {083002},
        pages = {083002},
          doi = {10.1103/PhysRevD.102.083002},
archivePrefix = {arXiv},
       eprint = {2007.05264},
 primaryClass = {gr-qc},
       adsurl = {https://ui.adsabs.harvard.edu/abs/2020PhRvD.102h3002P}
}

@ARTICLE{Maselli2017,
       author = {{Maselli}, Andrea and {Pnigouras}, Pantelis and {Nielsen}, Niklas Gr{\o}nlund and {Kouvaris}, Chris and {Kokkotas}, Kostas D.},
        title = "{Dark stars: Gravitational and electromagnetic observables}",
      journal = {\prd},
         year = 2017,
        month = jul,
       volume = {96},
       number = {2},
          eid = {023005},
        pages = {023005},
          doi = {10.1103/PhysRevD.96.023005},
archivePrefix = {arXiv},
       eprint = {1704.07286},
 primaryClass = {astro-ph.HE},
       adsurl = {https://ui.adsabs.harvard.edu/abs/2017PhRvD..96b3005M}
}

@ARTICLE{Seidel1994,
       author = {{Seidel}, Edward and {Suen}, Wai-Mo},
        title = "{Formation of solitonic stars through gravitational cooling}",
      journal = {\prl},
         year = 1994,
        month = apr,
       volume = {72},
       number = {16},
        pages = {2516-2519},
          doi = {10.1103/PhysRevLett.72.2516},
archivePrefix = {arXiv},
       eprint = {gr-qc/9309015},
 primaryClass = {gr-qc},
       adsurl = {https://ui.adsabs.harvard.edu/abs/1994PhRvL..72.2516S}
}

@ARTICLE{Bezares2018,
       author = {{Bezares}, Miguel and {Palenzuela}, Carlos},
        title = "{Gravitational waves from dark boson star binary mergers}",
      journal = {Classical and Quantum Gravity},
         year = 2018,
        month = dec,
       volume = {35},
       number = {23},
          eid = {234002},
        pages = {234002},
          doi = {10.1088/1361-6382/aae87c},
archivePrefix = {arXiv},
       eprint = {1808.10732},
 primaryClass = {gr-qc},
       adsurl = {https://ui.adsabs.harvard.edu/abs/2018CQGra..35w4002B}
}

@ARTICLE{Cardoso2017,
       author = {{Cardoso}, Vitor and {Franzin}, Edgardo and {Maselli}, Andrea and {Pani}, Paolo and {Raposo}, Guilherme},
        title = "{Testing strong-field gravity with tidal Love numbers}",
      journal = {\prd},
         year = 2017,
        month = apr,
       volume = {95},
       number = {8},
          eid = {084014},
        pages = {084014},
          doi = {10.1103/PhysRevD.95.084014},
archivePrefix = {arXiv},
       eprint = {1701.01116},
 primaryClass = {gr-qc},
       adsurl = {https://ui.adsabs.harvard.edu/abs/2017PhRvD..95h4014C}
}

@ARTICLE{Benhar2004,
       author = {{Benhar}, Omar and {Ferrari}, Valeria and {Gualtieri}, Leonardo},
        title = "{Gravitational wave asteroseismology reexamined}",
      journal = {\prd},
         year = 2004,
        month = dec,
       volume = {70},
       number = {12},
          eid = {124015},
        pages = {124015},
          doi = {10.1103/PhysRevD.70.124015},
archivePrefix = {arXiv},
       eprint = {astro-ph/0407529},
 primaryClass = {astro-ph},
       adsurl = {https://ui.adsabs.harvard.edu/abs/2004PhRvD..70l4015B}
}

@ARTICLE{Ho2020,
       author = {{Ho}, Wynn C.~G. and {Jones}, D.~I. and {Andersson}, Nils and {Espinoza}, Crist{\'o}bal M.},
        title = "{Gravitational waves from transient neutron star f -mode oscillations}",
      journal = {\prd},
         year = 2020,
        month = may,
       volume = {101},
       number = {10},
          eid = {103009},
        pages = {103009},
          doi = {10.1103/PhysRevD.101.103009},
archivePrefix = {arXiv},
       eprint = {2003.12082},
 primaryClass = {gr-qc},
       adsurl = {https://ui.adsabs.harvard.edu/abs/2020PhRvD.101j3009H}
}

@ARTICLE{PradhanPathak2023,
       author = {{Pradhan}, Bikram Keshari and {Pathak}, Dhruv and {Chatterjee}, Debarati},
        title = "{Constraining Nuclear Parameters Using Gravitational Waves from f-mode Oscillations in Neutron Stars}",
      journal = {\apj},
         year = 2023,
        month = oct,
       volume = {956},
       number = {1},
          eid = {38},
        pages = {38},
          doi = {10.3847/1538-4357/acef1f},
archivePrefix = {arXiv},
       eprint = {2306.04626},
 primaryClass = {astro-ph.HE},
       adsurl = {https://ui.adsabs.harvard.edu/abs/2023ApJ...956...38P}
}

@ARTICLE{Olivares2020,
       author = {{Olivares}, Hector and {Younsi}, Ziri and {Fromm}, Christian M. and {De Laurentis}, Mariafelicia and {Porth}, Oliver and {Mizuno}, Yosuke and {Falcke}, Heino and {Kramer}, Michael and {Rezzolla}, Luciano},
        title = "{How to tell an accreting boson star from a black hole}",
      journal = {\mnras},
         year = 2020,
        month = sep,
       volume = {497},
       number = {1},
        pages = {521-535},
          doi = {10.1093/mnras/staa1878},
archivePrefix = {arXiv},
       eprint = {1809.08682},
 primaryClass = {gr-qc},
       adsurl = {https://ui.adsabs.harvard.edu/abs/2020MNRAS.497..521O}
}

@ARTICLE{Sanchis-Gual2019,
       author = {{Sanchis-Gual}, N. and {Di Giovanni}, F. and {Zilh{\~a}o}, M. and {Herdeiro}, C. and {Cerd{\'a}-Dur{\'a}n}, P. and {Font}, J.~A. and {Radu}, E.},
        title = "{Nonlinear Dynamics of Spinning Bosonic Stars: Formation and Stability}",
      journal = {\prl},
         year = 2019,
        month = nov,
       volume = {123},
       number = {22},
          eid = {221101},
        pages = {221101},
          doi = {10.1103/PhysRevLett.123.221101},
archivePrefix = {arXiv},
       eprint = {1907.12565},
 primaryClass = {gr-qc},
       adsurl = {https://ui.adsabs.harvard.edu/abs/2019PhRvL.123v1101S}
}

@ARTICLE{Rogers2021,
       author = {{Rogers}, Keir K. and {Peiris}, Hiranya V.},
        title = "{Strong Bound on Canonical Ultralight Axion Dark Matter from the Lyman-Alpha Forest}",
      journal = {\prl},
         year = 2021,
        month = feb,
       volume = {126},
       number = {7},
          eid = {071302},
        pages = {071302},
          doi = {10.1103/PhysRevLett.126.071302},
archivePrefix = {arXiv},
       eprint = {2007.12705},
 primaryClass = {astro-ph.CO},
       adsurl = {https://ui.adsabs.harvard.edu/abs/2021PhRvL.126g1302R}
}

@ARTICLE{Nadler2021,
       author = {{Nadler}, E.~O. and {Drlica-Wagner}, A. and {Bechtol}, K. and {Mau}, S. and {Wechsler}, R.~H. and {Gluscevic}, V. and {Boddy}, K. and {Pace}, A.~B. and {Li}, T.~S. and {McNanna}, M. and {Riley}, A.~H. and {Garc{\'\i}a-Bellido}, J. and {Mao}, Y. -Y. and {Green}, G. and {Burke}, D.~L. and {Peter}, A. and {Jain}, B. and {Abbott}, T.~M.~C. and {Aguena}, M. and {Allam}, S. and {Annis}, J. and {Avila}, S. and {Brooks}, D. and {Carrasco Kind}, M. and {Carretero}, J. and {Costanzi}, M. and {da Costa}, L.~N. and {De Vicente}, J. and {Desai}, S. and {Diehl}, H.~T. and {Doel}, P. and {Everett}, S. and {Evrard}, A.~E. and {Flaugher}, B. and {Frieman}, J. and {Gerdes}, D.~W. and {Gruen}, D. and {Gruendl}, R.~A. and {Gschwend}, J. and {Gutierrez}, G. and {Hinton}, S.~R. and {Honscheid}, K. and {Huterer}, D. and {James}, D.~J. and {Krause}, E. and {Kuehn}, K. and {Kuropatkin}, N. and {Lahav}, O. and {Maia}, M.~A.~G. and {Marshall}, J.~L. and {Menanteau}, F. and {Miquel}, R. and {Palmese}, A. and {Paz-Chinch{\'o}n}, F. and {Plazas}, A.~A. and {Romer}, A.~K. and {Sanchez}, E. and {Scarpine}, V. and {Serrano}, S. and {Sevilla-Noarbe}, I. and {Smith}, M. and {Soares-Santos}, M. and {Suchyta}, E. and {Swanson}, M.~E.~C. and {Tarle}, G. and {Tucker}, D.~L. and {Walker}, A.~R. and {Wester}, W. and {DES Collaboration}},
        title = "{Constraints on Dark Matter Properties from Observations of Milky Way Satellite Galaxies}",
      journal = {\prl},
         year = 2021,
        month = mar,
       volume = {126},
       number = {9},
          eid = {091101},
        pages = {091101},
          doi = {10.1103/PhysRevLett.126.091101},
archivePrefix = {arXiv},
       eprint = {2008.00022},
 primaryClass = {astro-ph.CO},
       adsurl = {https://ui.adsabs.harvard.edu/abs/2021PhRvL.126i1101N}
}

@ARTICLE{Winch2024,
       author = {{Winch}, Harrison and {Rogers}, Keir K. and {Hlo{\v{z}}ek}, Ren{\'e}e and {Marsh}, David J.~E.},
        title = "{High-redshift, small-scale tests of ultralight axion dark matter using Hubble and Webb galaxy UV luminosities}",
      journal = {arXiv e-prints},
         year = 2024,
        month = apr,
          eid = {arXiv:2404.11071},
        pages = {arXiv:2404.11071},
          doi = {10.48550/arXiv.2404.11071},
archivePrefix = {arXiv},
       eprint = {2404.11071},
 primaryClass = {astro-ph.CO},
       adsurl = {https://ui.adsabs.harvard.edu/abs/2024arXiv240411071W}
}

@ARTICLE{Zimmerman2024,
       author = {{Zimmermann}, Tim and {Alvey}, James and {Marsh}, David J.~E. and {Fairbairn}, Malcolm and {Read}, Justin I.},
        title = "{Dwarf galaxies imply dark matter is heavier than $\mathbf{2.2 \times 10^{-21}} \, \mathbf{eV}$}",
      journal = {arXiv e-prints},
         year = 2024,
        month = may,
          eid = {arXiv:2405.20374},
        pages = {arXiv:2405.20374},
          doi = {10.48550/arXiv.2405.20374},
archivePrefix = {arXiv},
       eprint = {2405.20374},
 primaryClass = {astro-ph.CO},
       adsurl = {https://ui.adsabs.harvard.edu/abs/2024arXiv240520374Z}
}

@ARTICLE{Afzal2023Nanograv,
       author = {{Afzal}, Adeela and {Agazie}, Gabriella and {Anumarlapudi}, Akash and {Archibald}, Anne M. and {Arzoumanian}, Zaven and {Baker}, Paul T. and {B{\'e}csy}, Bence and {Blanco-Pillado}, Jose Juan and {Blecha}, Laura and {Boddy}, Kimberly K. and {Brazier}, Adam and {Brook}, Paul R. and {Burke-Spolaor}, Sarah and {Nanograv Collaboration}},
        title = "{The NANOGrav 15 yr Data Set: Search for Signals from New Physics}",
      journal = {\apjl},
         year = 2023,
        month = jul,
       volume = {951},
       number = {1},
          eid = {L11},
        pages = {L11},
          doi = {10.3847/2041-8213/acdc91},
archivePrefix = {arXiv},
       eprint = {2306.16219},
 primaryClass = {astro-ph.HE},
       adsurl = {https://ui.adsabs.harvard.edu/abs/2023ApJ...951L..11A}
}

@ARTICLE{Smarra2023,
       author = {{Smarra}, Clemente and {Goncharov}, Boris and {Barausse}, Enrico and {Antoniadis}, J. and {Babak}, S. and {Nielsen}, A. -S. Bak and {Bassa}, C.~G. and {Berthereau}, A. and {Bonetti}, M. and {Bortolas}, E. and {Brook}, P.~R. and {Burgay}, M. and {Caballero}, R.~N. and {Chalumeau}, A. and {Champion}, D.~J. and {Chanlaridis}, S. and {Chen}, S. and {Cognard}, I. and {European Pulsar Timing Array}},
        title = "{Second Data Release from the European Pulsar Timing Array: Challenging the Ultralight Dark Matter Paradigm}",
      journal = {\prl},
         year = 2023,
        month = oct,
       volume = {131},
       number = {17},
          eid = {171001},
        pages = {171001},
          doi = {10.1103/PhysRevLett.131.171001},
archivePrefix = {arXiv},
       eprint = {2306.16228},
 primaryClass = {astro-ph.HE},
       adsurl = {https://ui.adsabs.harvard.edu/abs/2023PhRvL.131q1001S}
}

@ARTICLE{Ryan1997,
       author = {{Ryan}, Fintan D.},
        title = "{Spinning boson stars with large self-interaction}",
      journal = {\prd},
         year = 1997,
        month = may,
       volume = {55},
       number = {10},
        pages = {6081-6091},
          doi = {10.1103/PhysRevD.55.6081},
       adsurl = {https://ui.adsabs.harvard.edu/abs/1997PhRvD..55.6081R}
}

@ARTICLE{Vaglio2023,
       author = {{Vaglio}, Massimo and {Pacilio}, Costantino and {Maselli}, Andrea and {Pani}, Paolo},
        title = "{Bayesian parameter estimation on boson-star binary signals with a coherent inspiral template and spin-dependent quadrupolar corrections}",
      journal = {\prd},
         year = 2023,
        month = jul,
       volume = {108},
       number = {2},
          eid = {023021},
        pages = {023021},
          doi = {10.1103/PhysRevD.108.023021},
archivePrefix = {arXiv},
       eprint = {2302.13954},
 primaryClass = {gr-qc},
       adsurl = {https://ui.adsabs.harvard.edu/abs/2023PhRvD.108b3021V}
}

@ARTICLE{Yang2024,
       author = {{Yang}, Jianzhi and {Cunha}, Pedro V.~P. and {Herdeiro}, Carlos A.~R.},
        title = "{Analytical proxy to families of numerical solutions: the case study of spherical mini-boson stars}",
      journal = {\jcap},
         year = 2024,
        month = aug,
       volume = {2024},
       number = {8},
          eid = {055},
        pages = {055},
          doi = {10.1088/1475-7516/2024/08/055},
archivePrefix = {arXiv},
       eprint = {2405.15651},
 primaryClass = {gr-qc},
       adsurl = {https://ui.adsabs.harvard.edu/abs/2024JCAP...08..055Y}
}

@ARTICLE{Yagi2017,
       author = {{Yagi}, Kent and {Yunes}, Nicol{\'a}s},
        title = "{Approximate universal relations for neutron stars and quark stars}",
      journal = {\physrep},
         year = 2017,
        month = apr,
       volume = {681},
        pages = {1-72},
          doi = {10.1016/j.physrep.2017.03.002},
archivePrefix = {arXiv},
       eprint = {1608.02582},
 primaryClass = {gr-qc},
       adsurl = {https://ui.adsabs.harvard.edu/abs/2017PhR...681....1Y}
}

@ARTICLE{Kokkotas2001,
       author = {{Kokkotas}, K.~D. and {Apostolatos}, T.~A. and {Andersson}, N.},
        title = "{The inverse problem for pulsating neutron stars: a `fingerprint analysis' for the supranuclear equation of state}",
      journal = {\mnras},
         year = 2001,
        month = jan,
       volume = {320},
       number = {3},
        pages = {307-315},
          doi = {10.1046/j.1365-8711.2001.03945.x},
archivePrefix = {arXiv},
       eprint = {gr-qc/9901072},
 primaryClass = {gr-qc},
       adsurl = {https://ui.adsabs.harvard.edu/abs/2001MNRAS.320..307K}
}

@ARTICLE{Echeverria1989,
       author = {{Echeverria}, Fernando},
        title = "{Gravitational-wave measurements of the mass and angular momentum of a black hole}",
      journal = {\prd},
         year = 1989,
        month = nov,
       volume = {40},
       number = {10},
        pages = {3194-3203},
          doi = {10.1103/PhysRevD.40.3194},
       adsurl = {https://ui.adsabs.harvard.edu/abs/1989PhRvD..40.3194E}
}

@ARTICLE{Robson2019,
       author = {{Robson}, Travis and {Cornish}, Neil J. and {Liu}, Chang},
        title = "{The construction and use of LISA sensitivity curves}",
      journal = {Classical and Quantum Gravity},
         year = 2019,
        month = may,
       volume = {36},
       number = {10},
          eid = {105011},
        pages = {105011},
          doi = {10.1088/1361-6382/ab1101},
archivePrefix = {arXiv},
       eprint = {1803.01944},
 primaryClass = {astro-ph.HE},
       adsurl = {https://ui.adsabs.harvard.edu/abs/2019CQGra..36j5011R}
}

@ARTICLE{GWDetection2016,
       author = {{Abbott}, B.~P. and {Abbott}, R. and {Abbott}, T.~D. and {Abernathy}, M.~R. and {Acernese}, F. and {Ackley}, K. and {Adams}, C. and {Adams}, T. and {Addesso}, P. and {Adhikari}, R.~X. and {Adya}, V.~B. and {Affeldt}, C. and {Agathos}, M. and {Agatsuma}, K. and {Aggarwal}, N. and {Aguiar}, O.~D. and {Aiello}, L. and {Ain}, A. and {Ajith}, P. and {Allen}, B. and {Allocca}, A. and {Altin}, P.~A. and {Anderson}, S.~B. and {Anderson}, W.~G. and {Arai}, K. and {Arain}, M.~A. and {Araya}, M.~C. and {Arceneaux}, C.~C. and {Areeda}, J.~S. and {Arnaud}, N. and {Arun}, K.~G. and {Ascenzi}, S. and {LIGO Scientific Collaboration} and {Virgo Collaboration}},
        title = "{Observation of Gravitational Waves from a Binary Black Hole Merger}",
      journal = {\prl},
         year = 2016,
        month = feb,
       volume = {116},
       number = {6},
          eid = {061102},
        pages = {061102},
          doi = {10.1103/PhysRevLett.116.061102},
archivePrefix = {arXiv},
       eprint = {1602.03837},
 primaryClass = {gr-qc},
       adsurl = {https://ui.adsabs.harvard.edu/abs/2016PhRvL.116f1102A}
}

@ARTICLE{Kain2021,
       author = {{Kain}, Ben},
        title = "{Boson stars and their radial oscillations}",
      journal = {\prd},
         year = 2021,
        month = jun,
       volume = {103},
       number = {12},
          eid = {123003},
        pages = {123003},
          doi = {10.1103/PhysRevD.103.123003},
archivePrefix = {arXiv},
       eprint = {2106.01740},
 primaryClass = {gr-qc},
       adsurl = {https://ui.adsabs.harvard.edu/abs/2021PhRvD.103l3003K}
}

@ARTICLE{Ralegankar2024,
       author = {{Ralegankar}, Pranjal and {Perri}, Daniele and {Kobayashi}, Takeshi},
        title = "{Gravothermalizing into primordial black holes, boson stars, and cannibal stars}",
      journal = {arXiv e-prints},
         year = 2024,
        month = oct,
          eid = {arXiv:2410.18948},
        pages = {arXiv:2410.18948},
          doi = {10.48550/arXiv.2410.18948},
archivePrefix = {arXiv},
       eprint = {2410.18948},
 primaryClass = {astro-ph.CO},
       adsurl = {https://ui.adsabs.harvard.edu/abs/2024arXiv241018948R}
}

@ARTICLE{Wheeler1955,
       author = {{Wheeler}, John Archibald},
        title = "{Geons}",
      journal = {Physical Review},
         year = 1955,
        month = jan,
       volume = {97},
       number = {2},
        pages = {511-536},
          doi = {10.1103/PhysRev.97.511},
       adsurl = {https://ui.adsabs.harvard.edu/abs/1955PhRv...97..511W}
}

@ARTICLE{Rosa2024,
       author = {{Rosa}, Jo{\~a}o Lu{\'\i}s and {Pelle}, Joaqu{\'\i}n and {P{\'e}rez}, Daniela},
        title = "{Accretion disks and relativistic line broadening in boson star spacetimes}",
      journal = {\prd},
         year = 2024,
        month = oct,
       volume = {110},
       number = {8},
          eid = {084068},
        pages = {084068},
          doi = {10.1103/PhysRevD.110.084068},
archivePrefix = {arXiv},
       eprint = {2403.11540},
 primaryClass = {gr-qc},
       adsurl = {https://ui.adsabs.harvard.edu/abs/2024PhRvD.110h4068R}
}

@ARTICLE{Evstafyeva2024,
       author = {{Evstafyeva}, Tamara and {Sperhake}, Ulrich and {Romero-Shaw}, Isobel M. and {Agathos}, Michalis},
        title = "{Gravitational-Wave Data Analysis with High-Precision Numerical Relativity Simulations of Boson Star Mergers}",
      journal = {\prl},
         year = 2024,
        month = sep,
       volume = {133},
       number = {13},
          eid = {131401},
        pages = {131401},
          doi = {10.1103/PhysRevLett.133.131401},
archivePrefix = {arXiv},
       eprint = {2406.02715},
 primaryClass = {gr-qc},
       adsurl = {https://ui.adsabs.harvard.edu/abs/2024PhRvL.133m1401E}
}

@ARTICLE{Perot2021,
       author = {{Perot}, L. and {Chamel}, N.},
        title = "{Role of dense matter in tidal deformations of inspiralling neutron stars and in gravitational waveforms with unified equations of state}",
      journal = {\prc},
         year = 2021,
        month = feb,
       volume = {103},
       number = {2},
          eid = {025801},
        pages = {025801},
          doi = {10.1103/PhysRevC.103.025801},
archivePrefix = {arXiv},
       eprint = {2102.02004},
 primaryClass = {gr-qc},
       adsurl = {https://ui.adsabs.harvard.edu/abs/2021PhRvC.103b5801P}
}

@ARTICLE{M87EHT2019,
       author = {{Event Horizon Telescope Collaboration} and {Akiyama}, Kazunori and {Alberdi}, Antxon and {Alef}, Walter and {Asada}, Keiichi and {Azulay}, Rebecca and {Baczko}, Anne-Kathrin and {Ball}, David and {Balokovi{\'c}}, Mislav and {Barrett}, John and {Bintley}, Dan},
        title = "{First M87 Event Horizon Telescope Results. I. The Shadow of the Supermassive Black Hole}",
      journal = {\apjl},
         year = 2019,
        month = apr,
       volume = {875},
       number = {1},
          eid = {L1},
        pages = {L1},
          doi = {10.3847/2041-8213/ab0ec7},
archivePrefix = {arXiv},
       eprint = {1906.11238},
 primaryClass = {astro-ph.GA},
       adsurl = {https://ui.adsabs.harvard.edu/abs/2019ApJ...875L...1E}
}

@ARTICLE{SagittariusEHT2022,
       author = {{Event Horizon Telescope Collaboration} and {Akiyama}, Kazunori and {Alberdi}, Antxon and {Alef}, Walter and {Algaba}, Juan Carlos and {Anantua}, Richard and {Asada}, Keiichi and {Azulay}, Rebecca and {Bach}, Uwe and {Baczko}, Anne-Kathrin and {Ball}, David and {Balokovi{\'c}}, Mislav and {Barrett}, John and {Baub{\"o}ck}, Michi and {Benson}, Bradford A. and {Bintley}, Dan and {Blackburn}, Lindy and {Blundell}, Raymond and {Bouman}, Katherine L. and {Bower}, Geoffrey C.},
        title = "{First Sagittarius A* Event Horizon Telescope Results. I. The Shadow of the Supermassive Black Hole in the Center of the Milky Way}",
      journal = {\apjl},
         year = 2022,
        month = may,
       volume = {930},
       number = {2},
          eid = {L12},
        pages = {L12},
          doi = {10.3847/2041-8213/ac6674},
       adsurl = {https://ui.adsabs.harvard.edu/abs/2022ApJ...930L..12E}
}

@ARTICLE{Celato2025,
       author = {{Celato}, Mariachiara and {Kr{\"u}ger}, Christian J. and {Kokkotas}, Kostas D.},
        title = "{Probing dark star parameters through f-mode gravitational wave signals}",
      journal = {\prd},
         year = 2025,
        month = jan,
       volume = {111},
       number = {2},
          eid = {023034},
        pages = {023034},
          doi = {10.1103/PhysRevD.111.023034},
archivePrefix = {arXiv},
       eprint = {2501.12031},
 primaryClass = {gr-qc},
       adsurl = {https://ui.adsabs.harvard.edu/abs/2025PhRvD.111b3034C}
}

@ARTICLE{Wu2023,
       author = {{Wu}, Jing-Yi and {Li}, Wei and {Huang}, Xin-Han and {Zhang}, Kilar},
        title = "{Dark I-Love-Q}",
      journal = {arXiv e-prints},
         year = 2023,
        month = sep,
          eid = {arXiv:2309.07971},
        pages = {arXiv:2309.07971},
          doi = {10.48550/arXiv.2309.07971},
archivePrefix = {arXiv},
       eprint = {2309.07971},
 primaryClass = {astro-ph.HE},
       adsurl = {https://ui.adsabs.harvard.edu/abs/2023arXiv230907971W}
}

@ARTICLE{tiango2020,
       author = {{Kuns}, Kevin A. and {Yu}, Hang and {Chen}, Yanbei and {Adhikari}, Rana X.},
        title = "{Astrophysics and cosmology with a decihertz gravitational-wave detector: TianGO}",
      journal = {\prd},
         year = 2020,
        month = aug,
       volume = {102},
       number = {4},
          eid = {043001},
        pages = {043001},
          doi = {10.1103/PhysRevD.102.043001},
archivePrefix = {arXiv},
       eprint = {1908.06004},
 primaryClass = {gr-qc},
       adsurl = {https://ui.adsabs.harvard.edu/abs/2020PhRvD.102d3001K}
}

@ARTICLE{decigo2011,
       author = {{Kawamura}, Seiji and {Ando}, Masaki and {Seto}, Naoki and {Sato}, Shuichi and {Nakamura}, Takashi and {Tsubono}, Kimio and {Kanda}, Nobuyuki and {Tanaka}, Takahiro and {Yokoyama}, Jun'ichi and {Funaki}, Ikkoh and {Numata}, Kenji and {Ioka}, Kunihito and {Takashima}, Takeshi and {Agatsuma}, Kazuhiro and {Akutsu}, Tomotada and {Aoyanagi}, Koh-suke and {Arai}, Koji and {Araya}, Akito and {Asada}, Hideki and {Aso}, Yoichi and {Chen}, Dan and {Chiba}, Takeshi and {Ebisuzaki}, Toshikazu and {Ejiri}, Yumiko and {Enoki}, Motohiro and {Eriguchi}, Yoshiharu and {Fujimoto}, Masa-Katsu and {Fujita}, Ryuichi and {Fukushima}, Mitsuhiro and {Futamase}, Toshifumi and {Harada}, Tomohiro and {Hashimoto}, Tatsuaki and {Hayama}, Kazuhiro and {Hikida}, Wataru and {Himemoto}, Yoshiaki and {Hirabayashi}, Hisashi and {Hiramatsu}, Takashi and {Hong}, Feng-Lei and {Horisawa}, Hideyuki and {Hosokawa}, Mizuhiko and {Ichiki}, Kiyotomo and {Ikegami}, Takeshi and {Inoue}, Kaiki T. and {Ishidoshiro}, Koji and {Ishihara}, Hideki and {Ishikawa}, Takehiko and {Ishizaki}, Hideharu and {Ito}, Hiroyuki and {Itoh}, Yousuke and {Izumi}, Kiwamu and {Kawano}, Isao and {Kawashima}, Nobuki and {Kawazoe}, Fumiko and {Kishimoto}, Naoko and {Kiuchi}, Kenta and {Kobayashi}, Shiho and {Kohri}, Kazunori and {Koizumi}, Hiroyuki and {Kojima}, Yasufumi and {Kokeyama}, Keiko and {Kokuyama}, Wataru and {Kotake}, Kei and {Kozai}, Yoshihide and {Kunimori}, Hiroo and {Kuninaka}, Hitoshi and {Kuroda}, Kazuaki and {Kuroyanagi}, Sachiko and {Maeda}, Kei-ichi and {Matsuhara}, Hideo and {Matsumoto}, Nobuyuki and {Michimura}, Yuta and {Miyakawa}, Osamu and {Miyamoto}, Umpei and {Miyoki}, Shinji and {Morimoto}, Mutsuko Y. and {Morisawa}, Toshiyuki and {Moriwaki}, Shigenori and {Mukohyama}, Shinji and {Musha}, Mitsuru and {Nagano}, Shigeo and {Naito}, Isao and {Nakamura}, Kouji and {Nakano}, Hiroyuki and {Nakao}, Kenichi and {Nakasuka}, Shinichi and {Nakayama}, Yoshinori and {Nakazawa}, Kazuhiro and {Nishida}, Erina and {Nishiyama}, Kazutaka and {Nishizawa}, Atsushi and {Niwa}, Yoshito and {Noumi}, Taiga and {Obuchi}, Yoshiyuki and {Ohashi}, Masatake and {Ohishi}, Naoko and {Ohkawa}, Masashi and {Okada}, Kenshi and {Okada}, Norio and {Oohara}, Kenichi and {Sago}, Norichika and {Saijo}, Motoyuki and {Saito}, Ryo and {Sakagami}, Masaaki and {Sakai}, Shin-ichiro and {Sakata}, Shihori and {Sasaki}, Misao and {Sato}, Takashi and {Shibata}, Masaru and {Shinkai}, Hisaaki and {Shoda}, Ayaka and {Somiya}, Kentaro and {Sotani}, Hajime and {Sugiyama}, Naoshi and {Suwa}, Yudai and {Suzuki}, Rieko and {Tagoshi}, Hideyuki and {Takahashi}, Fuminobu and {Takahashi}, Kakeru and {Takahashi}, Keitaro and {Takahashi}, Ryutaro and {Takahashi}, Ryuichi and {Takahashi}, Tadayuki and {Takahashi}, Hirotaka and {Akiteru}, Takamori and {Takano}, Tadashi and {Tanaka}, Nobuyuki and {Taniguchi}, Keisuke and {Taruya}, Atsushi and {Tashiro}, Hiroyuki and {Torii}, Yasuo and {Toyoshima}, Morio and {Tsujikawa}, Shinji and {Tsunesada}, Yoshiki and {Ueda}, Akitoshi and {Ueda}, Ken-ichi and {Utashima}, Masayoshi and {Wakabayashi}, Yaka and {Yagi}, Kent and {Yamakawa}, Hiroshi and {Yamamoto}, Kazuhiro and {Yamazaki}, Toshitaka and {Yoo}, Chul-Moon and {Yoshida}, Shijun and {Yoshino}, Taizoh and {Sun}, Ke-Xun},
        title = "{The Japanese space gravitational wave antenna: DECIGO}",
      journal = {Classical and Quantum Gravity},
         year = 2011,
        month = may,
       volume = {28},
       number = {9},
          eid = {094011},
        pages = {094011},
          doi = {10.1088/0264-9381/28/9/094011},
       adsurl = {https://ui.adsabs.harvard.edu/abs/2011CQGra..28i4011K}
}

@ARTICLE{Teodori2025,
       author = {{Teodori}, Luca and {Caputo}, Andrea and {Blum}, Kfir},
        title = "{Ultra-Light Dark Matter Simulations and Stellar Dynamics: Tension in Dwarf Galaxies for $m < 5\times10^{-21} $ eV}",
      journal = {arXiv e-prints},
         year = 2025,
        month = jan,
          eid = {arXiv:2501.07631},
        pages = {arXiv:2501.07631},
          doi = {10.48550/arXiv.2501.07631},
archivePrefix = {arXiv},
       eprint = {2501.07631},
 primaryClass = {astro-ph.GA},
       adsurl = {https://ui.adsabs.harvard.edu/abs/2025arXiv250107631T}
}

@ARTICLE{Dalal2022,
       author = {{Dalal}, Neal and {Kravtsov}, Andrey},
        title = "{Excluding fuzzy dark matter with sizes and stellar kinematics of ultrafaint dwarf galaxies}",
      journal = {\prd},
         year = 2022,
        month = sep,
       volume = {106},
       number = {6},
          eid = {063517},
        pages = {063517},
          doi = {10.1103/PhysRevD.106.063517},
archivePrefix = {arXiv},
       eprint = {2203.05750},
 primaryClass = {astro-ph.CO},
       adsurl = {https://ui.adsabs.harvard.edu/abs/2022PhRvD.106f3517D}
}

@ARTICLE{Marsh2019,
       author = {{Marsh}, David J.~E. and {Niemeyer}, Jens C.},
        title = "{Strong Constraints on Fuzzy Dark Matter from Ultrafaint Dwarf Galaxy Eridanus II}",
      journal = {\prl},
         year = 2019,
        month = aug,
       volume = {123},
       number = {5},
          eid = {051103},
        pages = {051103},
          doi = {10.1103/PhysRevLett.123.051103},
archivePrefix = {arXiv},
       eprint = {1810.08543},
 primaryClass = {astro-ph.CO},
       adsurl = {https://ui.adsabs.harvard.edu/abs/2019PhRvL.123e1103M}
}

@ARTICLE{Benito2020,
       author = {{Benito}, Mar{\'\i}a and {Criado}, Juan Carlos and {H{\"u}tsi}, Gert and {Raidal}, Martti and {Veerm{\"a}e}, Hardi},
        title = "{Implications of Milky Way substructures for the nature of dark matter}",
      journal = {\prd},
         year = 2020,
        month = may,
       volume = {101},
       number = {10},
          eid = {103023},
        pages = {103023},
          doi = {10.1103/PhysRevD.101.103023},
archivePrefix = {arXiv},
       eprint = {2001.11013},
 primaryClass = {astro-ph.CO},
       adsurl = {https://ui.adsabs.harvard.edu/abs/2020PhRvD.101j3023B}
}

@ARTICLE{Powell2023,
       author = {{Powell}, Devon M. and {Vegetti}, Simona and {McKean}, J.~P. and {White}, Simon D.~M. and {Ferreira}, Elisa G.~M. and {May}, Simon and {Spingola}, Cristiana},
        title = "{A lensed radio jet at milli-arcsecond resolution - II. Constraints on fuzzy dark matter from an extended gravitational arc}",
      journal = {\mnras},
         year = 2023,
        month = sep,
       volume = {524},
       number = {1},
        pages = {L84-L88},
          doi = {10.1093/mnrasl/slad074},
archivePrefix = {arXiv},
       eprint = {2302.10941},
 primaryClass = {astro-ph.CO},
       adsurl = {https://ui.adsabs.harvard.edu/abs/2023MNRAS.524L..84P}
}

@ARTICLE{Nebrin2019,
       author = {{Nebrin}, Olof and {Ghara}, Raghunath and {Mellema}, Garrelt},
        title = "{Fuzzy dark matter at cosmic dawn: new 21-cm constraints}",
      journal = {\jcap},
         year = 2019,
        month = apr,
       volume = {2019},
       number = {4},
          eid = {051},
        pages = {051},
          doi = {10.1088/1475-7516/2019/04/051},
archivePrefix = {arXiv},
       eprint = {1812.09760},
 primaryClass = {astro-ph.CO},
       adsurl = {https://ui.adsabs.harvard.edu/abs/2019JCAP...04..051N}
}

@ARTICLE{Laroche2022,
       author = {{Laroche}, Alexander and {Gilman}, Daniel and {Li}, Xinyu and {Bovy}, Jo and {Du}, Xiaolong},
        title = "{Quantum fluctuations masquerade as haloes: bounds on ultra-light dark matter from quadruply imaged quasars}",
      journal = {\mnras},
         year = 2022,
        month = dec,
       volume = {517},
       number = {2},
        pages = {1867-1883},
          doi = {10.1093/mnras/stac2677},
archivePrefix = {arXiv},
       eprint = {2206.11269},
 primaryClass = {astro-ph.CO},
       adsurl = {https://ui.adsabs.harvard.edu/abs/2022MNRAS.517.1867L}
}

@ARTICLE{Kulkarni2022,
       author = {{Kulkarni}, Mihir and {Ostriker}, Jeremiah P.},
        title = "{What is the halo mass function in a fuzzy dark matter cosmology?}",
      journal = {\mnras},
         year = 2022,
        month = feb,
       volume = {510},
       number = {1},
        pages = {1425-1430},
          doi = {10.1093/mnras/stab3520},
archivePrefix = {arXiv},
       eprint = {2011.02116},
 primaryClass = {astro-ph.CO},
       adsurl = {https://ui.adsabs.harvard.edu/abs/2022MNRAS.510.1425K}
}

@ARTICLE{Stott2018,
       author = {{Stott}, Matthew J. and {Marsh}, David J.~E.},
        title = "{Black hole spin constraints on the mass spectrum and number of axionlike fields}",
      journal = {\prd},
         year = 2018,
        month = oct,
       volume = {98},
       number = {8},
          eid = {083006},
        pages = {083006},
          doi = {10.1103/PhysRevD.98.083006},
archivePrefix = {arXiv},
       eprint = {1805.02016},
 primaryClass = {hep-ph},
       adsurl = {https://ui.adsabs.harvard.edu/abs/2018PhRvD..98h3006S}
}

@ARTICLE{Davoudiasl2019,
       author = {{Davoudiasl}, Hooman and {Denton}, Peter B.},
        title = "{Ultralight Boson Dark Matter and Event Horizon Telescope Observations of M 87$^{*}$}",
      journal = {\prl},
         year = 2019,
        month = jul,
       volume = {123},
       number = {2},
          eid = {021102},
        pages = {021102},
          doi = {10.1103/PhysRevLett.123.021102},
archivePrefix = {arXiv},
       eprint = {1904.09242},
 primaryClass = {astro-ph.CO},
       adsurl = {https://ui.adsabs.harvard.edu/abs/2019PhRvL.123b1102D}
}

@ARTICLE{Hlozek2015,
       author = {{Hlozek}, Ren{\'e}e and {Grin}, Daniel and {Marsh}, David J.~E. and {Ferreira}, Pedro G.},
        title = "{A search for ultralight axions using precision cosmological data}",
      journal = {\prd},
         year = 2015,
        month = may,
       volume = {91},
       number = {10},
          eid = {103512},
        pages = {103512},
          doi = {10.1103/PhysRevD.91.103512},
archivePrefix = {arXiv},
       eprint = {1410.2896},
 primaryClass = {astro-ph.CO},
       adsurl = {https://ui.adsabs.harvard.edu/abs/2015PhRvD..91j3512H}
}

@ARTICLE{Hayashi2021,
       author = {{Hayashi}, Kohei and {Ferreira}, Elisa G.~M. and {Chan}, Hei Yin Jowett},
        title = "{Narrowing the Mass Range of Fuzzy Dark Matter with Ultrafaint Dwarfs}",
      journal = {\apjl},
         year = 2021,
        month = may,
       volume = {912},
       number = {1},
          eid = {L3},
        pages = {L3},
          doi = {10.3847/2041-8213/abf501},
archivePrefix = {arXiv},
       eprint = {2102.05300},
 primaryClass = {astro-ph.CO},
       adsurl = {https://ui.adsabs.harvard.edu/abs/2021ApJ...912L...3H}
}

@article{Benito2025,
    author = {Benito, Mar\'\i{}a and H\"utsi, Gert and M\"u\"ursepp, Kristjan and S\'anchez\textasciitilde{}Almeida, Jorge and Urrutia, Juan and Vaskonen, Ville and Veerm\"ae, Hardi},
    title = "{Fuzzy dark matter fails to explain the dark matter cores}",
    eprint = "2502.12030",
    archivePrefix = "arXiv",
    primaryClass = "astro-ph.CO",
    month = "2",
    year = "2025"
}

@ARTICLE{Rosa2022,
       author = {{Rosa}, Jo{\~a}o Lu{\'\i}s and {Rubiera-Garcia}, Diego},
        title = "{Shadows of boson and Proca stars with thin accretion disks}",
      journal = {\prd},
         year = 2022,
        month = oct,
       volume = {106},
       number = {8},
          eid = {084004},
        pages = {084004},
          doi = {10.1103/PhysRevD.106.084004},
archivePrefix = {arXiv},
       eprint = {2204.12949},
 primaryClass = {gr-qc},
       adsurl = {https://ui.adsabs.harvard.edu/abs/2022PhRvD.106h4004R}
}

@ARTICLE{Rosa2023,
       author = {{Rosa}, Jo{\~a}o Lu{\'\i}s and {Macedo}, Caio F.~B. and {Rubiera-Garcia}, Diego},
        title = "{Imaging compact boson stars with hot spots and thin accretion disks}",
      journal = {\prd},
         year = 2023,
        month = aug,
       volume = {108},
       number = {4},
          eid = {044021},
        pages = {044021},
          doi = {10.1103/PhysRevD.108.044021},
archivePrefix = {arXiv},
       eprint = {2303.17296},
 primaryClass = {gr-qc},
       adsurl = {https://ui.adsabs.harvard.edu/abs/2023PhRvD.108d4021R}
}

@ARTICLE{LiddleMadsen1992,
       author = {{Liddle}, Andrew R. and {Madsen}, Mark S.},
        title = "{The Structure and Formation of Boson Stars}",
      journal = {International Journal of Modern Physics D},
         year = 1992,
        month = jan,
       volume = {1},
       number = {1},
        pages = {101-143},
          doi = {10.1142/S0218271892000057},
       adsurl = {https://ui.adsabs.harvard.edu/abs/1992IJMPD...1..101L}
}

@ARTICLE{shirke_BS_fmodes_short,
       author = {{Shirke}, Swarnim and {Keshari Pradhan}, Bikram and {Chatterjee}, Debarati and {Sagunski}, Laura and {Schaffner-Bielich}, J{\"u}rgen},
        title = "{Fundamental Oscillations of Massive Boson Stars and Distinguishability}",
      journal = {arXiv e-prints},
         year = 2025,
        month = feb,
          eid = {arXiv:2502.04059},
        pages = {arXiv:2502.04059},
          doi = {10.48550/arXiv.2502.04059},
archivePrefix = {arXiv},
       eprint = {2502.04059},
 primaryClass = {gr-qc},
       adsurl = {https://ui.adsabs.harvard.edu/abs/2025arXiv250204059S}
}

\end{document}